\documentclass[12pt,preprint]{aastex}
\newcommand{\kms}{km~s$^{-1}$}

\shorttitle{Kinematics of the Globular Clusters in M31}
\shortauthors{Lee et al.}

\begin{document}

\title{Wide-Field Survey of Globular Clusters in M31. II. Kinematics of the Globular Cluster System\altaffilmark{1}}

\author{Myung Gyoon Lee\altaffilmark{2},
Ho Seong Hwang\altaffilmark{2,3}, Sang Chul Kim\altaffilmark{2,4}, 
Hong Soo Park\altaffilmark{2}, \\Doug Geisler\altaffilmark{5},
Ata Sarajedini\altaffilmark{6}, \& William E. Harris\altaffilmark{7}}

\altaffiltext{1}{Based on observations with the Kitt Peak National Observatory,
National Optical Astronomy Observatory, which is operated by the Association
of Universities for Research in Astronomy, Inc. (AURA) under cooperative
agreement with the National Science Foundation.}
\altaffiltext{2}{Astronomy Program, Department of Physics and Astronomy, Seoul National University,
   Seoul 151-742, Korea; mglee@astro.snu.ac.kr \& hspark@astro.snu.ac.kr}
\altaffiltext{3}{Current address: Korea Institute for Advanced Study,
Seoul, Korea; hshwang@kias.re.kr}
\altaffiltext{4}{Korea Astronomy and Space Science Institute, Daejeon 305-348,
   Korea; sckim@kasi.re.kr}
\altaffiltext{5}{Grupo de Astronomia, Departamento de Fisica, Universidad de
   Concepci\'{o}n, Casilla 160-C, Concepci\'{o}n, Chile; dgeisler@astro-udec.cl}
\altaffiltext{6}{Department of Astronomy, University of Florida,
   Gainesville, FL 32611, ata@astro.ufl.edu}
\altaffiltext{7}{Department of Physics \& Astronomy, McMaster University,
   Hamilton ON L8S 4M1, Canada; harris@physics.mcmaster.ca}


\begin{abstract}
We present a kinematic analysis of the globular cluster (GC)
system in M31. 
Adding new velocity data for 150 GCs in our wide-field
survey to those in the literature, 
we increase the number of M31 GCs with measured velocities
by 42 per cent, to 504. Using the photometric and spectroscopic
database of these 504 GCs, we have investigated
the kinematics of the M31 GC system. We find that the all GC system
shows strong rotation, with rotation amplitude of $v_{\rm rot}\sim190$ km s$^{-1}$,
and that a weak rotation persists even for the outermost samples at $|Y|\ge 5$
kpc where $Y$ represents the projected distance from the major axis. 
The rotation-corrected velocity dispersion for the GC system
is estimated to be $\sigma_{p,r}\sim130$ km s$^{-1}$, and it increases from 
$\sigma_{p,r}\sim120$ km s$^{-1}$ at  $|Y|< 1$ kpc to 
$\sigma_{p,r}\sim150$ km s$^{-1}$ at  $|Y|\ge 5$ kpc. These results are very similar
to those for the metal-poor GCs. 
This shows that there is a dynamically hot halo in M31
that is rotating but primarily pressure-supported. 
We have identified 50 
``friendless" GCs, and they appear to rotate around the major axis of
M31, unlike M31's disk rotation around the minor axis. For the
subsamples of metal-poor and metal-rich GCs,
we have found that the metal-rich GCs are more centrally concentrated
than the metal-poor GCs,
and both subsamples show strong rotation.
For the subsamples of
bright and faint GCs, it is found that the rotation for the faint GCs 
is stronger than that for the bright GCs. 
We have identified
56 GCs and GC candidates with X-ray detection including 39 genuine
GCs with measured velocities. It is found that the majority of
X-ray emitting GCs follow the disk rotation, and that the redder, more
metal-rich, and brighter GCs are more likely to be detected as
X-ray emitting GCs, as seen for GCs   in early-type galaxies.
We have derived a rotation curve of M31 using the GCs at $|Y|\le 0.6$ kpc,
and it agrees well for the range of $R=20-45\arcmin$ to that  based on other tracers
except for the planetary nebulae.
We have estimated the dynamical mass of M31 using `Projected Mass
Estimator (PME)' and `Tracer Mass Estimator (TME)' as $M_{\rm PME}=
5.5_{-0.3}^{+0.4} \times 10^{11} M_\odot$ out to a radius of $\sim55$ kpc
and $M_{\rm TME}= 19.2_{-1.3}^{+1.4} \times 10^{11} M_\odot$ for
a radius of $\sim100$ kpc, respectively. 
We finally discuss the implication of these results and compare the kinematics of
GCs with that of planetary nebulae in M31.

\end{abstract}



\section{Introduction}

Globular clusters (GCs) are the oldest stellar systems in the universe and thus
maintain a
fossil record of the early history of galaxies. 
They are much brighter than stars and the spatial distribution of the GCs 
is much more extended than that of the halo stars in galaxies. 
Therefore they are an excellent tool to investigate the kinematics of the main components
(stellar and dark matter halo, bulge, and disk) of galaxies and to estimate the mass of their host galaxies. 

M31 is a unique object as               the nearest spiral galaxy in the Local Group,
providing an excellent place to study GCs. It has the largest GC system of
any galaxy within several Mpc.
Therefore, there are numerous studies of GCs in M31 since the first study
by \citet{hub32} (e.g., \citep{van69,sar77,bat87,huc91,bar00,per02,mor04,bur04,hux05,bea05,puz05,gal06,kim07} and
see extensive lists of earlier references in \citet{hod92,van00}).
In addition, there were  several surveys and studies of the M31 GCs with X-ray
detection, finding that the X-ray emitting GCs are 
commonly associated with low-mass X-ray binaries
\citep{sup01,kong02,di02,kaa02,wil04,tru04,pie05,fan05,tru05,tru06}.

Kinematics is one of the notable features for the M31 GCs,
and has long been a focus of study.
It is known that the kinematics of the M31 GCs is much different from that
of the Galactic GCs in the sense 
that the former is dominated by rotation, while
the latter shows weak rotation \citep{per02} 

The first detailed kinematic analysis of the M31 GC system was 
given by \citet{har74} using radial velocities for
44 GCs in \citet{van69}. They estimated the total mass of M31,
M$_{\rm tot}=3.4\pm1.4\times10^{11} M_\odot$, and found that the
metal-poor GCs have a larger velocity dispersion than the metal-rich
GCs.
\citet{huc82} obtained digital, low-dispersion spectra of 61 GCs,
and found that the GC system rotates with an amplitude of $160\pm40$ km s$^{-1}$.
They also found  a measurable but small metallicity gradient and no
significant spectroscopic difference between the GCs with X-ray detection
and those without X-ray detection.
\citet{fed90} using the velocity data of 81 GCs
obtained a total mass of $3.2\times 10^{11} M_\odot$ for a radius of 16 kpc
using a projected mass estimator. A comparison of the data with the model of 
\citet{ric84} showed that the GCs have circular or isotropic orbits, and
lead to a total mass of $\sim 7\times 10^{11} M_\odot$ based on circular orbits.
Later, \citet{fed93} updated the total mass of M31 to be
$5-8\times 10^{11} M_\odot$ for a radius of 30 kpc
using velocity data for 176 GCs.

In the 1990's, metallicity information for a large sample of GCs ($N>100$) became available,
and it was possible to investigate the kinematics of GC subsamples
based on the metallicity.
\citet{huc91} estimated the radial velocities and metallicities for 150 GCs, and presented
the detailed kinematics of the GC system using   expanded subsamples of GCs based on metallicity.
They identified a weak metallicity gradient as a function of projected radius.
They found that, in the inner region ($<2$ kpc), the metal-rich ([Fe/H]$\ge-0.8$) GCs
exhibit a rapid rotation of $100-200$ km s$^{-1}$, while the metal-poor GCs
show no  significant rotation. However, the rotation of the two subsamples
could  not be easily measured at larger radii.
Later, \citet{per02} showed, using higher precision data of over 200 GCs
( typical velocity measurement error of 12 km s$^{-1}$), that the
entire M31 GC system is highly rotation-supported.  
This is in strong contrast to the case of the Milky Way where the GC system composed mostly of halo GCs is pressure-supported, although a minor system of the bulge GCs shows some rotation.
They also found that the centrally concentrated
metal-rich GCs appear not to be flattened, but are   analogous to
the bulge population with a strong rotation, and that the less concentrated
metal-poor GCs also show a rapid rotation.
\citet{gal06} obtained velocity data for 42 genuine GCs, and produced
a large data set of 349 M31 GCs with up-to-date velocities
by combining all information in the literature.

Recently, large datasets of velocities for M31 GCs and several deep photometric studies
 of the halo fields in M31 have provided important clues
for understanding the evolution of M31. 
\citet{bro03} presented very deep photometry of stars in a halo field along the minor axis of M31 at a projected distance of 11 kpc (corresponding to a deprojected distance of 53 kpc) and found that the M31 halo includes a major intermediate-age metal-rich population and a significant old metal-poor population. They suggested from this result that M31 may have undergone a major merger $6-8$ Gyr ago. Later \citet{bro07} suggested, from photometry
of another minor-axis halo field at a projected distance of 21 kpc, that the outer spheroid is dominated by the debris of earlier merging events that might have occurred more than 8 Gyr ago.

\citet{mor04} suggested the existence of an old thin disk composed of GCs in M31, using the data for 321 GCs presented in \citet{per02}. Since the GCs in the thin disk have a metallicity distribution from [Fe/H] $\sim-2.2$ to
above solar, and the metal-poor GCs are regarded as being as old as their
Galactic counterparts,
they concluded that the thin disk has remained undisturbed for $\sim10$ Gyr or more.
This is contradictory with 
 the scenario of \citet{bro03} that M31 may have undergone a major merger $6-8$ Gyr ago.
Later \citet{bea04,bea05} suggested that the majority of M31 thin disk GCs
may be significantly younger than 10 Gyr, up to $\sim 1$ Gyr and metal-rich ($-0.20\le$[Fe/H]$\le0.35$).
The existence of young and metal-rich GCs in the disk is also confirmed in other studies as well \citep{bur04,puz05}.

On the other hand \citet{ash93} first tried to identify subgroups in the M31 GC
system using the positional and velocity information for 144 GCs in
\citet{huc91}. They suggested that the clustering of GCs in 
position and velocity space may be a surviving signature of
gaseous clumps from which the galaxy halo of M31 formed. Later,
\citet{per03}, using 301 GCs with measured velocity and
metallicity in \citet{per02}, searched for the subclustering of M31
GCs in the parameter space of projected radius, radial velocity,
and metallicity. 
Interestingly \citet{mor03} found a possible tidally distorted
satellite of M31, Andromeda VIII, from the identification of a
group of $5-12$ planetary nebulae (PNe), $1-3$ GCs, and two HI clouds. 
Recent kinematic studies of M31 streams \citep{iba01,mcc04} started
examining the GC groups including Andromeda VIII as a possible
progenitor of the stream (e.g., \citealt{fon06a,far06a}).

Although there were numerous studies to obtain the radial velocities
and the metallicities for M31 GCs, there are still many GC candidates
for which spectroscopic information is needed. For example,
there are 569 GC candidates for which radial velocities are not available
in a combined sample of M31 GCs in \citet{gal06}.
Recently we have performed a homogeneous photometric and spectroscopic,
wide-field survey of GCs in M31 covering a field of $\sim$ 3\arcdeg $\times$ 3\arcdeg~
centered on M31. 
We presented the first result of our survey: a catalog of new GCs in M31 (\citealt{kim07}, Paper I). This is the second in the series of papers on
this survey.

In this paper, we present the results of a kinematic study of the GC
system in M31 using 
the velocity data of 150 GCs newly obtained in our survey
and 354 GCs previously known in the literature. Section 2
gives a brief description of the data used in this analysis, and
section 3 investigates the kinematic properties for various subsamples of M31 GCs.
We discuss primary results 
in \S 4. A summary is given in the final section.

We adopted a distance to M31  of $780$ kpc derived by \citet{mcc05}
based on the method of the tip of the red giant branch, which is similar
to the value given by \citet{fre90}, 770 kpc. 
At this distance one arcmin in the sky corresponds to 0.227 kpc.
We adopted the center position of M31 as 
$\alpha_0=00^{\rm h}42^{\rm m}44.^{\rm s}3$,
$\delta_0=+41^\circ16\arcmin09\arcsec$ (J2000; \citealt{cra92}).
Following the geometric transformation of \citet{huc91},
  we define the $X$ and $Y$ coordinates,
  which are the projected distances along the major and minor axis, respectively.
$X$ and $Y$ increase towards the northeast and the northwest, respectively,
  and are given by,
\begin{eqnarray}
x^\prime=\sin{(\alpha - \alpha_0)} \cos{\delta}, \\
y^\prime=\sin{\delta}\cos{\delta_0} - \cos{(\alpha - \alpha_0)}
  \cos{\delta}\sin{\delta_0},  \\
X = x^\prime\,\sin{({\rm PA})} + y^\prime\,\cos{({\rm PA})},  \\
Y = -x^\prime\,\cos{({\rm PA})} + y^\prime\,\sin{({\rm PA})},
\end{eqnarray}
where 
$x'$ and $y'$ are Cartesian coordinates for $\alpha$ and $\delta$ with a center at
($\alpha_0$, $\delta_0$), and 
PA$=37.7^\circ$ is the position angle for the major axis \citep{dev58}.

\section{Data}\label{data}

\subsection{Master catalog of M31 GCs}

First we used the data of GCs in M31 given in Paper I,
which describes the details of our survey's photometric and spectroscopic observations,
data reduction, and the data set. 
Here we only give a brief summary of the data set of M31 GCs.

We selected GC candidates in Washington $CMT_1$ CCD images covering
a $\sim 3\arcdeg \times 3\arcdeg$ field centered on M31, obtained at the KPNO 0.9 m telescope.
Spectroscopic observations were made using the Hydra multifiber bench spectrograph at the WIYN 3.5 m telescope
in September 2000 and November 2001 for 748 GC candidates including 106 previously known GCs.
We determined the radial velocities for GC candidates by cross-correlating
the candidate spectrum with those of two bright bonafide GCs in M31.
Finally, we have independently found 1164 GCs and GC candidates,
of which 559 are previously known GCs or GC candidates
and 605 are newly found GC candidates.
Among the GCs and GC candidates in our survey, there are 211 genuine GCs (class 1 in
Paper I) with measured velocity.
We have derived new velocity data for 150 GCs in our wide-field survey.

\begin{figure}
\includegraphics [width=165mm] {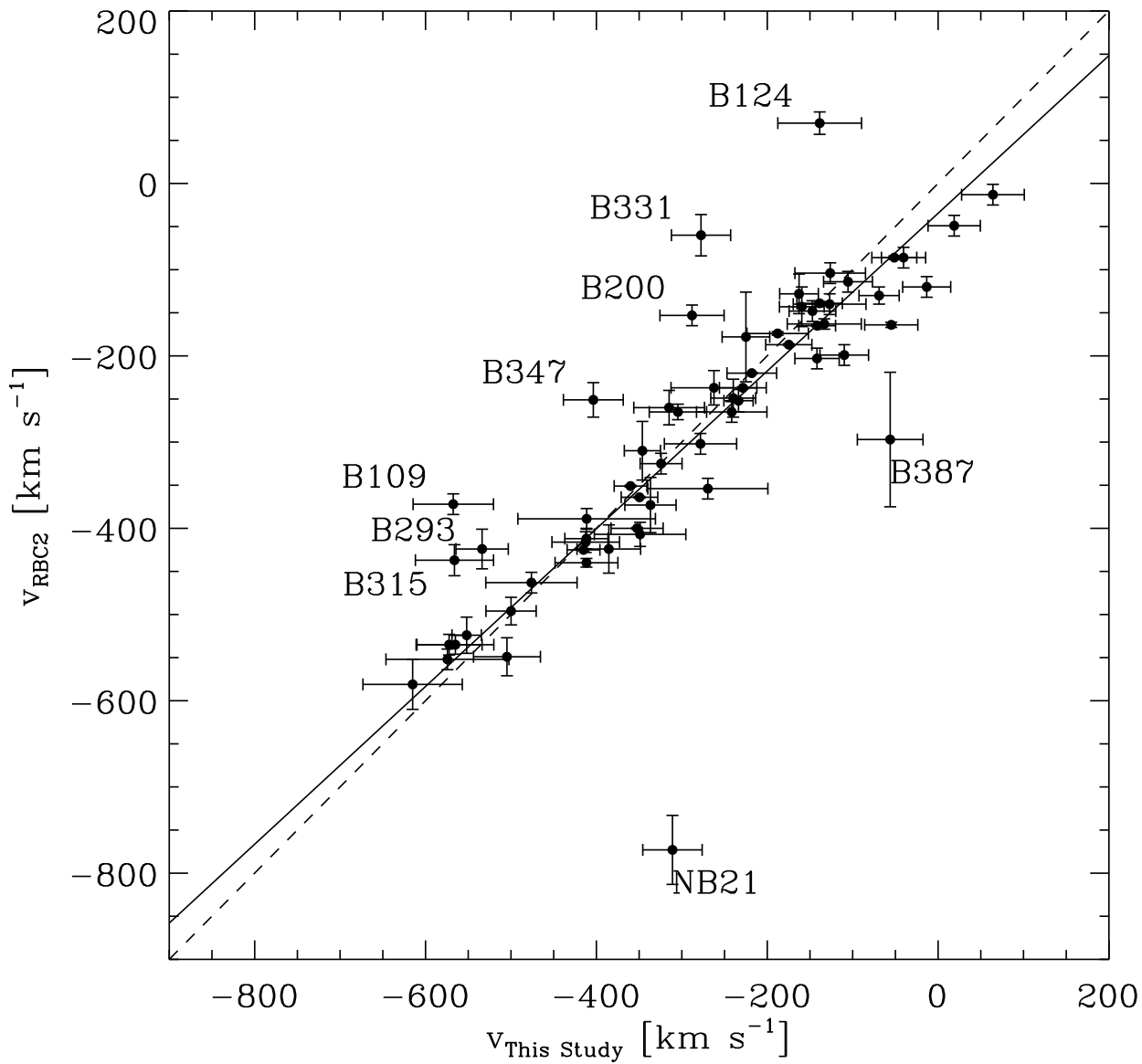}
\caption{Comparison of the radial velocities for M31 GCs measured
in this study with those in the literature (RBC2). The solid line
indicates the linear least-squares fit, and the dashed line
denotes the one-to-one relation. We labeled the ID   of GCs in
RBC2 for the GCs showing a velocity difference of over $110$ km
s$^{-1}$ between this study and other studies. } \label{fig-velcomp}
\end{figure}

\citet{gal04} compiled all known GCs and GC candidates in M31 in
the literature, and published the Revised Bologna Catalogue (RBC)
with 1035 objects. Later, \citet{gal06} published an updated version
of the Revised Bologna Catalogue V2.0 (RBC2) with 1164 objects. In
order to produce a master catalog of radial velocities for M31
genuine GCs, we have secured, from the RBC2, 349 genuine GCs (class
1 in RBC2) with measured velocities and 5 GC candidates (ID B366,
G137, B102, B430, B134 in RBC2) with measured velocities that were
classified as genuine GCs in our survey. There are 61 objects in
common between this study and the RBC2, if we use a matching tolerance
of 4\arcsec. We have compared our velocity measurements with
published velocity measurements (references are in RBC2) for these
common objects, as displayed in Figure \ref{fig-velcomp}. This shows
that our measurements, with a typical error of 35.3 km/s, agree with the published values within the combined errors
except for nine objects showing velocity differences of over $110$ km
s$^{-1}$. We present the identification in RBC2 for these GCs in Figure
\ref{fig-velcomp}. 
We tried to find any cause of the large difference for the nine objects, finding
some explanation for five of them (B109, B124, B293, B315, and NB 21), 
but none for the other four(B347, B200, B331, and B387).
The reasons for large velocity differences for five of these 
nine objects are as follows:
\begin{enumerate}

\item \textbf{B109} There are two velocity measurements for this object:
$-613\pm24$ km s$^{-1}$ in \citet{huc91} and $-372\pm12$ km s$^{-1}$ in \citet{per02}.
Since the velocity difference between two measurements is large and
\citet{per02} gives higher precision data, RBC2 quoted $-372\pm12$ km s$^{-1}$.
However, the measured velocity in our study is $-568\pm47$ km s$^{-1}$,
which is consistent with the value in \citet{huc91}.

\item \textbf{B124} There are two velocity measurements for this object:
$-75\pm22$ km s$^{-1}$ in \citet{jab98} and $70\pm13$ km s$^{-1}$ in \citet{bar00}.
Since the velocity difference between these two measurements is large and
\citet{bar00} gives the value with the smaller error, RBC2 quoted $70\pm13$ km s$^{-1}$.
However, the measured velocity in our study is $-138\pm49$ km s$^{-1}$,
which is consistent with the value in \citet{jab98}.

\item \textbf{B293} There are two velocity measurements for this object:
$-345\pm35$ km s$^{-1}$ in \citet{huc91} and $-467$ km s$^{-1}$ in \citet{fed93}.
RBC2 presented the value of a weighted mean ($-424\pm23$ km s$^{-1}$),
although the velocity difference is about $130$ km s$^{-1}$.
However, the measured velocity in our study is $-534\pm31$ km s$^{-1}$,
which is closer to the value of \citet{fed93}.

\item \textbf{B315} There are three velocity measurements for this object:
$-291\pm108$ km s$^{-1}$ in \citet{huc91}, $-434$ km s$^{-1}$ in \citet{fed93},
and $-559\pm12$ km s$^{-1}$ in \citet{per02}.
RBC2 presented the value of a weighted mean ($-437\pm18$ km s$^{-1}$)
excluding the measurement in \citet{per02}.
However, the measured velocity in our study is $-566\pm46$ km s$^{-1}$,
which is consistent with the value of \citet{per02}.

\item \textbf{NB21} The velocity measurement in our study is
$-311\pm35$ km s$^{-1}$, while that in \citet{jab98} is $-773\pm40$ km s$^{-1}$.
Since the pointing coordinate of our spectroscopic observation is different
from the photometric center of this object by about 1.72 $\arcsec$,
and this object is located in Field 25 in Paper I that is close to the M31 center,
the observed spectrum in this study might be significantly contaminated by the
M31 stellar light.

\end{enumerate}

Since our measurements for B109, B124, B293, and B315 are similar to those of
at least one value in the literature, we prefer to keep our measurements for those GCs
at this time. 
We keep our measurement for NB21 for further analysis,
since there is no convincing evidence to change our measurement at this moment.

From the weighted linear fit discarding these nine objects with large velocity differences,
we derive a transformation relation between the two systems,      

\begin{equation}
v_{\rm This~ study} = 1.09 (\pm0.03)~v_{\rm RBC2} +37.8 (\pm8.8) {\rm~km~s}^{-1},
~{\rm rms =37.8 ~km~s^{-1}, ~N=52} . \label{eq-trans}
\end{equation}

The radial velocities for the GCs in RBC2 were transformed onto our
velocity system using equation (\ref{eq-trans}), and the
transformed velocities were used for further analysis. We used our
velocity measurements for the above nine objects.
Finally, we produced a master catalog of radial
velocities for 504 genuine GCs in M31. 
We derived $T_1$ magnitudes for 483 out of 504 GCs in the master catalog
from the CCD images used for our survey as described in \citet{kim07}.

Figure 2 displays the errors of the measured velocities for the M31 GCs versus $T_1$ magnitude (a) and versus $R$, a galactocentric radial distance corrected for the inclination of M31 using the equation $R=[X^2+(Y/{\rm cos}~i)^2]^{1/2}$ where $i$ is the inclination angle,
$i=77.7 \arcdeg$ (b). 
In Figure 2(a), the errors for 136 metal-poor GCs and 63 metal-rich GCs derived in this study are plotted with those for 174 metal-poor GCs  and 58 metal-rich GCs derived in the previous studies.
In Figure 2(b), the errors for 46 bright GCs and 99 faint GCs derived in this study
are plotted with those for 155 bright GCs and 104 faint GCs derived in the previous studies
(Division of the entire sample of M31 GCs into subclasses depending on the metallicity and magnitude will be described in detail in the following section).
Mean errors for the measured velocities are
 $27\pm18$ km~s$^{-1}$ for all 504  GCs,  $40\pm16$ km~s$^{-1}$ for 211 GCs measured in this study,
and $17\pm12$ km~s$^{-1}$ for 293 GCs measured in the previous studies.
Mean errors for the measured velocities of the bright and faint GCs
are $35\pm12$ km~s$^{-1}$ and $40\pm17$ km~s$^{-1}$ for 46 bright GCs and 99 faint GCs, respectively, measured in this study, and
$17\pm13$ km~s$^{-1}$ and $17\pm9$ km~s$^{-1}$ for 155 bright GCs and 104 faint GCs, respectively, measured in the previous studies.
In the case of the previous measurements, a significant fraction of the measurements for
faint GCs have typical errors of 12 km~s$^{-1}$ given by \citet{per02}, 
much smaller than the values in other studies, which 
leads to a smaller mean value for faint GCs.
Excluding the data by \citet{per02}, the mean errors are 
$18\pm15$ km~s$^{-1}$ and $22\pm13$ km~s$^{-1}$ for bright GCs and faint GCs, respectively. 
Figure 2 shows that the mean errors for the measured velocities increase
as the GCs get fainter, and that they are larger in the inner region where the background level is higher than that in the outer region of M31.

The mean values of the radial velocities are derived using the biweight location of
\citet{beers90}: $\overline{v_p}=-281\pm13$ 
km~s$^{-1}$ for 211 GCs measured in this study, and
$\overline{v_p}=-285\pm8$ 
km~s$^{-1}$ for all 504 GCs in the master catalog. 
These values are consistent with those in
the previous studies ($\overline{v_p}=-284\pm9$ km~s$^{-1}$ in
\citealt{per02} and $\overline{v_p}=-296\pm12$ km~s$^{-1}$ in
\citealt{gal06}) within the uncertainty, and are also compatible
with the M31 systemic velocity measured in this study
($v_p = -290\pm69 $ km~s$^{-1}$) and that in the
literature ($v_p = -300\pm4$ km~s$^{-1}$ in \citealt{dev91}) within the uncertainty.
The larger error in our estimate of the M31 systemic velocity
is primarily due to the fact that we used, as templates, the spectra
of GCs with small velocity dispersion to estimate the velocity of
the nucleus of M31 that has much larger velocity dispersion.

Figure \ref{fig-velhist} shows
the radial velocity histogram for all 504 GCs as well as 211 GCs measured in this study.
It appears that the majority of all GCs are in the range
$-750$ km s$^{-1}$ $\le v_p  < +100$ km s$^{-1}$, and
the velocity distribution for all GCs is roughly
symmetric with respect to the mean value of the radial velocities.
Interestingly, there is  one GC with $v_p \le -850$ km s$^{-1}$
measured in this study (ID 106 in Table 5 of Paper I).
This GC is in the northern disk of M31 (Field 13 in Fig. 1 of Paper I),
and has T$_1$=18.94, C$-$T$_1$=1.83, and $v_p = -890\pm38$ km s$^{-1}$.
Since the signal-to-noise ratio (S/N) of the spectrum for this GC is only $\sim$5
(measured using H$\beta$ and Fe5015 lines), a higher S/N spectrum is needed
to verify its radial velocity and  membership clearly.

The velocity dispersion derived using the biweight scale
of \citet{beers90} is estimated to be 
$\sigma_p= 185_{-6}^{+7}$ km~s$^{-1}$ for 211 GCs measured in this study, and
$\sigma_p=178_{-4}^{+4}$ km~s$^{-1}$ for all 504 GCs in the master
catalog. Both agree well within the errors.

In Figure \ref{fig-img}, we show the spatial distribution of 504
GCs with measured velocities overlaid on a $4\arcdeg \times
4\arcdeg$ optical image of M31 from the Digitized Sky Survey. The
spatial distribution of 211 GCs measured in this study
is globally similar to that of the 293 GCs measured in the literature.
However, the GCs measured in this
  study are mainly located in the outer disk of M31 since the
  spectroscopic observation for the outer disk was done repeatedly
  (see Paper I for the configuration of the spectroscopic observations).

\subsection{Subsamples}

Metallicities of 199 GCs in M31 were derived from the line indices
measured in the WIYN/Hydra spectra by Lee et al. (2007, in preparation, Paper III).
We made subsamples based on the metallicities and $T_1$ magnitudes
for 504 GCs in the master catalog for further analysis.
We combined the metallicity data of 176 GCs in \citet{bar00} and
194 GCs in \citet{per02} with our data of 199 GCs  (Paper III)
in order to include
the metallicity data in the master catalog. There are 32 objects in
common between this study and \citet{bar00}, and 34 objects in
common between this study and \citet{per02}. We have compared our
metallicity measurements with those in these two previous studies for
the common objects, as displayed in Figure \ref{fig-fehcomp}. It
shows that our measurements agree well with the published
ones overall, while the scatter in the
comparison between this study and \citet{bar00}
is smaller than that in the
comparison between this study and \citet{per02}. We also derive
transformation relations between the measurement in this study,
and those in \citet{bar00} and in \citet{per02} using the GCs
with small measurement errors,

\begin{equation}
{\rm [Fe/H]}_{\rm This~study} = 0.92 (\pm0.02)~{\rm [Fe/H]}_{\rm Barmby~et~al.} -0.11 (\pm0.02),~{\rm rms=0.19, N=30}, \label{eq-transb00}
\end{equation}
\begin{equation}
{\rm [Fe/H]}_{\rm This~ study} = 0.87 (\pm0.03)~{\rm [Fe/H]}_{\rm Perrett~et~al.} -0.12 (\pm0.03),~{\rm rms=0.20, N=25}. \label{eq-transp00}
\end{equation}

The measured metallicities for the GCs in \citet{bar00} and
\citet{per02} were transformed into our metallicity system using
equations (\ref{eq-transb00}) and (\ref{eq-transp00}), and the
transformed metallicities were used for further analysis. We used
our metallicity measurements for the common objects. Finally, we
obtained metallicities for 431 GCs in total. The details of our
metallicity data will be given in Paper III.

We present the metallicity distribution
of M31 GCs in Figure \ref{fig-subsam}(a). The KMM test based on
the algorithm of \citet{ash94} yields the result that the hypothesis that
a unimodal distribution fits the data better rather than a bimodal distribution can be
rejected at a confidence level of $99.9\%$. The bimodal test
results in the following two groups: the metal-poor ([Fe/H]$\le-0.905$) GCs ($N=310$) with a
mean value of [Fe/H]$=-1.47$ ($\sigma=0.37$) and the metal-rich ([Fe/H]$>-0.905$) GCs ($N=121$)
with a mean value of [Fe/H]$=-0.62$ ($\sigma=0.32$). 
The KMM test with a trimodal
distribution also rejects the hypothesis of a unimodal metallicity
distribution at a confidence level of $99.8\%$. The trimodal
test results in the following three groups:   
  the metal-poor GCs ($N=183$ GCs) with a mean value of [Fe/H]$=-1.67$ ($\sigma=0.29$), 
  the intermediate metallicity GCs ($N=155$ GCs) with a mean value of [Fe/H]$=-1.14$ ($\sigma=0.26$), 
  and the metal-rich GCs ($N=93$ GCs) with a mean value of [Fe/H]$=-0.52$ ($\sigma=0.27$). 
Although there is no reason to reject the trimodal metallicity distribution for
M31 GCs (to be discussed in detail in Paper III), we use the two
subsamples of metal-poor and metal-rich GCs from the bimodal
distribution to increase the number of GCs per subsample for kinematic analysis
in this study.

We derived the mean values of the radial velocities:
$\overline{v_p}=-294_{-11}^{+11}$ km s$^{-1}$ for 310 metal-poor
GCs, and $\overline{v_p}=-250_{-14}^{+15}$ km s$^{-1}$ for 121
metal-rich GCs. The value for the metal-poor GCs is consistent
with that for the entire 504 GC  sample ($\overline{v_p}=-285_{-8}^{+8}$
km~s$^{-1}$), but the value for the metal-rich GCs is significantly smaller than
that for the metal-poor GCs, which was seen earlier in \citet{per02}. 

For the velocity dispersion, we found that the dispersion 
  for the metal-poor GCs ($\sigma_p= 183_{-5}^{+5}$ km s$^{-1}$) is significantly
larger than 
  that of the metal-rich GCs ($\sigma_p=154_{-9}^{+11}$ km s$^{-1}$).
This is in contrast with    the result of \citet{per02} that the
dispersions for the metal-poor and metal-rich GCs agree within the
quoted errors ($\sigma_p= 155\pm7$ km s$^{-1}$ for the metal-poor GCs
and $\sigma_p= 146\pm12$ km s$^{-1}$ for the metal-rich GCs).
We find that this difference may be due to the fact that the sample used by \citet{per02}
was  derived mainly from  the disk of M31, 
while ours are based on a much wider area, including a larger fraction of probable
halo GCs.

The luminosity function for M31 GCs is shown in Figure
\ref{fig-subsam} (b). We used only 483 GCs for which $T_1$ magnitudes
are available in our photometric catalog of M31 survey. Since the
survey depth is not uniform across the galaxy due to varying background
of stellar light, we use only the GCs brighter than $T_1=18.5$ to make
subsamples based on $T_1$ magnitudes. Finally, we divide the GCs into
201 bright ($T_1\le16.9$) and 203 faint ($16.9<T_1\le18.5$) GCs
so that the number of GCs in each subsample is similar.

\section{Results}\label{results}

Using the master catalog of 504 GCs in M31,
we have investigated the kinematic properties of the M31 GC system
for several cases of subsamples: all GCs, metal-poor and metal rich GCs,
bright and faint GCs, and X-ray emitting GCs.
Primary results of the kinematics of the M31 GC system derived in this study
are summarized in Table 1.

\subsection {All GCs}\label{egc}

\subsubsection {Kinematic Properties}\label{egckin}

In Figure \ref{fig-spatvel}, we present the spatial distribution
of M31 GCs with the measured velocity. 
Here we adopted as the systemic velocity of M31
the value derived for all GCs in the previous section,
$-285\pm8$ km~s$^{-1}$. 
Figure \ref{fig-spatvel} shows clearly a rotation signature.
The receding GCs, with velocities greater than the systemic
velocity of M31, and the approaching GCs,  with  velocities
less than the systemic velocity of M31, are spatially separated
effectively by the minor axis, indicating a disk rotation. Interestingly,
there are some GCs showing a retrograde rotation (filled circles at
$+X$ and open circles at $-X$) that will be discussed in \S \ref{egcnofof}.
We present the radial velocities for M31 GCs
versus the projected radii along the major axis (X) and along the
minor axis (Y) in Figure \ref{fig-spatvel} (b) and (c),
respectively. 
A linear least-squares fit, passing through $(X,v_p-v_{M31})=(0,0)$, along
the major axis results in 
$v_p-v_{M31}=4.69_{-0.55}^{+0.61}X $ km s$^{-1}$ and
$v_p-v_{M31}=0.31_{-1.17}^{+1.28}Y $ km s$^{-1}$, where $X$ and $Y$ are units of arcmin.
The large value of the slope between $v_p-v_{M31}$ and $X$ indicates a strong
rotation of the M31 GC system around the minor axis, while the
small value of the slope between $v_p-v_{M31}$ and $Y$ indicates no significant 
rotation around the major axis.

In Figure \ref{fig-rotfit}, we present the radial velocities of
GCs as a function of the projected distance along the major axis of
M31 and the velocity histogram of the GCs for the samples of
different distance bins along the minor axis in order to
investigate any rotational variation along the minor axis. 
The mean radial velocity in a
distance bin of 10$\arcmin$ is overlaid by a large square with an
errorbar representing the velocity dispersion. We fit the mean
radial velocities with a straight line representing a solid-body
rotation for the inner region. We select the appropriate boundary
that gives the minimum $\chi$ value by moving the fitting boundary
from 10$\arcmin$ to 50$\arcmin$ symmetrically. 
Then, a flat rotation curve is
connected to the solid-body rotation curve at that radius. The
rotation amplitude that is inclination-corrected,
is estimated by one-half of the velocity
difference between the flat rotation curves. We also estimate the
rotation amplitude using the velocity histogram for M31 GCs. We
fit the observed velocity histogram using the sum of two or three
Gaussian functions.
Then, the rotation amplitude is represented by one-half of the
velocity difference between the two Gaussian peaks - one is the
largest positive value and the other is the smallest negative value.   
We derive finally the rotation velocity corrected for inclination, 
$v_{\rm rot}=v_{\rm p,rot}/{\rm sin}~i$ where $v_{\rm p,rot}$ represents
a projected rotation velocity.
The rotation amplitude of all    GCs in M31 estimated by fitting the
rotation curve is $v_{\rm rot}\simeq188_{-28}^{+34}$ km s$^{-1}$, 
while that from the velocity histogram is $v_{\rm rot}\sim188_{-33}^{+39}$ km s$^{-1}$.
These two values agree very well.
The uncertainties on these (and following) estimates 
represent 68\% (1 $\sigma$) confidence
intervals. We computed the uncertainties using 1000 artificial data sets constructed
by randomly choosing GCs up to the number of GCs in the
real data. The fitting procedure is performed on the 1000 trial data sets, the results
are sorted, and the values corresponding to the 16th and 84th percentiles are identified.
The uncertainties are defined as the offsets between these values and the values
computed using the real data.
We have also derived the rotation-corrected velocity dispersion of the all GC
sample, that is, the dispersion about the rotation curve shown in Figure 8:
 $\sigma_{p,r} = 134^{+5}_{-5}$ km s$^{-1}$.

The velocity histogram for the thin disk $|Y|< 1$ kpc (d) shows not only two
rotation peaks but also a velocity dispersion peak due to the central bulge of M31.
The rotation amplitudes decrease as the distance along the minor axis of M31
increases. Interestingly, the rotation amplitude is not zero even for
the outermost samples at $|Y|\ge 5$ kpc, although the small size of the sample
for outer GCs leads to a large uncertainty (Figure \ref{fig-rotfit} (i) and (j)).

This result indicates that the halo (or extended bulge)
of M31 may be not totally pressure-supported but is rotating as well. 
This result is consistent with
the idea that the M31 halo is dominated 
by a moderately rotating large bulge rather than by
a non-rotating halo far from the plane, which is based on
the kinematics of PNe in M31 \citep{hur04}.
However, this does not necessarily indicate that there is no classical halo in M31
(see \citet{irw05, kal06}), as discussed in the next section.

On the other hand,
the rotation-corrected velocity dispersion of the GCs increases from 
$\sigma_{p,r}=119^{+9}_{-8}$ km s$^{-1}$ at  $|Y|< 1$ kpc to 
$\sigma_{p,r}=151^{+9}_{-9}$ km s$^{-1}$ at  $|Y|\ge 5$ kpc. 
Thus the velocity dispersion becomes about twice as large as the nominal rotation
amplitude at $|Y|\ge 5$ kpc, showing that pressure support is much larger than
rotation support in the outer halo.

\subsubsection {Friendless GCs}\label{egcnofof}

Although the GC system in M31 is dominated by disk rotation, there
are also seen some GCs that do not follow the disk rotation.
To identify the GCs that do not follow the disk rotation, we
select `friendless' GCs adopting the method used for the study of the kinematics
of M31 PNe in \citet{mer03}. 

In summary, we select the friendless GCs that deviate from the
mean velocity of neighbor GCs on the sky more than $n \times
\sigma$, where $\sigma$ is the velocity dispersion of neighbor
GCs. The number of GCs that defines the local environment is taken
to be $\sim [N_{\rm all~GCs}(=504)]^{1/2}$, and $n=2$ is used in
this study. Among all 504 GCs, we find 50 friendless GCs. The
spatial distribution of the friendless GCs is shown in Figure
\ref{fig-spatnofof}. It is seen that they are located 
mostly in the disk. 
However, most of them are located in the
quadrants of retro-grade rotation (the first and third) in the $v_p - v_{M31}$ 
versus $X$ diagram.
The friendless GCs in the second and fourth quadrants have mostly 
velocities larger than the largest velocities of the GCs following
the major disk rotation.
These friendless GCs may have different origin from the main population 
of GCs following disk rotation.

To highlight the kinematics of the friendless GCs, we plot, in Figure \ref{fig-spatvelnofof},
the spatial distribution of only the friendless GCs with the symbol size scaled according to their velocity deviations from the systemic velocity of M31.
Interestingly, it appears that the friendless GCs rotate around the major axis of M31
unlike the rotation around the minor axis for the disk population.

\citet{mer03} suggested that the friendless PNe in their sample may link to the Southern Stream and
the Northern Spur since the positions and velocities of the friendless PNe
are consistent with  the two streams. 
Recently, other stellar substructures in M31
were found by several authors \citep{fer02,mcc04,far06b}.
Since the kinematic information on stellar substructures in M31 is available,
it would be interesting to compare the kinematics of the friendless GCs with
those of stellar substructures, which may provide a clue
to understand the motion and origin of the friendless GCs, but it is beyond
the scope of this paper.

In Figure \ref{fig-photnofof}, we show the photometric
properties of $T_1$ magnitudes and ($C-T_1$) colors,
and spectroscopic metallicities for the friendless GCs.
The friendless GCs have a mean $T_1$ magnitude of $<T_1>=17.13$ mag with a deviation
of $\sigma_{T_1}=1.47$ mag, which is similar to
the values of $<T_1>=17.09$ mag with a deviation of $\sigma_{T_1}=1.30$ mag for the normal GCs.
The mean metallicities are also similar between the friendless and normal GCs,
but the mean ($C-T_1$) color for the friendless GCs ($<C-T_1>=1.66$ mag with $\sigma_{C-T_1}=0.42$)
is slightly redder than that for the normal GCs ($<C-T_1>=1.55$ mag with $\sigma_{C-T_1}=0.60$).
We used a Kolmogorov-Smirnov (KS) test to determine whether the friendless and
normal GC populations were drawn from the same distribution. The hypothesis
that the two distributions are extracted from the same parent population can be rejected
at the confidence level of $\sim86\%$ for the $T_1$ magnitude distribution,
$\sim91\%$ for ($C-T_1$) color distribution, and $\sim1\%$ for the metallicity distribution.
Therefore, only $T_1$ magnitudes and ($C-T_1$) colors appear 
to be slightly different.  However,
this problem needs to be investigated further using the reddening-corrected magnitudes and colors.

\subsection {Metal-Poor and Metal-Rich GCs}

Figure \ref{fig-spathistfeh} shows the spatial and radial distributions of the 310 metal-poor and 121 metal-rich  GCs. 
The most striking difference in the radial distribution between the two subpopulations is
that the metal-rich GCs show a much  stronger central concentration than
the metal-poor GCs,  which is consistent with the results based on the smaller data set in
\citet{huc91} and \citet{per02}. 
In particular the radial distribution of the metal-rich GCs show a very narrow peak 
with a half width of about 10 arcmin, showing that they belong to the central
bulge of M31. 
Median values of the absolute
distances along the major axis for the metal-poor and metal-rich
GCs are $23.1\arcmin$ and $16.0\arcmin$, respectively, and those
along the minor axis for the metal-poor and metal-rich GCs are
$9.3\arcmin$ and $4.7\arcmin$, respectively. 
In addition, a small number of metal-rich GCs are found in the outer  region,
even at about $70\arcmin$ from the major axis.
There are 9 GCs within the boundary of the standard diameter in NGC 205.
All of these GCs are metal-poor, as found by earlier studies \citep{per03}.

To highlight the difference in spatial distribution between the metal-poor and
metal-rich GCs, we plot, in Figure \ref{fig-spatconfeh},
the number density contours for the metallicity subsamples.
Several features are noted in Figure \ref{fig-spatconfeh}.
First,  the spatial distribution of the metal-rich GCs shows a much  stronger central concentration than
that  of the  metal-poor GCs,  as seen before.
Second, the spatial distribution of the metal-rich GCs shows almost circular structure in the central region of M31, while
that of the metal-poor GCs shows an extended structure elongated along the major axis with
a similar ellipticity to that of the M31 disk.
This difference is less clearly seen in the sample of bright GCs ($T_1<17$).
Third, there are several substructures in both subpopulations.
The most notable metal-poor substructures are one 
in the region $-40\arcmin<X<-20\arcmin$ and $Y\approx -10\arcmin$, and one
in the region $-60\arcmin<X<-40\arcmin$ and $Y\approx 0\arcmin$. These two correspond to the GC groups IDs 10 and 11, respectively, in \citet{per03}.
These are close to, but not consistent with, the position of And VIII. 
One metal-rich GC and
four metal-poor GCs are found within the standard diameter of And VIII.
The substructure of metal-poor GCs in NGC 205 corresponds to the GC group ID 5
 in \citet{per03}.
Two notable substructures among the metal-rich GCs  are on the major axis: 
one at $X\approx -25\arcmin$ and the other at $X\approx 20\arcmin$. These correspond, 
respectively, to the GC group ID 8 and the sum of IDs 2 and 3, respectively,
 in \citet{per03}.
One substructure containing both metal-poor GCs and metal-rich GCs at $X\approx -10\arcmin$ and $Y\approx 20\arcmin$ (seen only in the bright GCs)
corresponds to the GC group ID 7
 in \citet{per03}.
Therefore most of the GC groups in \citet{per03} are confirmed in this study.

In Figure \ref{fig-spatvelfeh}, we present the radial velocities
for the metal-poor and metal-rich GCs versus the projected radii
along the major axis (X) and the minor axis (Y). 
A linear least-squares fit, passing through $(X,v_p-v_{M31})=(0,0)$, along
the major axis results in $v_p-v_{M31}=+5.12_{-0.49}^{+0.46}X$ km
s$^{-1}$ for the metal-poor GCs and
$v_p-v_{M31}=+5.28_{-1.54}^{+1.52}X $ km s$^{-1}$ for the
metal-rich GCs. 
A least-squares fit along the minor axis results
in $v_p-v_{M31}=-0.20_{-1.47}^{+1.60}Y$ km s$^{-1}$ for the
metal-poor GCs and $v_p-v_{M31}=+1.46_{-2.34}^{+3.10}Y$ km
s$^{-1}$ for the metal-rich GCs. 
As known previously, both the
metal-poor and metal-rich GCs show a strong and very similar rotation around the
minor axis and no or a weak rotation around the major axis. 
The kinematic difference between the two subsamples is
not significant at this stage.

To investigate any kinematic difference between the two subsamples in detail,
  we plot, in Figure \ref{fig-rotfitfeh},
  the radial velocities  for the metal-poor and metal-rich GCs as a function of the projected
  distance along the major axis of M31 and the velocity histograms.
We also plot the mean radial velocities of the GCs in a distance bin of $10\arcmin$ along the
major axis.
We estimate the rotation amplitudes for the two subsamples using
the same fitting procedure applied in Figure \ref{fig-rotfit}. 
The rotation amplitude estimated by fitting the rotation curves (left panels) shows that 
  the rotation of the metal-rich GCs ($v_{\rm rot}=191_{-37}^{+34}$ km s$^{-1}$) is indistinguishable from that of the
  metal-poor GCs ($v_{\rm rot}=193_{-41}^{+44}$ km s$^{-1}$). 
However, \citet{per02} found that the rotation of the metal-rich GCs 
  ($v_{\rm rot}=160\pm19$ km s$^{-1}$) is slightly stronger than that of the
  metal-poor GCs ($v_{\rm rot}=131\pm13$ km s$^{-1}$).

It is worth noting that the rotation curve in the metal-poor GCs is
asymmetric at $|X| >30 $ arcmin. The metal-poor GCs at $|X|<30 $ arcmin follow a
solid-body rotation, while those at $|X| >30 $ arcmin do not. 
Since the distribution of the metal-poor
GCs is more extended than the metal-rich GCs, the asymmetric rotation
curve is seen only for the metal-poor GCs. Contours in
Figure \ref{fig-rotfitfeh} show clearly this asymmetry at $-60\arcmin<X<-30\arcmin $ where the
mean radial velocities are larger than the opposite end (at $30\arcmin<X<60\arcmin $).
Note also that there is a significant substructure at $X\approx -25\arcmin$  
with $v_p - v_{M31} \approx -200$ km s$^{-1}$, found in a wide-field survey.
This corresponds to the most  notable metal-poor substructure described above.
Therefore these substructures are distinguishable not only in spatial distribution but also
in kinematics. This substructure may be related with the Southern Stream emanating from the southwest disk of M31 found in a wide field survey of red giants by \citet{iba01}.

In addition, we derived the rotation amplitude  from the velocity
histogram of the metal-poor GCs (panel (b)).
The rotation amplitude of the metal-poor GCs is estimated to be
$v_{\rm rot}=151_{-8}^{+8}$ km s$^{-1}$, 
which is slightly smaller than, but agrees within the error to,
the value derived from the  $v_p - v_{M31}$ vs $X$ diagram above. 
The velocity histogram of the metal-rich GCs shows a single strong peak, 
showing that the sample is dominated by the GCs in the central bulge.
So we did not try to derive a rotation velocity for the metal-rich GCs
from the velocity histogram. 
We have derived the rotation-corrected velocity dispersions:
$\sigma_{p,r}=129^{+7}_{-6}$ km s$^{-1}$ for the metal-poor GCs, 
and $\sigma_{p,r}=121^{+9}_{-10}$ km s$^{-1}$ for the metal-rich GCs.
Thus the rotation-corrected velocity dispersion for the metal-poor GCs
is very similar to         that for the metal-rich GCs.

In Figures \ref{fig-rotfitmp} and \ref{fig-rotfitmr}, 
we also present the radial velocities of the metal-poor and metal-rich GCs 
as a function of the projected distance along the major axis of
M31 and the velocity histogram of the GCs for the samples of
different distance bins along the minor axis in order to
investigate the rotational variation along the minor axis. 
The metal-poor GCs show that the rotation velocity is roughly decreasing with
increasing distance from the major axis, but the uncertainties are too large
to derive a reliable rotational velocity for the outer region at $|Y|>3$ kpc.
The metal-rich GCs show a similar trend to that of the metal-poor GCs 
at $|Y|<3$ kpc, but they
suffer more from the small number statistics problem.

The rotation-corrected velocity dispersion of the metal-poor GCs increases from 
$\sigma_{p,r}=119^{+17}_{-15}$ km s$^{-1}$ at  $|Y|< 1$ kpc to 
$\sigma_{p,r}=145^{+11}_{-11}$ km s$^{-1}$ at  $|Y|\ge 5$ kpc. 
Thus the velocity dispersion becomes about twice as large as   the rotation
amplitude at $|Y|\ge 5$ kpc, showing that pressure support is much larger than
rotation support in the outer halo of M31.

\subsection {Bright and Faint GCs}

In Figure \ref{fig-spathistmag},
we show the spatial distribution of the bright and faint GCs.
It is seen that the bright GCs are more centrally concentrated
than the faint GCs. 
Median values of the absolute distances along the major axis for the 201 bright and 203 faint GCs are
$14.9\arcmin$ and $22.2\arcmin$, respectively,
 and those along the minor axis for the bright and faint GCs are
$7.1\arcmin$ and $8.0\arcmin$, respectively.
Thus, the difference between the two subsamples in the spatial distribution along the major
axis is significant, while that along the minor axis is not.
This is mainly due to the fact that a significant fraction of bright GCs are 
located in the central bulge of M31 
and that a large number of faint GCs in the central region of M31 were not 
detected in the current surveys.

We present the radial velocities of the bright and faint GCs versus
the projected radii along the major axis (X) and along the minor
axis (Y) in Figure \ref{fig-spatvelmag}. The linear least-squares
fits, passing through $(X,v_p-v_{M31})=(0,0)$, along the major
axis result in $v_p-v_{M31}=+4.98_{-0.96}^{+0.98}X$ km s$^{-1}$
for the bright GCs and $v_p-v_{M31}=+5.06_{-0.30}^{+0.30}X$ km
s$^{-1}$ for the faint GCs. The least-squares fit along the minor
axis results in $v_p-v_{M31}=+0.18_{-1.55}^{+1.66}Y$ km s$^{-1}$
for the bright GCs and $v_p-v_{M31}=-0.79_{-1.16}^{+1.17}Y$ km
s$^{-1}$ for the faint GCs. Both the bright and faint GCs show a
strong rotation around the minor axis and no or weak rotation
around the major axis. The kinematic difference between the
two subsamples is not significant.

We also plot the radial velocities for the bright and faint GCs as
a function of the projected distance along the major axis of M31
and the velocity histograms in Figure \ref{fig-rotfitmag}. We
estimate the rotation amplitudes for the two subsamples using the
same fitting procedure applied in Figure \ref{fig-rotfit}. The
rotation amplitudes estimated by fitting the rotation curves (left
panels) show that 
  the rotation of the faint GCs ($v_{\rm rot}=209_{-15}^{+21}$ km s$^{-1}$) is stronger than 
  that of the bright GCs ($v_{\rm rot}=129_{-35}^{+35}$ km s$^{-1}$). 
In addition, the rotation amplitude from the velocity histogram shows again
  that the rotation of the faint GCs ($v_{\rm rot}=178_{-23}^{+23}$ km s$^{-1}$) is stronger 
  than that of the bright GCs ($v_{\rm rot}=144_{-46}^{+43}$ km s$^{-1}$). 
The difference in  rotation amplitudes between the bright and faint GCs is closely related to
the difference in their spatial distributions. Since the faint GCs
are more extended along the major axis than the bright GCs, the rotation amplitude of
the faint GCs is estimated to be larger than that of the bright
GCs.
It is noted that the rotation-corrected velocity dispersion for the bright GCs,
$\sigma_{p,r}=146^{+7}_{-7}$ km s$^{-1}$ is much larger than that for the faint GCs, 
$\sigma_{p,r}=107^{+8}_{-8}$ km s$^{-1}$.

\subsection {GCs with X-ray detection}

To investigate the kinematic properties of X-ray emitting GCs in
M31, we cross-correlated the GCs and GC candidates 
from the optical survey with the X-ray sources in M31. We used as
the optical source list, the catalog of 1778 GC candidates prepared combining 
our survey, RBC2, \citet{hux05} and \citet{mac07}. 

We made a
master catalog of X-ray sources in M31 by combining 560 sources in
the {\it ROSAT} survey \citep{sup01}, 204 sources in the {\it
Chandra} survey \citep{kong02}, 856 sources in the {\it
XMM-Newton} survey \citep{pie05}, and 62 X-ray emitting GCs in
\citet{fan05}. If a GC candidate in the optical survey lies within
a circle of $2\sigma$ positional uncertainty of an X-ray source, we
regard it as a match. We used $1\sigma$ positional uncertainty for
the {\it ROSAT} sources due to its large positional uncertainty
(5$-$48 arcsec of $1\sigma$ with a median value of 7 arcsec). In
addition, we matched optical GC candidates with 62 X-ray emitting
GCs in \citet{fan05} using the identification of GCs, since
\citet{fan05} presented only a catalog of X-ray emitting GCs.

Finally, we identify 56 GCs and GC candidates (classes 1 and 2 in
Paper I and RBC2) with X-ray detections (40 genuine GCs including
39 GCs with measured velocities, and 16 GC candidates). There are
13 GCs that were identified as X-ray emitting GCs in
\citet{fan05}, but were not included in our combined optical GC
catalog. Therefore, we increase the number of X-ray emitting GCs
in M31 by 69, including 13 GCs that were not presented in our
GC catalog. For further kinematic analysis, we use only
the homogeneous data set of 56 X-ray emitting GCs and GC candidates
presented in our optical GC catalog.

We show, in Figure \ref{fig-spatxgc}, the spatial distribution of
39 genuine GCs with X-ray detection and measured velocity in
company with 1 GC and 16 GC candidates with X-ray detection but no
measured velocity. It is difficult to make strong conclusions about the spatial
distribution of the X-ray emitting GCs since the X-ray surveys
did not   uniformly cover the whole region in M31. 
However, it is clearly seen that a large number of the X-ray emitting GCs are
located in the central bulge region of M31 and that they show a large velocity
dispersion.
In addition, it appears that the majority of X-ray emitting GCs
follow the disk rotation of M31. 

In Figure \ref{fig-photxgc}, we
show the photometric properties of the X-ray emitting GCs compared
with that of all 504 GCs. 
We used the KS test to determine
whether the X-ray emitting and the entire GC population were drawn
from the same distribution. Interestingly, the hypothesis that the
two distributions are extracted from the same parent population
can be rejected at the confidence level of $\sim99\%$ for the
distribution of $T_1$ magnitude, ($C-T_1$) color, and
the metallicity. This result based on the analysis of the expanded
sample of M31 X-ray emitting GCs confirms the previous results
\citep{bel95,di02,tru04,fan05}. The properties of X-ray emitting
GCs in M31 are consistent with those of GCs in early-type
galaxies: redder, more metal-rich, and brighter GCs are more
likely to be detected as X-ray emitting GCs
\citep{sar03,kim06,siv06}.

\subsection {Rotation Curve}

Rotation curves are useful for inferring the amount and
distribution of underlying dark matter in galaxies (e.g.,
\citealt{sof01}). For M31, there are several studies to derive
the rotation curve using various tracers: HII regions
\citep{rub70,kent89}, HI \citep{bra91,car06,car07}, CO
\citep{loi95}, PNe \citep{mer06}, and GCs \citep{per02}. 

To obtain
the rotation curve of M31 using the sample of 504 GCs in this
study, we first selected 454 GCs following the M31 disk rotation by
rejecting 50 friendless GCs defined in \S \ref{egcnofof}.
Then, we selected 75 GCs with $|Y/X|<0.5$ and $|Y|<0.6$ kpc that
are dominated by the rotational motion, since the inclination of
M31 to the line of sight is as high as $i=77.7 \arcdeg$.

 We
display the rotation curve using the selected GCs in Figure
\ref{fig-rotcurve}. 
We plot the rotation velocity corrected for inclination, 
$v_{\rm rot}=(v_p-v_{M31})(R/X)/{\rm sin}~i$, where $X$ is the
projected distance along the major axis of M31 and $R$ is the
deprojected radius in the plane of the disk, $R=[X^2+(Y/{\rm
cos}~i)^2]^{1/2}$. For each GC, we use the nearest 21 GCs 
to estimate the mean velocity, taken as the rotation
velocity at that radius, and to estimate the velocity dispersion.
We stop the calculation when the number of the nearest GCs is less
than 21 ($R<8.9\arcmin$ and $R>46.4\arcmin$). Typical 1$\sigma$
uncertainties using the bootstrap procedure for the rotation
velocity and for the dispersion are 12 km s$^{-1}$ and 8 km
s$^{-1}$, respectively. For comparison, we derive also the rotation
curve of PNe using the velocity data in \citet{mer06} by applying
the same procedure as above. For PNe, typical 1$\sigma$
uncertainties for the rotation velocity and for the dispersion are
both 4 km s$^{-1}$.

It appears that the rotation curve of the GCs in this study agrees
well with those of other tracers in the range $R=20\arcmin-45\arcmin$.
However, the rotation curve of PNe derived using the velocity data
of \citet{mer06} is different from the other profiles in the similar radial
range. The rotation velocity for PNe at $R=20-60\arcmin$ 
is   on  average about 50 \kms ~smaller compared with
those of other tracers, as seen in Figure 33 of \citet{mer06}. 
They showed that the presence of significant
random velocities at all radii account for the asymmetric drift
between the rotational motion and the local circular speed for
PNe. 
In the inner region ($R<20\arcmin$) of M31, the rotation curves
are significantly different from each other.
A careful study is needed for the rotation curves in the inner
region due to the contribution of the non-circular motion of the
bulge and the projection effect of an inclined M31 disk.

The velocity dispersion profiles for the GCs and the PNe are similar
in general except for the inner region at $R<15\arcmin$. 
The velocity dispersion decreases from $\sim120$ \kms at $R=10\arcmin$
to $\sim80$ \kms at $R=30\arcmin$, and changes little in the outermost area. 
The velocity dispersion of the GCs seems to increase again 
at $R\approx 45\arcmin$, while that of the PNe keeps slowly decreasing with
$R$.
 More data on the GCs further out in the halo are needed to confirm this difference, 
which can probe whether
a pressure-supported hot halo exists or not in M31.

\subsection {Mass of M31}

We have estimated the mass of M31 from the kinematic information
of the GCs in this study using two methods:
 `Projected Mass Estimator (PME)' and `Tracer Mass Estimator (TME)'.
PME was introduced by \citet{bah81} for the case of 
test particles orbiting a point mass, 
and later was extended by \citet{hei85} for those tracing the total mass.
The mass based on PME is given by

\begin{equation}
M_{\rm PME} = \frac{f_p}{NG} \sum_{i=1}^N v_i^2 R_{p,i} \mbox{,}
\label{eq-pme}
\end{equation}

where $N$ is the number of objects, $G$ is the gravitational constant,
$R_p$ is the projected galactocentric radius, and
$v$ is the line-of-sight velocity relative to the systemic velocity of the galaxy.
The constant $f_p=32/\pi$ is used for the objects with an isotropic orbit,
while $f_p=64/\pi$ for the objects with a radial orbit.
We obtain $M_{\rm PME}= 5.5_{-0.3}^{+0.4} \times 10^{11} M_\odot$
using all 504 GCs out to a radius of $242\arcmin\sim55$ kpc
with an assumption of isotropic orbits.

Later \citet{eva03} introduced a new method of TME for the
case that the tracer population such as GCs and PNe does not
follow the underlying mass distribution dominated by dark matter.
They applied the TME to 321 GCs of \citet{per02}, and obtained,
for $R_p \sim100$ kpc, $M_{\rm TME}= 1.2 \times 10^{12} M_\odot$ that
is composed of a rotation contribution, $M_{\rm rot}= 3
\times10^{11} M_\odot$ and of a pressure contribution, $M_{\rm
press}= 9 \times10^{11} M_\odot$. Similarly, \citet{gal06}
obtained a larger mass of $M_{\rm TME}= 2.4 \times 10^{12}
M_\odot$ using the TME with the expanded sample of 349 GCs. In
this study, we first select 301 metal-poor GCs considered to be
halo GCs, and choose a radial range where the surface number density of
the halo GCs falls off as a power law. 
The surface number density of our metal-poor GCs follows approximately a 
power law $R^{-3}_p$ beyond $30\arcmin$,
similar to that of \citet{eva03}. 
Finally, 146 halo GCs in the
range of projected radius, $R_p =30\arcmin-152\arcmin$ ($6.7-34.2$
kpc) are selected. We set $\alpha=0$ (isothermal-like galaxy),
$\gamma=4$, $R_{\rm p, in}=6.7$ kpc, and $R_{\rm p, out}=100$ kpc using
the same assumptions as in \citet{eva03}. Then, we obtain for
$R_p\sim100$ kpc, the total mass of M31, $M_{\rm TME}=
19.2_{-1.3}^{+1.4} \times 10^{11} M_\odot$ that is composed of a
rotation contribution, $M_{\rm rot}= 6.2_{-0.6}^{+0.6}
\times10^{11} M_\odot$, and of a pressure contribution, $M_{\rm
press}= 13.1_{-1.3}^{+1.4} \times10^{11} M_\odot$. 

To compare our
estimation with the results in the literature, we plot the various
dynamical mass estimates using GCs
\citep{har74,van81,fed93,eva00,per02,eva03,gal06}, satellite
galaxies \citep{cou99,eva00,cote00}, and streams \citep{far06a} of
M31 in Figure \ref{fig-mass}. In addition, several profiles of the
mass model for M31 are plotted \citep{wid03,iba04,sei06,gee06}. 
The mass estimate based on PME derived in this study agrees well
with other observational estimates by \citet{eva00, gal06} 
and two mass models by \citet{wid03, sei06}. 

However, the mass estimates based on TME derived in this study and in
\citet{gal06} are larger than other estimates, and they are also larger than any models.
If the TME is
correct, the mass model of M31 needs a revision to a larger mass.
This is also consistent with the \citet{maj07}'s suggestion based on
the study of a new dwarf galaxy, AndXIV.
Noting that AndXIV shows a large
velocity of $-203$ km s$^{-1}$ relative to M31 at a projected
galactocentric distance of 162 kpc, \citet{maj07} suggested that recent
mass models of M31 need a revision to larger, if AndXIV is bound to M31. 
Similar discussion on the increase of the Milky Way Galaxy's mass 
based on the most distant satellite ($d= 270$ kpc), the Leo I dwarf spheroidal galaxy, 
was given  in \citet{lee93}. 

\section{Discussion}\label{discussion}

\subsection{A dominant bulge versus a wimpy halo in M31?}

The existence of a halo in M31 has been controversial.
While \citet{huc91} suggested that the metal-poor GCs in M31 show
little evidence of rotation, indicating that they belong to the halo of M31,
other studies based on surface photometry, counts of red giant branch stars,
and counts of PNe along the minor axis of M31 found an extended bulge, but no evidence
of any halo component in M31 \citep{pri94,mou86,hur04,irw05,mer06}.
However, recent studies covering a very wide field of M31 are finding evidence for
a huge halo in M31 \citep{irw05, guh05, cha06, kal06, gil06, iba07} as well as a vast extended disk reaching
about 40 kpc from the center of M31 \citep{iba05}.

\citet{irw05} presented a minor-axis surface brightness profile reaching out to
a projected radius of $4^\circ$ ($\approx 55$ kpc), combining surface photometry
and faint red giant star counts. They found that the minor-axis profile can be
fit by a de Vaucouleurs law with effective bulge radius  $R_{\rm eff} = 0^\circ.10$ (1.4 kpc) 
out to a projected radius of $1^\circ.4$ ($\approx 20$ kpc),
and that there is a separate component beyond $1^\circ.4$ that can be fit
by either a power law of index about --2.3 or an exponential law of scale length 14 kpc.
One thing interesting about this component is that it has the same color as the inner region,
indicating that it
has a similar metallicity to that of the bulge  (assuming the two components have the same age).
In addition, the photometry of remote fields along the minor axis
of M31 showed that most stars in these fields have unexpectedly high metallicity,
[m/H]$\sim -0.5$ ([Fe/H]$\sim -0.8$) while there are some metal-poor stars 
as well \citep{mou86,dur01}. 
Thus there are two components in the outer region of M31 that are distinguished
in structure, but not in metallicity. 
We may say from the results based on stars
that M31 has a huge dominant bulge reaching out to $4^\circ$, 
or that M31 has two bulges: an
inner bulge (out to $1^\circ.4$) and an outer bulge (out to $4^\circ$).
Even the inner bulge is much larger than the central bulge in our Galaxy,
and the outer bulge may be called a halo.

The progenitors of PNe are low to intermediate mass stars so that they are
often considered as a good tracer for old stellar populations. For example,
the radial number density profiles of PNe in M31 show 
excellent agreement with the $R$-band surface brightness profiles of the stellar bulge and disk \citep{wal87,mer06}. 
In particular, the number density profile of PNe along the minor
axis of M31 is consistent with the $R^{1/4}$ profile 
out to one degree \citep{mer06, hur04}.
\citet{mer06} pointed out that there is  little excess of PNe over the  $R^{1/4}$ profile even at $\sim$10 $R_{eff}$ (one degree), indicating no separate halo component. This is also consistent with the result based on the surface brightness profile along
the minor axis \citep{pri94, irw05}.
Considering the rotating kinematics of the outer PNe at $|Y|>6$ kpc in M31 as well as the above points, \citet{hur04} suggested that the bulge is the most dominating component in M31, while
a nonrotating, pressure-supported halo is, if any, only a very minor component of M31.

In general, or at least in our Galaxy, metal-poor GCs are considered as a tracer of the halo, while metal-rich GCs belong to the bulge or thick disk
like the PNe, although the bulge does have some metal-poor GCs \citep{har96}.
Note that the kinematics of 17 old star clusters in M33, irrespective of their metallicity, show evidence for the existence of a hot halo in M33 \citep{cha02}. Therefore we may use metal-poor GCs to investigate the
properties of any halo in M31. 
The kinematics of the metal-poor GCs along the major axis
at $|Y|<1$ kpc show a rotation as strong as the metal-rich GCs. The rotation amplitude
of the metal-poor GCs decreases with increasing distance from the major axis, 
 and the metal-poor GCs show some rotation ($60\sim100$ \kms) even far from the major axis at $|Y|>5$ kpc.
These results are consistent with those of \citet{per02}, but not with
\citet{huc91}. 

However, the rotation-corrected velocity dispersion for the metal-poor GCs is about twice
as large as    the rotation amplitude in the outer region at $|Y|>5$ kpc.
The rotation-corrected velocity dispersion for the metal-poor GCs at $|Y|>5$ kpc in M31,
$\sigma_{p,r}=145^{+11}_{-11}$ km s$^{-1}$, is comparable
to  that of the metal-poor GCs in the Galactic halo at $R_{GC} >4$ kpc, $\sigma =123$ km s$^{-1}$ given by \citet{cot99}.
Considering that the spatial distribution of
the metal-poor GCs in M31 is much more extended than that of the metal-rich GCs and that
the velocity dispersion of the metal-poor GCs in the outer region of M31 is 
about two times larger than the rotation amplitude,
we conclude that there is a dynamically hot halo in M31 that is rotating but primarily pressure-supported. 
This is consistent with the discovery of a pressure-supported metal-poor stellar halo based on the observations of red giant branch stars at radii from 10 kpc to 165 kpc 
\citep{cha06,gil06, kal06}. 
\citet{cha06} found from the analysis of a large sample of red giant branch stars:
(a) there is seen a stellar halo component with low metallicity ( [Fe/H]$\sim -1.4$ ) 
at radii from 10 to 70 kpc, (b) there is little radial gradient of metallicity, 
(c) there is no significant evidence for rotation, and 
(d) the velocity dispersion of this metal-poor halo can be represented by
$\sigma_v (R_p ) = 152 - 0.90 R_p $ km s$^{-1}$ where $R_p$ is a projected galactocentric distance in units of kpc. 

In Figure 25 we display the radial variation of the mean metallicity for the metal-poor GCs
in comparison with that for the red giant branch stars given by \citet{kal06}.
A few features are noted in Figure 25: 
(a) the mean metallicity for the metal-poor GCs is much lower than that for 
the red giant branch stars at the same distance from the M31 center;
(b) the mean metallicity for the metal-poor GCs in the inner region is similar to that 
for the red giant branch stars much further out at $R_p \approx 80$ kpc that was considered to be the part of the stellar halo by \citet{kal06}; and
(c) the mean metallicity of the metal-poor GCs shows little radial dependence: 
[Fe/H]$=-1.48\pm0.27$ at $R=7.1$ kpc to [Fe/H]$=-1.62\pm0.25$ at $R=25.5$ kpc
(if we use all GCs, we obtain [Fe/H]$=-1.24\pm0.27$ at $R=6.7$ kpc to [Fe/H]$=-1.62\pm0.25$ at $R=25.5$ kpc).
These show that the metal-poor GCs are tracing the metal-poor halo, even in the inner region
of M31.

Recently \citet{iba07} found a faint, smooth and extremely extended halo component, reaching out to
150 kpc, about three times larger than the limit studied by \citet{irw05}.
In addition, they found, from the photometry of red giant branch stars, 
that the majority of the stars in this extended halo are metal-poor, in contrast to the \citet{irw05} finding that its integrated color is similar to the inner bulge.
In spite of its huge size, the total luminosity of this halo derived assuming that it is symmetric is found to be $\sim 10^9$ L$_\odot$, similar to the stellar halo of our Galaxy.
Therefore it will be interesting to find GCs and investigate their kinematics in this outer halo of M31.

\subsection{Comparison of GC and PN kinematics}

PNe are an excellent tool for studying the kinematics of the bulge and halo of a galaxy.
Recently huge data sets for the PNe in M31 were published by \citet{mer06, hal06}.
How do  the kinematics of the GCs compare with that of the PNe in M31?
We have derived the kinematics of about 2500 PNe using the same method 
to the data given by \citet{mer06}, as used for the analysis
of the GC data, 
which are summarized in Table 2.

Figure 26 displays 
radial velocities (dots) versus the projected distances along
the major axis ($X$) and the velocity histogram for the PNe in M31
based on the data given by \citet{mer06}, like Figure 8 for GCs. 
We plot also the mean radial velocity and its dispersion of the PNe in a distance bin of 10 $\arcmin$ along the major axis. 
Velocity histograms of PNe are displayed with Gaussian fits as in Figure 8.
Figure 26 shows that the rotational properties of the PNe are very similar to those
of the GCs.
The rotational amplitudes of all PNe are 
$v_{\rm rot}\simeq187^{+12}_{-13}$ km s$^{-1}$ from the rotation curve, 
and $v_{\rm rot}\simeq198^{+58}_{-33}$ km s$^{-1}$ from the velocity histogram,
which are similar to those of GCs derived in this study, 
$v_{\rm rot}\simeq188_{-28}^{+34}$ km s$^{-1}$, 
and $v_{\rm rot}\sim188_{-33}^{+39}$ km s$^{-1}$, respectively.
The rotation amplitudes of the PNe decrease as the distance along the minor axis of M31
increases, but the rotation amplitude is not zero even for
the outermost samples at $|Y|\ge 5$ kpc, which are consistent with the results
based on GCs derived in this study. 

 The rotation-corrected velocity dispersion for  all PNe  
is estimated to be $\sigma_{p,r}=94^{+1}_{-1}$ km s$^{-1}$, which is much smaller
than that for all GCs, $\sigma_{p,r}\sim 134 \pm 5$ km s$^{-1}$.
The rotation-corrected velocity dispersion for the PNe at $|Y|<1$ kpc is   $\sigma_{p,r}=97^{+2}_{-2}$ km s$^{-1}$, and it becomes smaller and
shows little variation with $|Y|$:
$\sigma_{p,r}=75^{+2}_{-2}$ km s$^{-1}$ for  $1\le|Y|<3$ kpc,
$\sigma_{p,r}=88^{+4}_{-4}$ km s$^{-1}$ for  $3\le|Y|<5$ kpc, and
 $\sigma_{p,r}=79^{+3}_{-3}$ km s$^{-1}$ for  $|Y|\ge5$ kpc.
Thus the rotation-corrected velocity dispersion for the PNe at $|Y|<1$ kpc is 
consistent  with that for the metal-rich GCs at $|Y|<1$ kpc, 
$\sigma_{p,r}=107^{+24}_{-22}$ km s$^{-1}$. 
We cannot compare the kinematics of the PNe at $|Y|>1$ kpc with that of
the metal-rich GCs at $|Y|>1$, because there are too few such GCs in this region.

The rotation-corrected velocity dispersions for the PNe at $|Y|>1$ kpc 
are significantly different from those for the metal-poor GCs. 
That is, the rotation-corrected velocity dispersion for the PNe at $|Y|>1$ kpc
are smaller than that for the metal-poor GCs at $|Y|>1$ kpc, 
and the former shows little variation depending on $|Y|$ for  $|Y|>1$ kpc, 
while the latter increases with $|Y|$.
These differences show that the PNe are tracing well the bulge, while the metal-poor GCs
are tracing better the metal-poor halo, although their spatial coverages in the sky are similar.
Therefore we conclude again that there do exist both an extended rotating bulge and
a dynamically hot metal-poor halo in M31.

\subsection {Formation and Evolution of M31}

Main features related with the kinematics of the M31 GC system revealed 
in this study can be summarized as follows.
Most of the GCs show a strong rotation around the minor axis.
The rotation velocity decreases with increasing distance from the major axis. 
It is still as large as $60\sim100$ \kms ~even at $|Y|>5$ kpc,
indicating the bulge of M31 is large and rotation-dominated, which is
consistent with the finding based on the PNe by \citet{hur04} and \citet{mer06}.
The rotation velocity of the metal-rich GCs is  similar to 
that of the metal-poor GCs in the region close to the major disk. 
The spatial distribution of metal-rich  GCs
shows a stronger central concentration than that of the metal-poor GCs.
The rotation-corrected velocity dispersion of the metal-poor GCs in the outer region of M31 is
about twice larger than the rotation amplitude.
There are several substructures in the spatial distribution of
both metal-poor GCs and metal-rich GCs.
There are about 50 friendless GCs all over the disk of M31, and most of them
show retrograde motion. 
Interestingly it appears that these GCs show a rotation around
the major axis, while most of M31 GCs show a rotation around the  minor axis.

What do these results in addition to the known properties of M31
tell us about the formation and evolution of M31
in comparison with our Galaxy?
The most distinguishable kinematic difference between the M31 GCs and the Galactic GCs is a difference in the bulge and halo kinematics. 
In our Galaxy metal-poor halo GCs are the most dominant population among the GCs, 
and metal-rich GCs in the bulge (or thick disk) are a minor population .
The bulge GCs show significant rotation ($v_{rot}=193\pm29$ \kms), while the halo GCs 
are pressure-supported ($v_{rot}=43\pm29$ \kms and $\sigma_v = 123$ km s$^{-1}$) \citep{arm89, cot99}.
In contrast, the disk and bulge GCs are the most dominant population in M31, 
and the halo GCs may be a minor population compared with the Galactic halo GCs.

The current picture of the formation and evolution of the halo in our Galaxy
involves basically two classical scenarios:
(a) The monolithic collapse model according to which the stellar halo is formed
by the rapid collapse of a monolithic protogalactic cloud in a dynamical time scale of 
about 0.1 Gyr
 \citep{egg62}, and
(b) the accretion/merging model according to which the stellar halo is formed via successive
accretion and merging of protogalactic fragments 
\citep{sea78, whi78}. The latter is related with
the current paradigm for explaining
the formation of the cosmological large scale structure: 
a hierarchical merging scenario in the $\Lambda$CDM universe.
It appears that both scenarios are involved in forming the halo of our Galaxy \citep{chi00}.
While the inner part of our Galaxy can be explained mainly by the first, the
outer part of our Galaxy is explained better by the latter.

Recently, \citet{ren05a} suggested from a semi-analytic multizone chemical evolution model
that the observed higher metallicity in the M31 halo compared with the Galactic halo can be explained
by two scenarios:
(a) a higher star formation efficiency for the halo that is expected from the accretion of
smaller satellites, or 
(b) a larger reservoir of infalling halo gas with a longer halo phase, producing the intermediate-age
population in the halo. These were also confirmed in the semi-cosmological spiral galaxy
simulation by \citet{ren05b}, indicating that the M31 halo may have 
a more protracted assembly history compared with the halo of our Galaxy.  
On the other hand, \citet{fon06b} found from the simulation of the stellar halo in our Galaxy based on the hierarchical scenario:
(a) the stellar halo of our Galaxy formed inside out via accretion of satellites;
(b) the inner halo at $R<20$ kpc assembled rapidly, and most of its mass were in place more than
8-9 Gyr ago;
(c) satellites surviving today were accreted within the past few Gyr; and
(d) a major fraction of the present-day stellar halo was originally from a few massive satellites
with $M\sim 10^8 - 10^{10} $ $M_\odot$. 
They also pointed out that the much lower metallicity of the Galactic halo compared with the M31 halo
indicates that our Galaxy could have formed mainly through smaller satellites.
Noting that our Galaxy has a stellar mass and angular momentum that are 2 to 3 times lower than M31, which is a typical spiral galaxy, \citet{ham07} suggested that our Galaxy has 
an exceptionally quiet formation history involving no major merger during the last 10-11 Gyr. 
However, it is noted that the term 'M31 halo' used in these studies corresponds to the extended bulge, 
not to the metal-poor halo in M31. Therefore these studies explain the formation of the
extended bulge in M31 and the halo in our Galaxy. 

M31 might have formed and evolved in a way  similar to our Galaxy, with some differences.
The existence of the rotating metal-rich central bulge, the radial gradient of the 
metallicity and the extended smooth metal-poor halo in M31 can be explained by a rapid dissipative collapse of a proto-Galactic cloud, 
while the extended bulge and its several substructures 
including giant streams and friendless GCs (\citet{iba05, gil07, iba07} and this study),
are evidence for accretion and merging of small scale fragments. 
Differences between M31 and our Galaxy are that the degree of merging was larger,
the scale of merging fragments was larger, and the duration of merging (or the frequency
of merging) was larger in M31 compared with our Galaxy.  
Remaining questions are: 
(a) Why does M31 have an extended metal-rich bulge in M31, which is not seen in our Galaxy?;
(b) Why is the M31 extended bulge a strong rotator?;
(c) Are there metal-poor GCs that do not show systematic rotation in the halo of M31?;
(d) Why do the metal-poor GCs in M31 show some rotation while those in our Galaxy 
do not?; and 
(e) Is there indeed a population of young disk GCs in M31 as posited by
\citet{mor04}, which has no counterpart in the Galaxy?

\section{Summary}

We produced a master catalog of  504 GCs with measured velocity in M31 
combining the data for 211 GCs measured in our survey \citep{kim07} 
and 354 GCs (including 61 common GCs) available in the literature.
Using the photometric and spectroscopic database of these GCs,
we have investigated the kinematics of the GC system of M31.
Our velocities 
are in good agreement with published values for objects in common, 
with a typical error of 35.3 km/s.  
Primary results are summarized below:

\begin{enumerate}
\item The mean value and dispersion of the radial velocities
for the 211 GCs measured in this study are
$\overline{v_p}=-281_{-13}^{+13}$ km s$^{-1}$ and $\sigma_p=
185_{-6}^{+7}$ km s$^{-1}$, respectively. Using all 504 GCs in the
master catalog, we obtain the mean value and dispersion of the
radial velocities as $\overline{v_p}=-285_{-8}^{+8}$ km~s$^{-1}$
and $\sigma_p=178_{-4}^{+4}$ km~s$^{-1}$, respectively.

\item For all    GCs,
the rotation amplitude estimated by fitting the rotation curve is 
  $v_{\rm rot}=188_{-28}^{+34}$ km s$^{-1}$,
while the one from the velocity histogram is 
  $v_{\rm rot}=188_{-33}^{+39}$ km s$^{-1}$.
The rotation amplitude decreases as the distance along the minor axis of M31
increases. 
However, the rotation amplitude is not zero for the outermost samples at $|Y|\ge 5$ kpc,
indicating an extended rotating bulge or a hot rotating halo. 

\item We have identified 50 friendless GCs among all 504 GCs.
It appears that the friendless GCs rotate around the major axis of
M31 unlike the rotation around the minor axis for the disk
population, which calls for more sophisticated investigation. A KS
test yields that the difference between the friendless and
normal GCs is significant for the distribution of $T_1$
magnitude and ($C-T_1$) color.

\item For the subsamples of the metal-poor GCs and metal-rich GCs,
it is found that the metal-rich GCs are more centrally
concentrated than the metal-poor GCs, and that the rotation amplitudes for
the metal-poor GCs and metal-rich GCs are similar.
For the subsamples of bright and faint GCs, it is
found that the bright GCs are more centrally concentrated than the
faint GCs, and that the rotation for the faint GCs is stronger
than that for bright GCs.

\item The rotation-corrected velocity dispersion for all GCs  
is estimated to be $\sigma_{p,r}\sim130$ km s$^{-1}$, and it increases from 
$\sigma_{p,r}\sim120$ km s$^{-1}$ at  $|Y|< 1$ kpc to 
$\sigma_{p,r}\sim150$ km s$^{-1}$ at  $|Y|\ge 5$ kpc.
 These results are very similar to those for the metal-poor GCs.
This shows that there is a dynamically hot halo in M31
that is rotating but primarily pressure-supported. 

\item We have identified 56 GCs and GC candidates with X-ray detection
including 39 genuine GCs with measured velocities. Kinematic
difference between the X-ray emitting and other GCs is not
significant at this stage, but further studies are needed with an
expanded sample of X-ray emitting GCs. The photometric properties
of the X-ray emitting GCs show that the GCs that are redder, more
metal-rich, and brighter are more likely to be detected as X-ray
emitting GCs, as seen for GCs in early-type galaxies

\item We derived the rotation curve of M31 using the metal-poor GCs at
$|Y|<0.6$ kpc 
from 8.9 to 46.4 arcmin. The rotation curve using the GCs in this study
is consistent with those using other tracers in the range $20-45\arcmin$
except for that based on PNe of \citet{mer06}.

\item The dynamical mass of M31 using the kinematic information of the GCs in this study
is estimated to be $M_{\rm PME}= 5.5_{-0.3}^{+0.4} \times 10^{11} M_\odot$
out to a radius of $R\sim55$ kpc using the PME, and
$M_{\rm TME}= 19.2_{-1.3}^{+1.4} \times 10^{11} M_\odot$ at $R\sim100$ kpc
using the TME.


\end{enumerate}

\acknowledgments
The authors are thankful to the anonymous referee for useful commnets that
improved significantly the original manuscript.
The authors are grateful to the staff members of the KPNO
  for their warm support during our observations and data reduction.
The WIYN Observatory is a joint facility of the University of 
 Wisconsin-Madison, Indiana University, Yale University, and 
  the National Optical Astronomy Observatory.
M.G.L. was supported in part 
by the grant R01-2004-000-10490-0 from the Basic Research Program of the Korea
Science and Engineering Foundation.
D.G. gratefully acknowledges support from the Chilean Centro de 
  Astrof{\'i}sica FONDAP No. 15010003.
A.S. was supported by NSF CAREER grant AST 00-94048.

\clearpage

\begin{deluxetable}{crccccc}
\tablewidth{0pc} 
\tablecaption{Kinematics of the M31 Globular Cluster System\label{tab-kin}}

\tablehead{
$|Y|$  & N   & $\overline{v_p}^a$   & $\sigma_p^b$      & ${\sigma}_{p,r}^c$ & $v_{\rm rot, 1}^d$  & $v_{\rm rot, 2}^e$ \\
(kpc)  &     & (km s$^{-1}$)      & (km s$^{-1}$)    & (km s$^{-1}$)     & (km s$^{-1}$)   & (km s$^{-1}$)}

\startdata

\multicolumn{7}{c}{All GCs} \\
\hline
0$\leq$ & 504 & $-285_{-8}^{+8}$      & $178_{-4}^{+4}$      &$134_{-   5}^{+   5}$ & $188_{-28}^{+34}$  & $188_{-33}^{+39}$ \\
$0-1$     &  149&$ -287_{-  16}^{+  19}$&$ 200_{-   8}^{+   8}$&$ 119_{-   8}^{+   9}$& $188_{-51}^{+59}$  & $234_{-17}^{+53}$ \\
$1-3$     &  203&$ -297_{-  11}^{+  13}$&$ 172_{-   5}^{+   5}$&$ 110_{-   9}^{+   9}$& $180_{-25}^{+23}$  & $147_{-17}^{+12}$ \\
$3-5$     &   65&$ -286_{-  19}^{+  19}$&$ 156_{-  13}^{+  14}$&$ 141_{-  11}^{+  11}$& $122_{-110}^{+148}$& $89_{-110}^{+148}$ \\
5$\leq$ &   87&$ -257_{-  18}^{+  19}$&$ 162_{-  12}^{+  12}$&$ 151_{-   9}^{+   9}$& $98_{-57}^{+70}$   & $59_{-24}^{+55}$ \\
\hline

\multicolumn{7}{c}{Metal-Poor GCs ([Fe/H]$<-0.905$)} \\
\hline
0$\leq$ & 310   & $-294_{-11}^{+11}$    & $183_{-5}^{+5}$      &$ 129_{-   6}^{+   7}$& $193_{-41}^{+44}$  & $151_{-8}^{+8}$    \\
$0-1$     &   79&$ -296_{-  24}^{+  27}$&$ 210_{-   9}^{+   9}$&$ 119_{-  15}^{+  17}$& $297_{-109}^{+124}$& $239_{-16}^{+17}$ \\
$1-3$     &  135&$ -306_{-  15}^{+  15}$&$ 175_{-   6}^{+   7}$&$ 111_{-  10}^{+  12}$& $206_{-16}^{+27}$  & $152_{-13}^{+13}$ \\
$3-5$     &   37&$ -302_{-  28}^{+  30}$&$ 167_{-  20}^{+  21}$&$ 201_{-  24}^{+  23}$& $301_{-169}^{+196}$& $64_{-47}^{+114}$ \\
5$\leq$ &   59  &$ -260_{-  22}^{+  23}$&$ 163_{-  13}^{+  14}$&$ 145_{-  11}^{+  11}$& $100_{-93}^{+110}$ & $66_{-36}^{+52}$ \\
\hline
\multicolumn{7}{c}{Metal-Rich GCs ([Fe/H]$>-0.905$)} \\
\hline
0$\leq$ & 121 & $-250_{-14}^{+15}$   & $154_{-9}^{+11}$     &$ 121_{-  10}^{+   9}$& $191_{-37}^{+34}$ & ...               \\
$0-1$     &  50&$ -258_{-  26}^{+  24}$&$ 176_{-  17}^{+  20}$&$ 107_{-  22}^{+  24}$& $315_{-91}^{+131}$& $195_{-48}^{+66}$ \\
$1-3$     &  43&$ -240_{-  25}^{+  28}$&$ 151_{-  16}^{+  16}$&$ 112_{-  15}^{+  16}$& $215_{-76}^{+92}$ & $72_{-52}^{+88}$  \\
$3-5$     &  18&$ -216_{-  37}^{+  33}$&$ 135_{-  21}^{+  21}$&...                   & ...               & ...               \\
5$\leq$ &  10&$ -274_{-  33}^{+  31}$&$  92_{-  25}^{+  40}$&...                   & ...               & ...               \\

\hline
\multicolumn{7}{c}{Bright GCs ($T_1 \leq 16.9$)} \\
\hline
0$\leq$ & 201 & $-286_{-12}^{+12}$ & $163_{-7}^{+8}$ &$ 146_{-   7}^{+   7}$ & $129_{-35}^{+35}$ & $144_{-46}^{+43}$  \\

\hline
\multicolumn{7}{c}{Faint GCs ($16.9<T_1 \leq 18.5$)} \\
\hline
0$\leq$ & 203 & $-273_{-13}^{+13}$ & $186_{-6}^{+7}$ &$ 107_{-   8}^{+   8}$ & $209_{-15}^{+21}$ & $178_{-23}^{+23}$  \\

\enddata

\tablenotetext{a}{Mean velocity.}
\tablenotetext{b}{Velocity dispersion.}
\tablenotetext{c}{Rotation-corrected velocity dispersion.}
\tablenotetext{d}{Rotation velocity derived from the rotation curve 
in the $v_p - v_{M31}$ vs. $X$ diagram.}
\tablenotetext{e}{Rotation velocity derived from the velocity histogram.}
\end{deluxetable}
\begin{deluxetable}{crccccc}
\tablewidth{0pc} 
\tablecaption{Kinematics of the M31 Planetary Nebulae System\label{tab-kinpn}}

\tablehead{
$|Y|$  & N   & $\overline{v_p}$   & $\sigma_p$      & ${\sigma}_{p,r}$ & $v_{\rm rot, 1}$  & $v_{\rm rot, 2}$ \\
(kpc)  &     & (km s$^{-1}$)      & (km s$^{-1}$)    & (km s$^{-1}$)     & (km s$^{-1}$)   & (km s$^{-1}$)}

\startdata

0$\leq$ & 2526&$ -305_{-   3}^{+   3}$&$ 167_{-   1}^{+   1}$&$  94_{-   1}^{+   1}$&$ 187_{-13}^{+12}$&$198_{-33}^{+58}$\\
$0-1$   & 1016&$ -320_{-   6}^{+   6}$&$ 189_{-   2}^{+   2}$&$  97_{-   2}^{+   2}$&$ 238_{-10}^{+9}$&$213_{-7}^{+8}$\\
$1-3$   &  914&$ -312_{-   5}^{+   5}$&$ 165_{-   2}^{+   2}$&$  75_{-   2}^{+   2}$&$ 181_{-8}^{+8}$&$146_{-6}^{+4}$\\
$3-5$   &  350&$ -269_{-   7}^{+   7}$&$ 130_{-   3}^{+   4}$&$  88_{-   4}^{+   4}$&$ 187_{-16}^{+13}$&$106_{-57}^{+109}$\\
5$\leq$ &  246&$ -272_{-   6}^{+   6}$&$ 102_{-   5}^{+   5}$&$  79_{-   3}^{+   3}$&$ 76_{-26}^{+32}$&$29_{-19}^{+155}$\\
\enddata
\end{deluxetable}

\clearpage

\clearpage
\begin{figure}
\includegraphics [width=165mm] {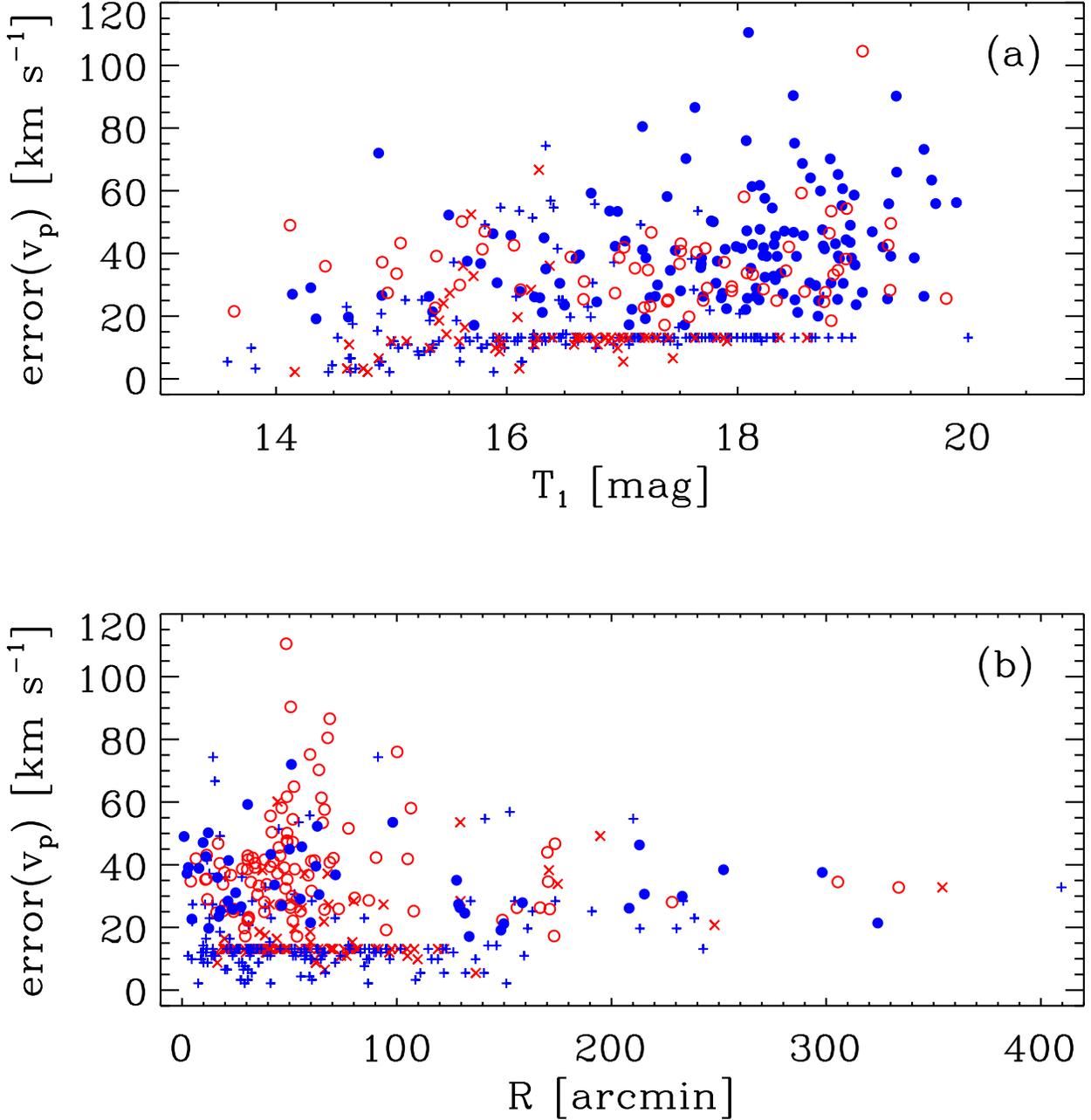}
\caption{Errors of the measured velocities of the M31 GCs versus $T_1$ magnitude (a) and versus $R$, a galactocentric radial distance corrected for the inclination of M31 (b).
In panel (a), filled circles for 136 metal-poor GCs derived in this study,
 open circles for 63 metal-rich GCs derived in this study,
plusses for 174 metal-poor GCs derived in the previous studies, and
crosses for 58 metal-rich GCs derived in the previous studies.
In panel (b), filled circles for 46 bright GCs derived in this study,
 open circles for 99 faint GCs derived in this study,
plusses for 155 bright GCs derived in the previous studies, and
crosses for 104 faint GCs derived in the previous studies.
} \label{fig-verr}
\end{figure}
\clearpage
\begin{figure}
\includegraphics [width=165mm] {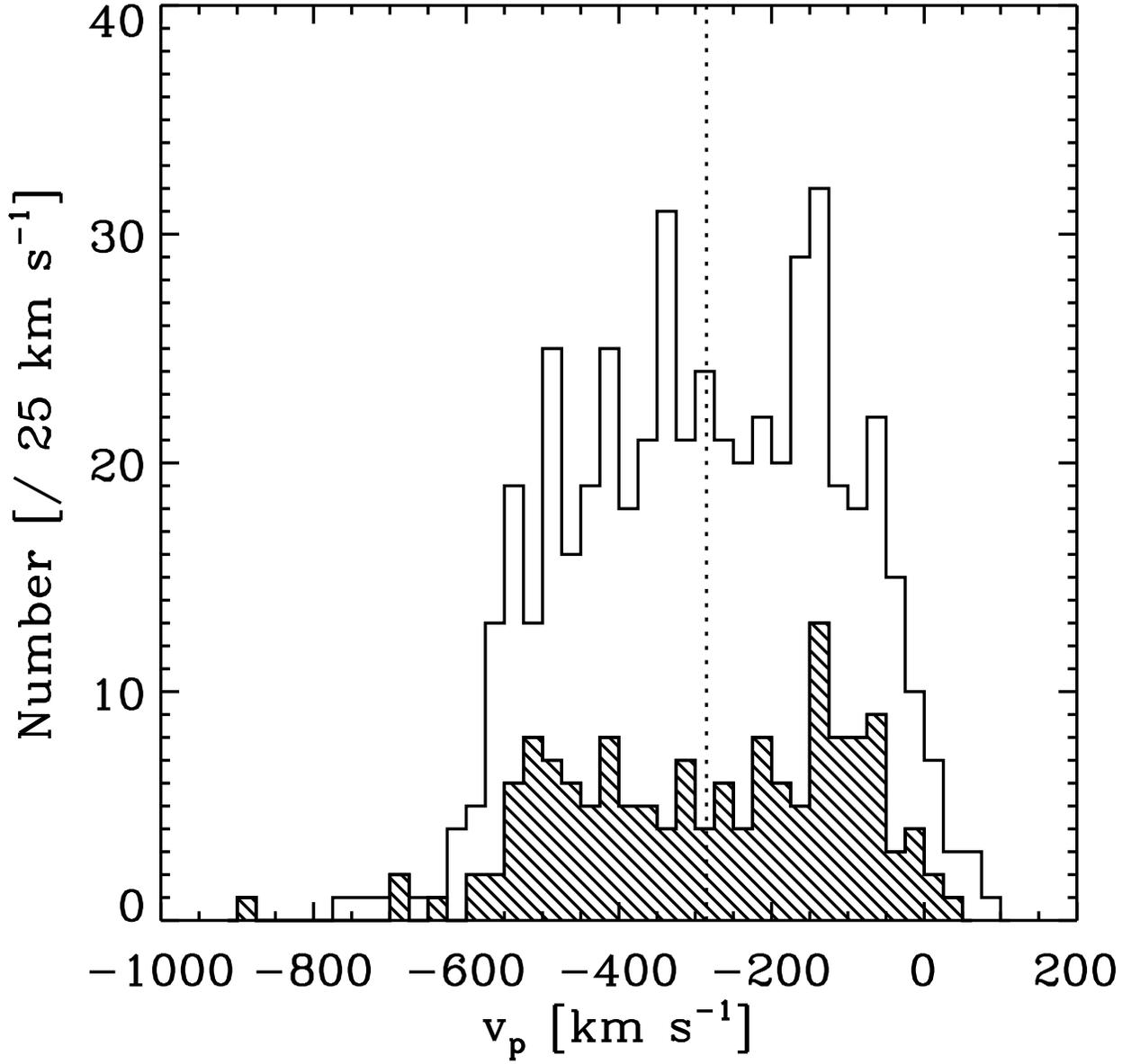}
\caption{Radial velocity histogram for all 504 GCs in the master
catalog (open histogram) compared with 211 GCs measured in
this study (hatched histogram). The vertical dotted line indicates
the mean value of the radial velocities, $\overline{v_p}=-285$
km~s$^{-1}$, for all GCs.} \label{fig-velhist}
\end{figure}
\clearpage

\begin{figure}
\includegraphics [width=165mm] {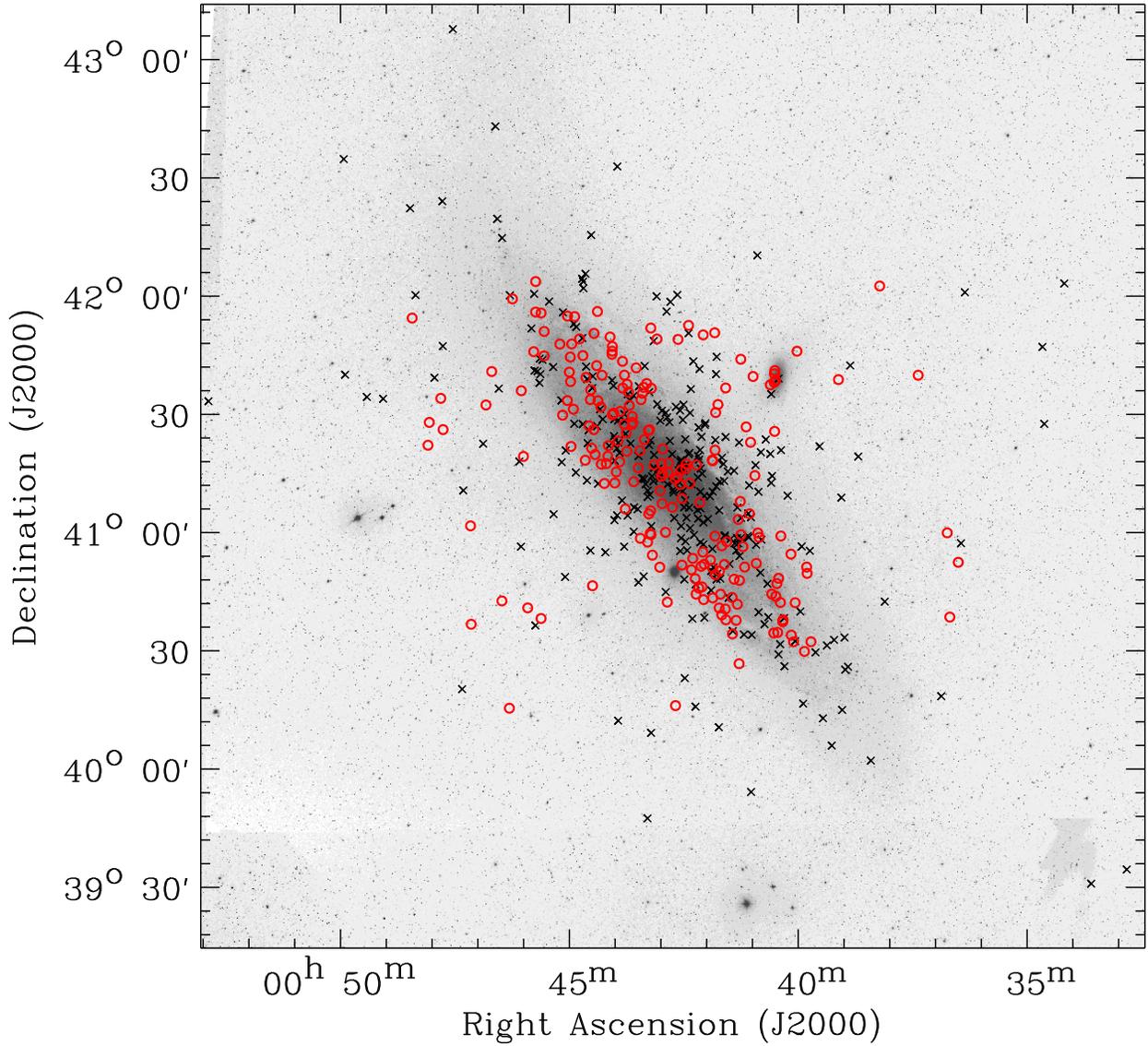}
\caption{Positions of 211 GCs measured in this study (open
circles) and of 293 GCs measured in the literature (crosses)
excluding 61 GCs overlapping with our survey, overlaid on a
$4\arcdeg \times 4\arcdeg$ optical image of M31 from the Digitized
Sky Survey. } \label{fig-img}
\end{figure}
\clearpage

\begin{figure}
\includegraphics [width=165mm] {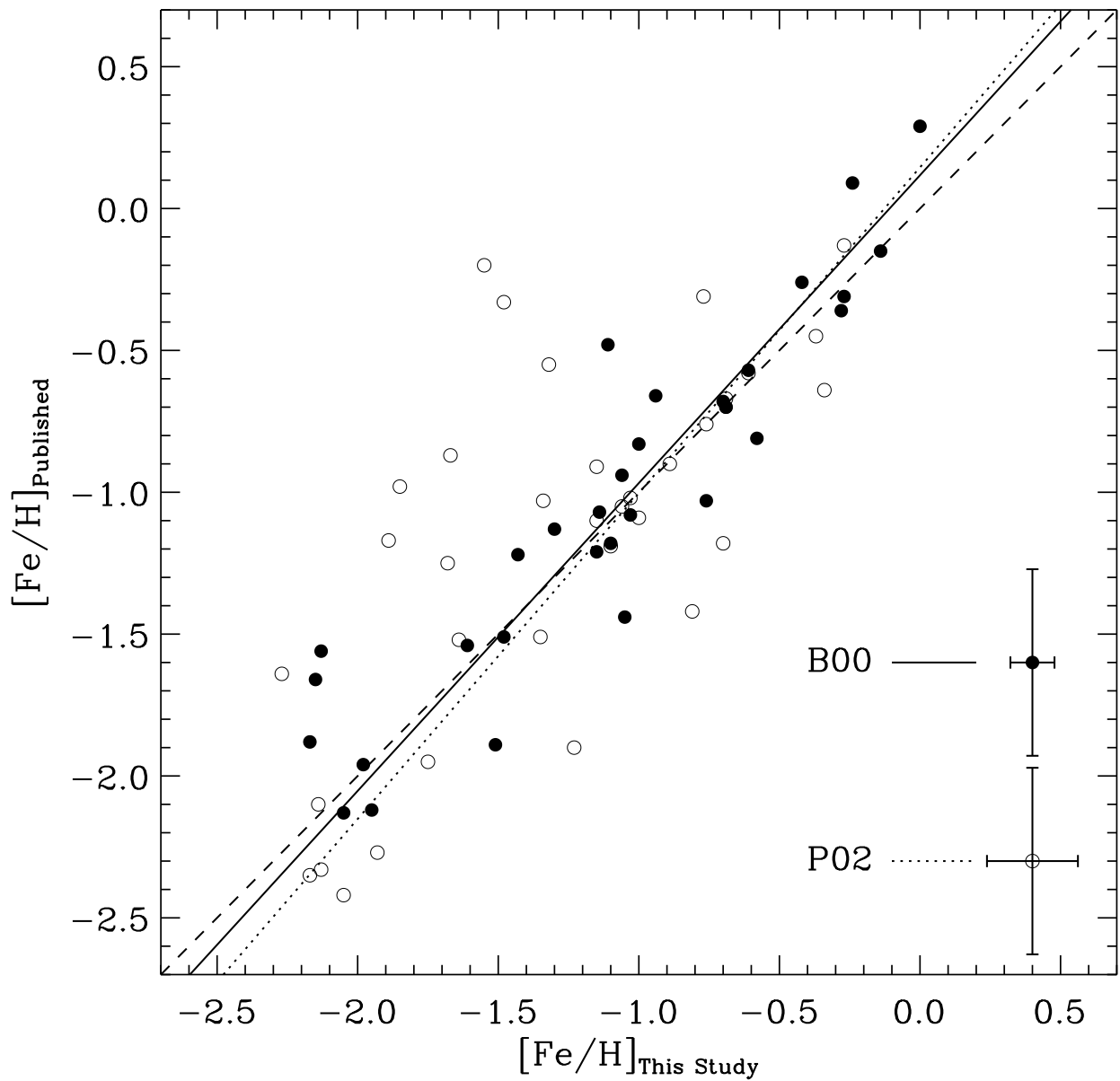}
\caption{Comparison of the measured metallicities for M31 GCs
measured in this study with those in the literature. Filled
circles indicate 32 GCs common between this study and
\citet[B00]{bar00}, and open circles 34 GCs common between this
study and \citet[P02]{per02}. The dashed line denotes the
one-to-one relation, and the solid line (between this study and B00)
and dotted line (between this study and P02)  indicate the linear
least-squares fits. Typical errors of measured metallicities are
shown by errorbars according to the reference.}
\label{fig-fehcomp}
\end{figure}

\begin{figure}
\includegraphics [width=165mm] {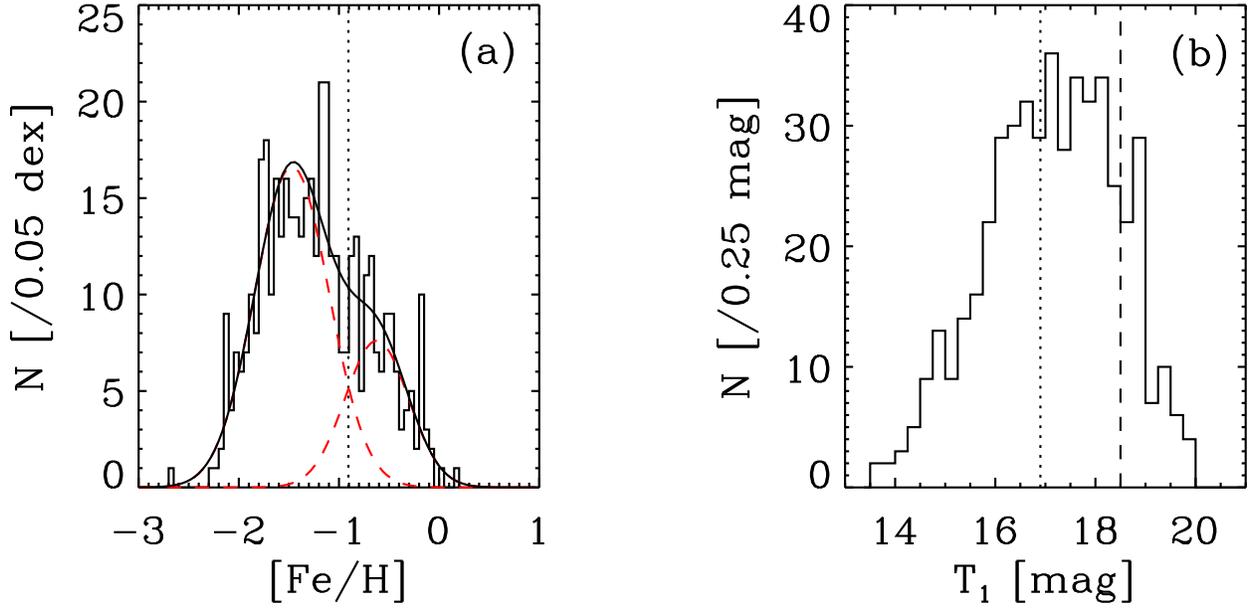}
\caption{(a) Metallicity distribution for 431 GCs with measured
metallicities. The vertical dotted line indicates the value, found
from the KMM test, dividing the GCs into metal-poor and
metal-rich subcomponents. The dashed lines represent the double Gaussian
fits, and the solid line indicates the sum of the two Gaussian fits.
(b) Luminosity function for 483 GCs found in the photometric
catalog of our survey. The vertical dashed line denotes the
magnitude limit to make subsamples of the bright and faint GCs,
and the vertical dotted line indicates the dividing magnitude for the subsamples.}
\label{fig-subsam}
\end{figure}
\clearpage

\begin{figure}
\includegraphics [width=165mm] {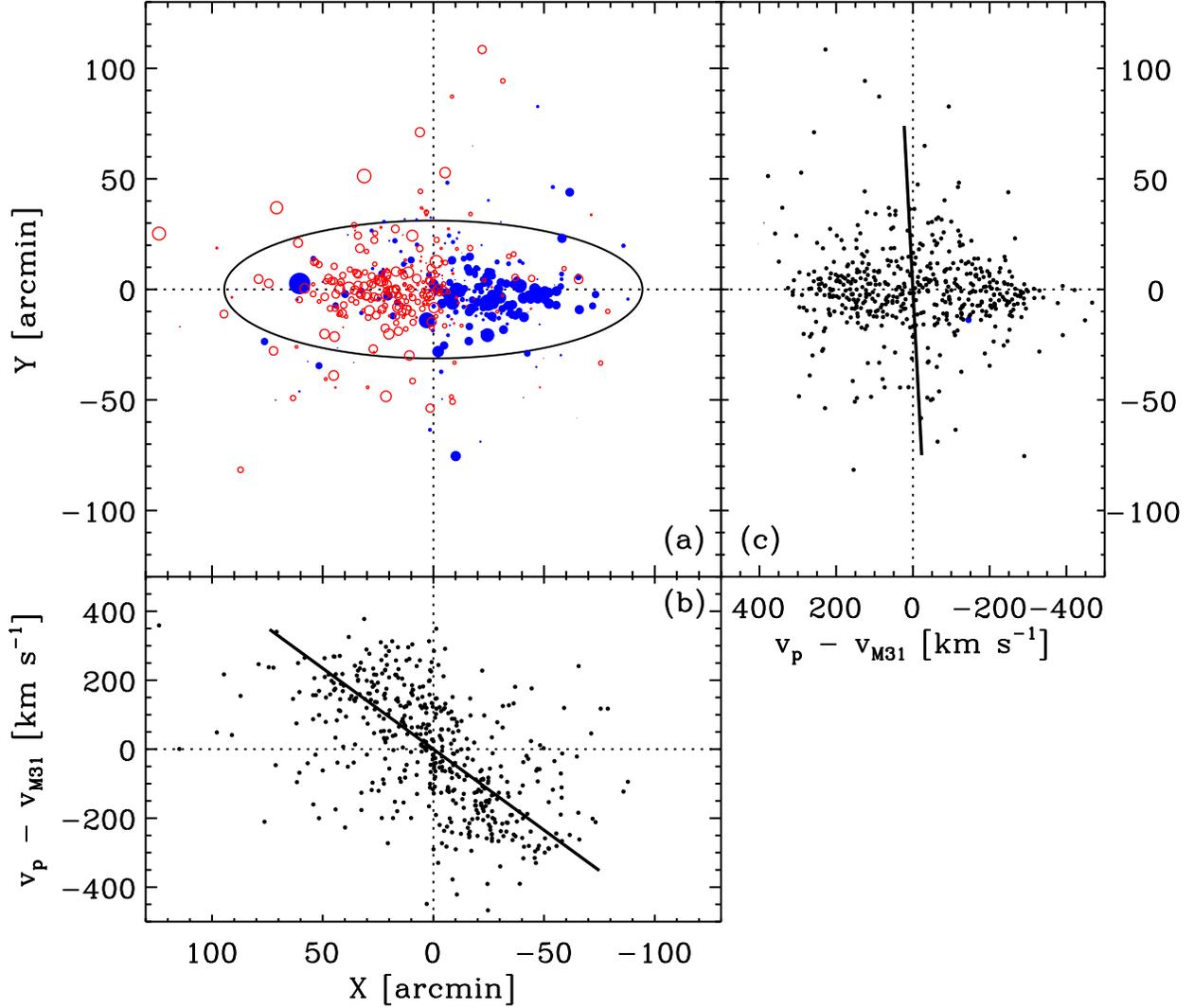}
\caption{Spatial distribution of all GCs with measured
velocity (a), and radial velocities for M31 GCs versus the
projected radii along the major axis (b) and along the minor axis
(c). The receding GCs are plotted by open symbols, while the
approaching GCs by filled symbols. The symbol size is relatively
proportional to the velocity deviation. The large ellipse represents the
optical extent of M31 based on the standard diameter measured
at a level of $25$ mag arcsec$^{-2}$ and ellipticity from
\citet{kar04}. Solid lines in (b) and (c) indicate the linear
least-squares fits. } \label{fig-spatvel}
\end{figure}

\begin{figure}
\includegraphics [width=165mm] {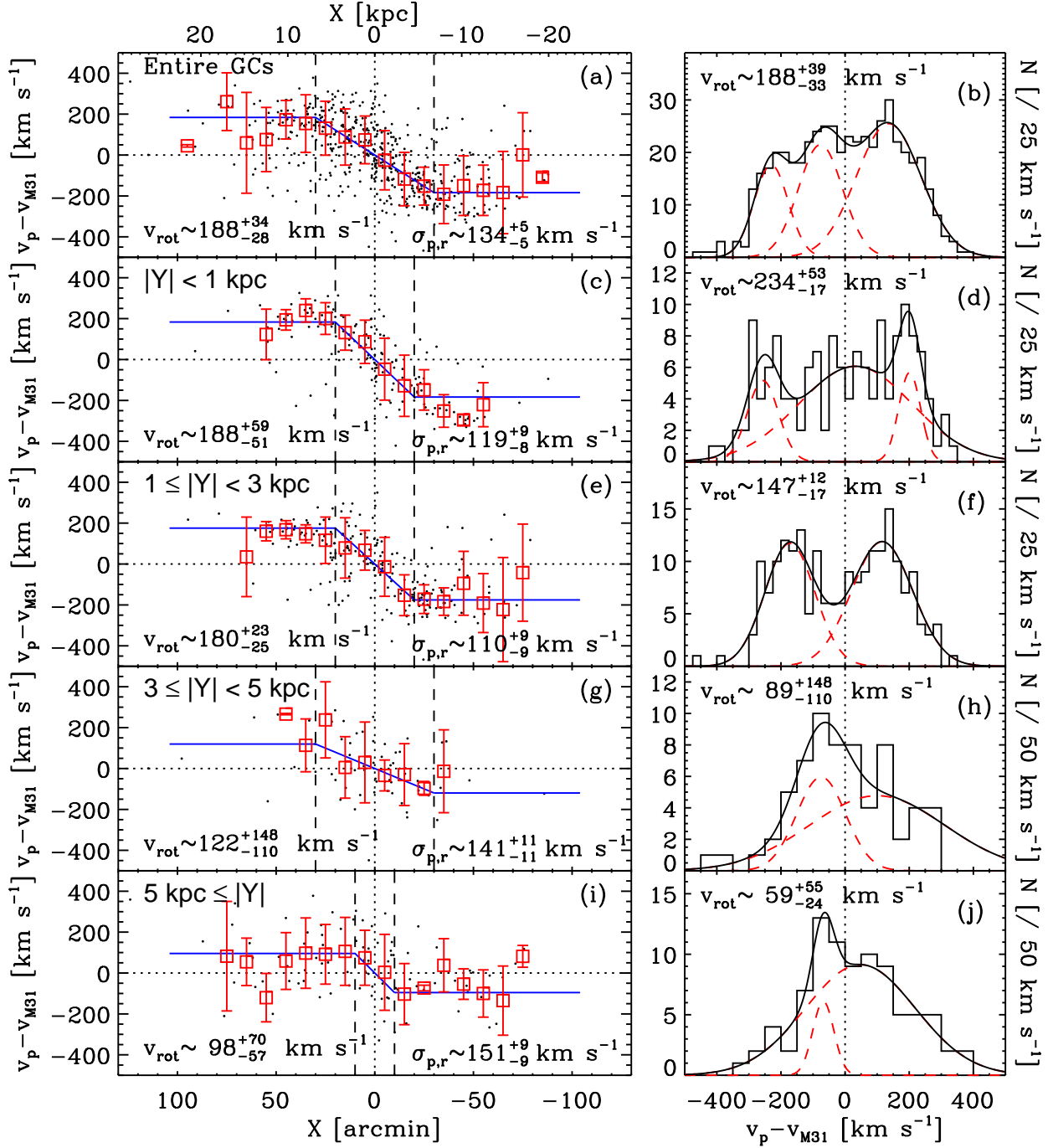}
\caption{Radial velocities for M31 GCs as a function of projected
distance along the major axis (left panels) and velocity
histogram (right panels). Large open squares indicate the mean
radial velocity of the GCs in a distance bin of 10 $\arcmin$ along
the major axis. The vertical errorbar denotes the velocity
dispersion of the GCs in the distance bin. In the velocity
histogram, the sum of individual Gaussian fits is shown by a solid
line, and the individual fits by dashed lines. The rotation fits
for the GCs over the whole region are in (a) and (b), those in the
range $|Y|< 1$ kpc in (c) and (d), those in the range $1\le|Y|< 3$
kpc in (e) and (f), those in the range $3\le|Y|< 5$ kpc in (g) and
(h), and those in the range 5 kpc $\le|Y|$ in (i) and (j).
Vertical dashed lines indicate the boundaries for the fit of a
solid-body rotation. } \label{fig-rotfit}
\end{figure}

\begin{figure}
\includegraphics [width=165mm] {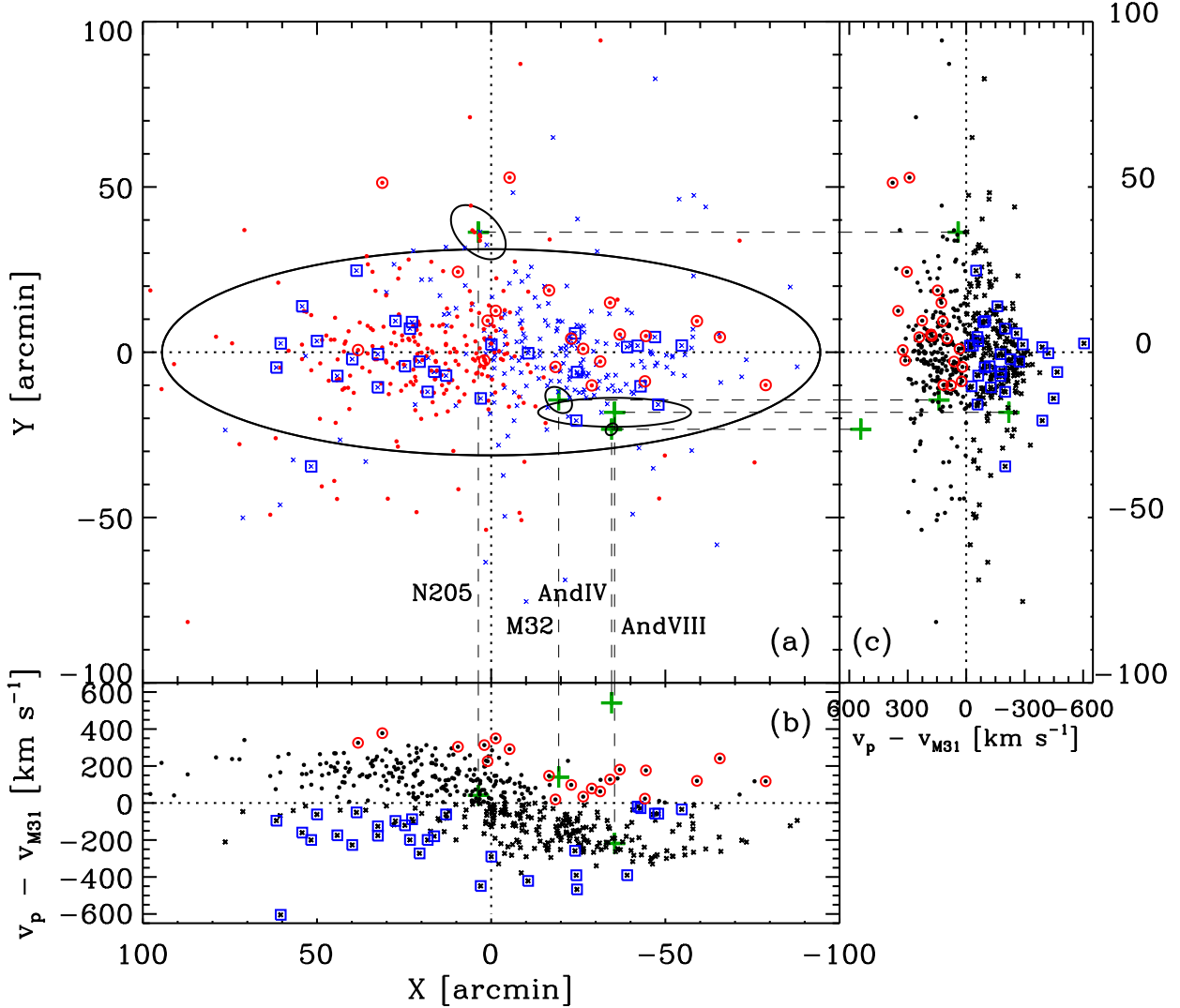}
\caption{Spatial distribution of the friendless GCs (a), and
radial velocities versus the projected radii along the major axis
(b), and along the minor axis (c). The receding GCs are plotted by
filled circles, while the approaching GCs by crosses. The
approaching friendless GCs are indicated by open squares, while
the receding friendless GCs by open circles. The largest ellipse
represents the optical extent of M31 based on the standard
diameter measured at a level of $25$ mag arcsec$^{-2}$ and
ellipticity. Satellite galaxies of M31 are indicated by 
small ellipses and large plus signs. } \label{fig-spatnofof}
\end{figure}

\begin{figure}
\includegraphics [width=165mm] {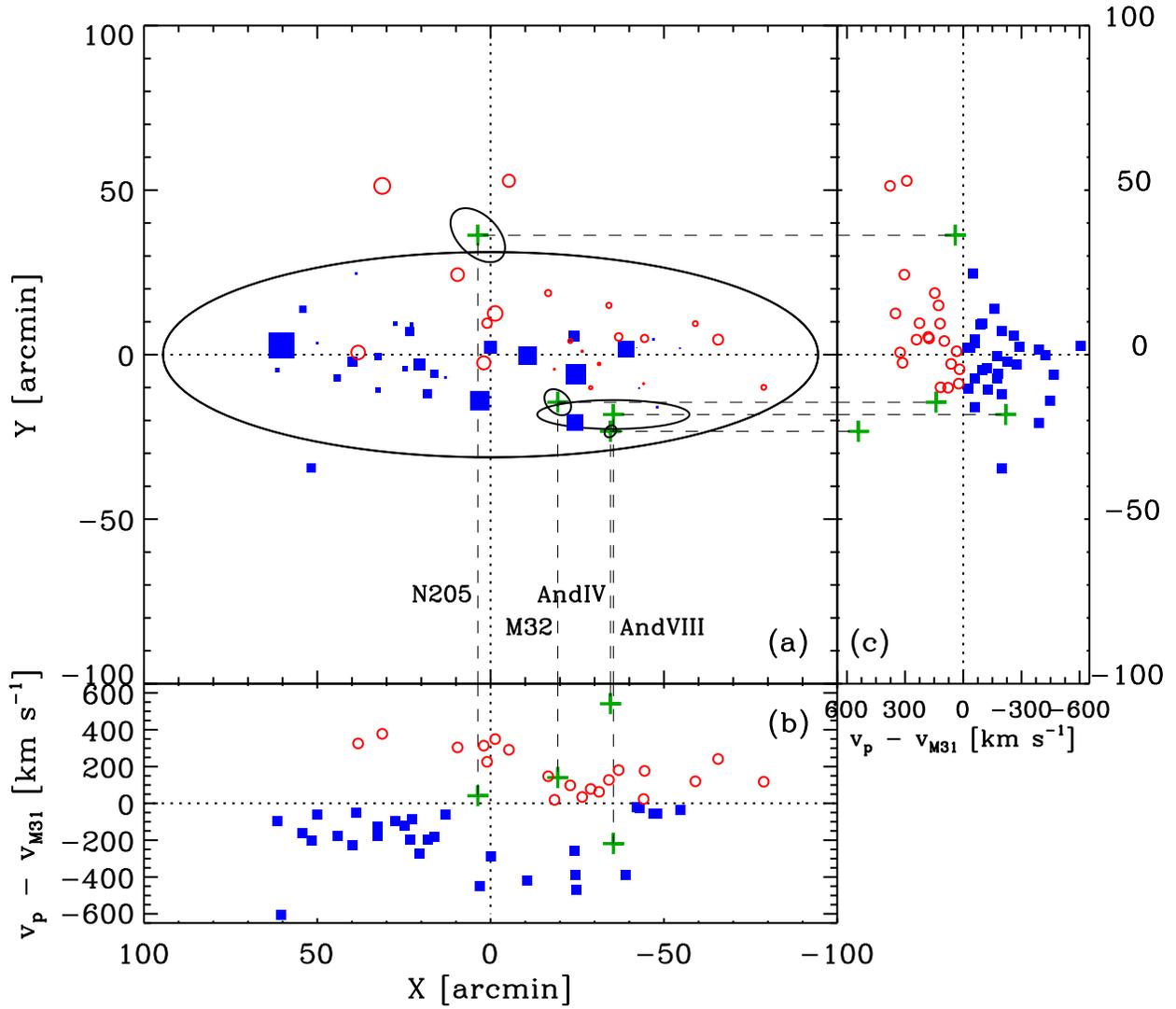}
\caption{Same as Fig. \ref{fig-spatnofof}, but for the friendless
GCs only. The symbol size in panel (a) is relatively proportional to the
velocity deviation. } \label{fig-spatvelnofof}
\end{figure}

\begin{figure}
\includegraphics [width=165mm] {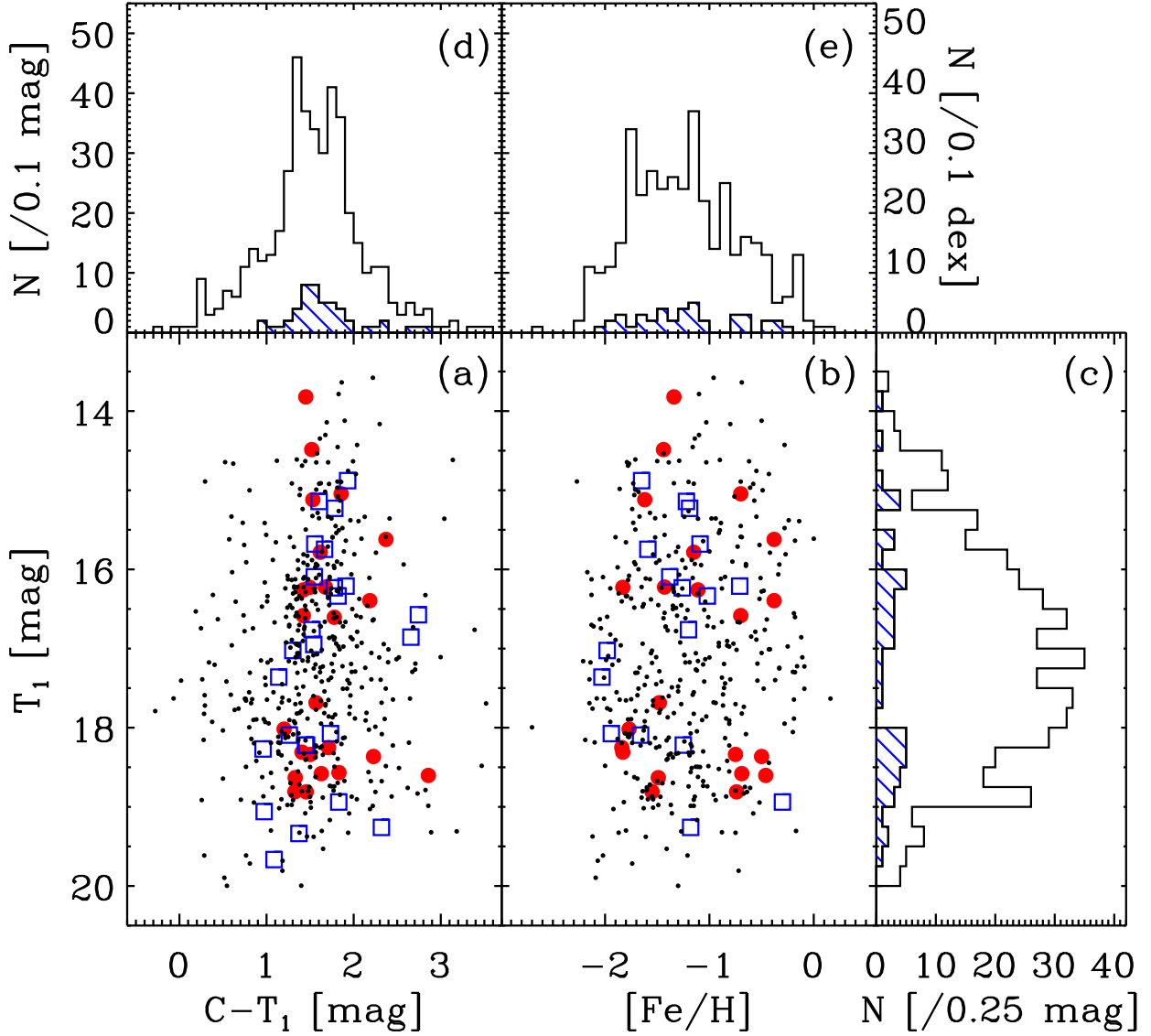}
\caption{$T_1$ magnitudes versus ($C-T_1$) colors (a) and versus
metallicities ([Fe/H]) (b), with histograms for $T_1$ magnitudes
(c), ($C-T_1$) colors (d), and metallicities (e) for the
friendless GCs. Normal         GCs are indicated by small dots.
The friendless GCs with smaller velocities than the rotation curve
shown in Fig. \ref{fig-spatvel} (b), are represented by open
rectangles, and the friendless GCs with larger velocities than the
rotation curve shown in Fig. \ref{fig-spatvel} (b) by filled
circles. The histogram for the friendless GCs is cross-hatched.
} \label{fig-photnofof}
\end{figure}

\begin{figure}
\includegraphics [width=165mm] {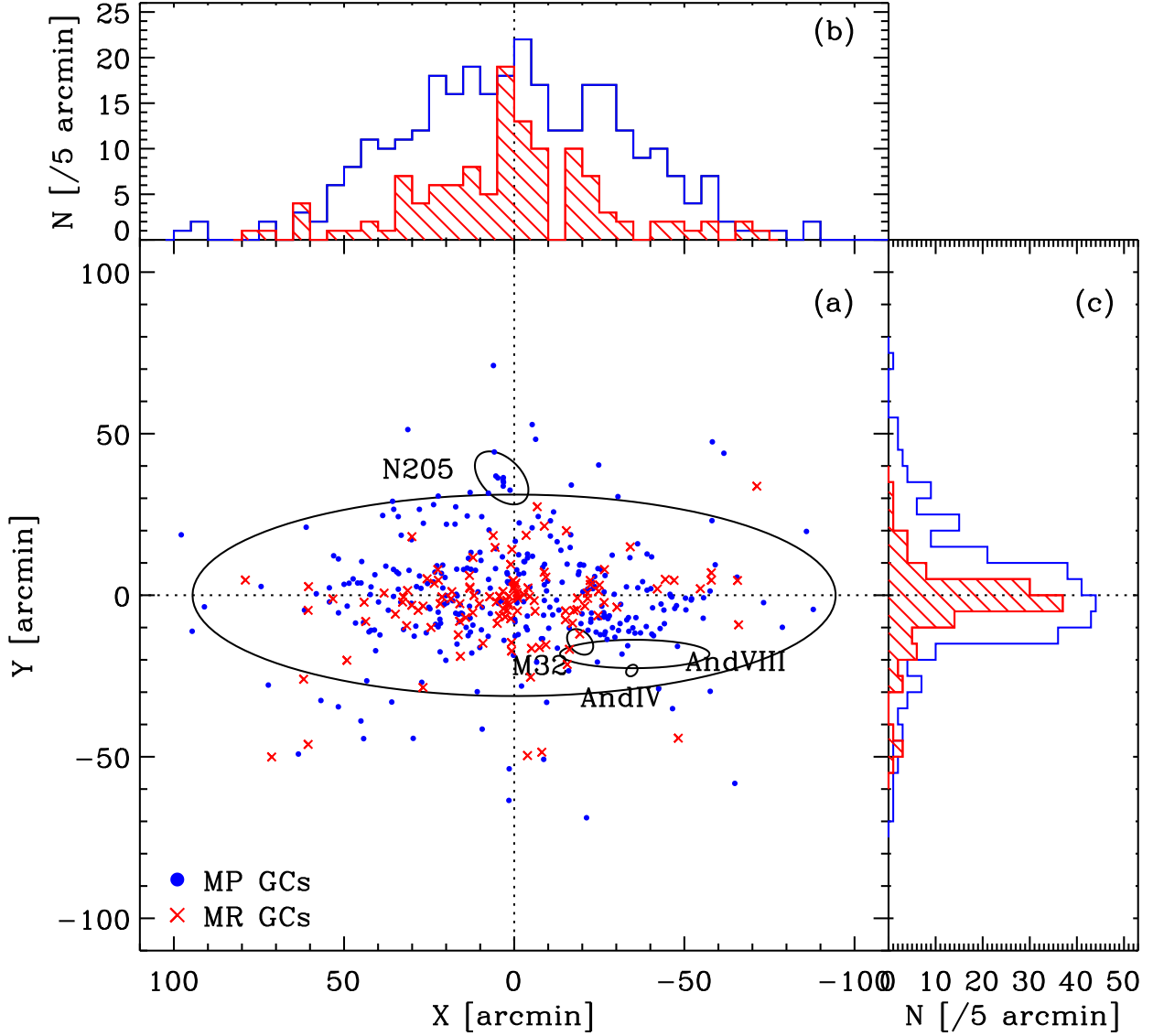}
\caption{Spatial distribution of the metal-poor (filled circles)
and metal-rich (crosses) GCs (a), and the histogram along the
major axis (b), and along the minor axis (c). The largest ellipse
represents the optical extent of M31 based on the standard
diameter measured at a level of $25$ mag arcsec$^{-2}$ and
ellipticity. Satellite galaxies of M31 are indicated by 
small ellipses. The histograms for the metal-poor
and metal-rich GCs are indicated by open and hatched histograms,
respectively. } \label{fig-spathistfeh}
\end{figure}
\clearpage

\begin{figure}
\includegraphics [width=165mm] {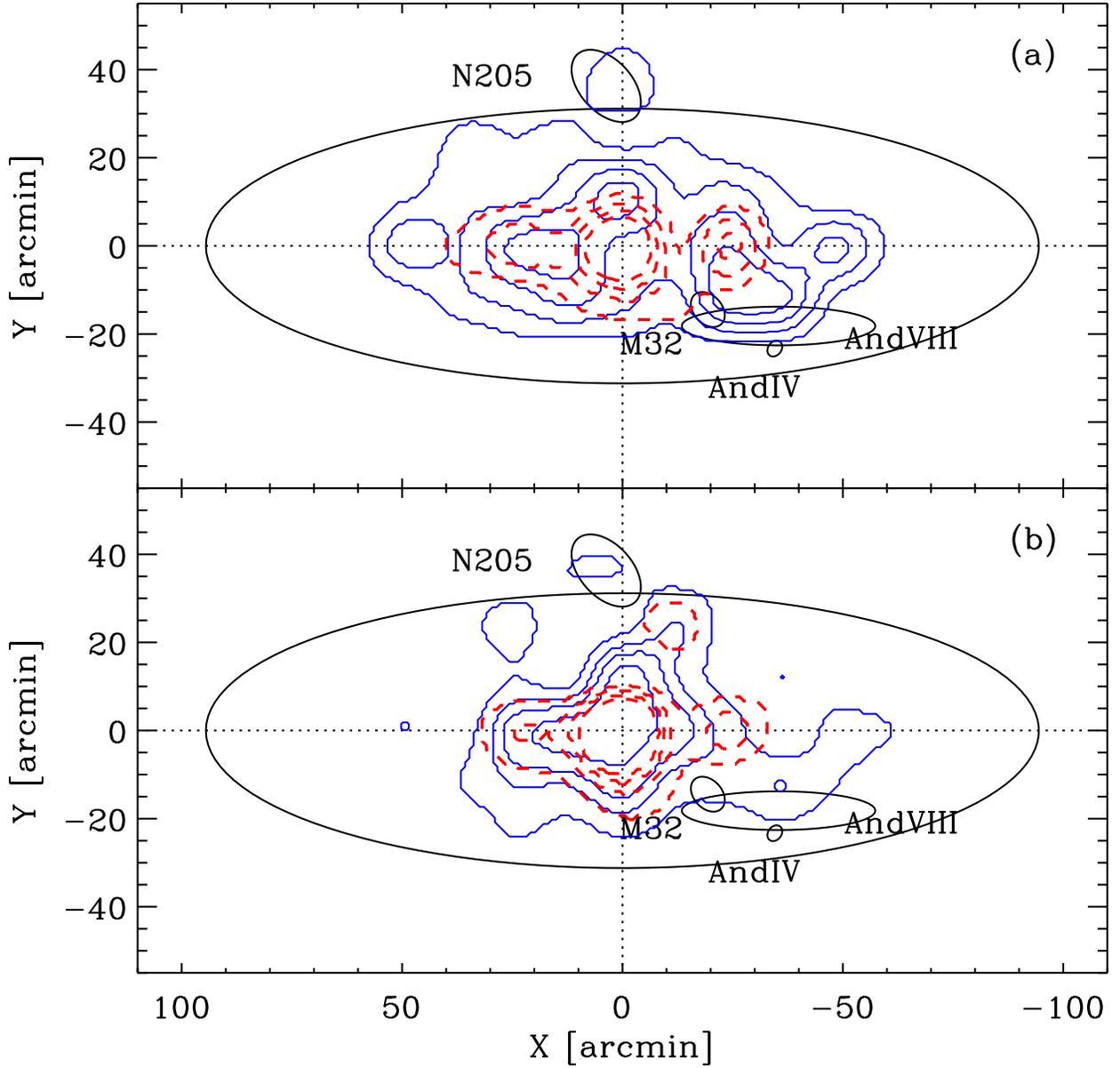} 
\caption{(a) Number density contours for all the metallicity
subsamples (solid lines for the metal-poor GCs and dashed lines for the metal-rich GCs), 
and (b) those for the 
the metallicity subsamples with $T_1<17$.
The contour levels are 0.01, 0.02, 0.03, 0.04 (GCs/arcmin$^2$) in (a)
and 0.02, 0.04, 0.06, 0.09 (GCs/arcmin$^2$) in (b).
The largest ellipse
represents the optical extent of M31 based on the standard
diameter measured at a level of $25$ mag arcsec$^{-2}$ and
ellipticity. Satellite galaxies of M31 are indicated by 
small ellipses. } \label{fig-spatconfeh}
\end{figure}

\begin{figure}
\includegraphics [width=165mm] {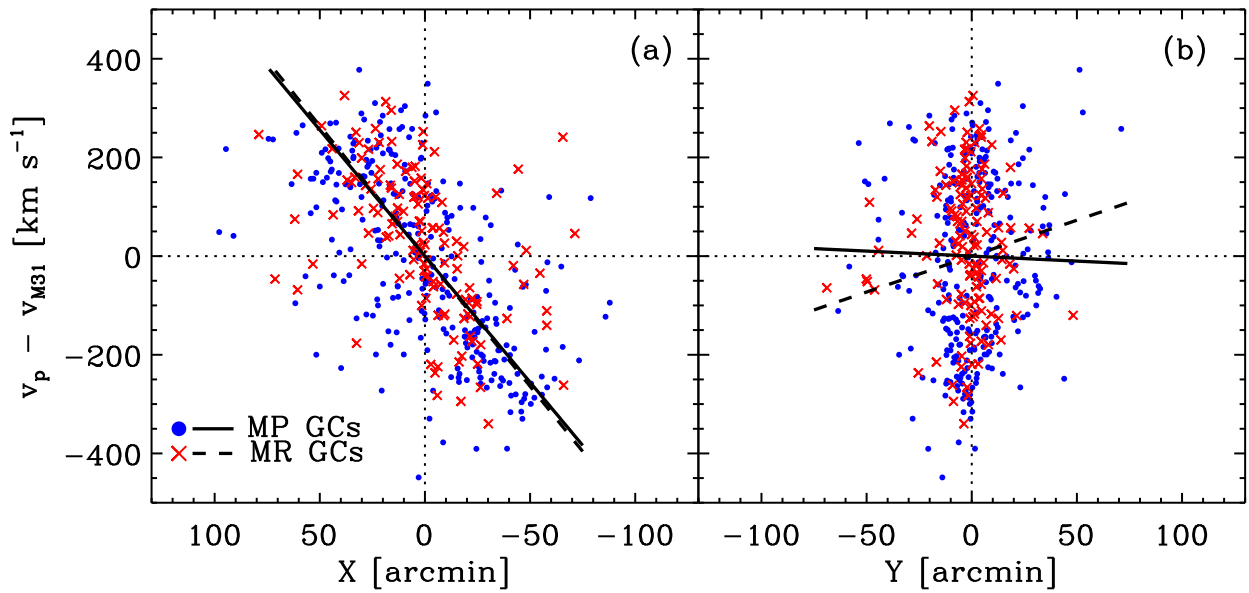}
\caption{Radial velocities for the metal-poor (MP, filled circles)
and metal-rich (MR, crosses) GCs versus
the projected radii along the major axis (a) and
along the minor axis (b).
The least-squares fits for the MP and MR GCs
are indicated by the solid and dashed lines, respectively.
} \label{fig-spatvelfeh}
\end{figure}

\begin{figure}
\includegraphics [width=165mm] {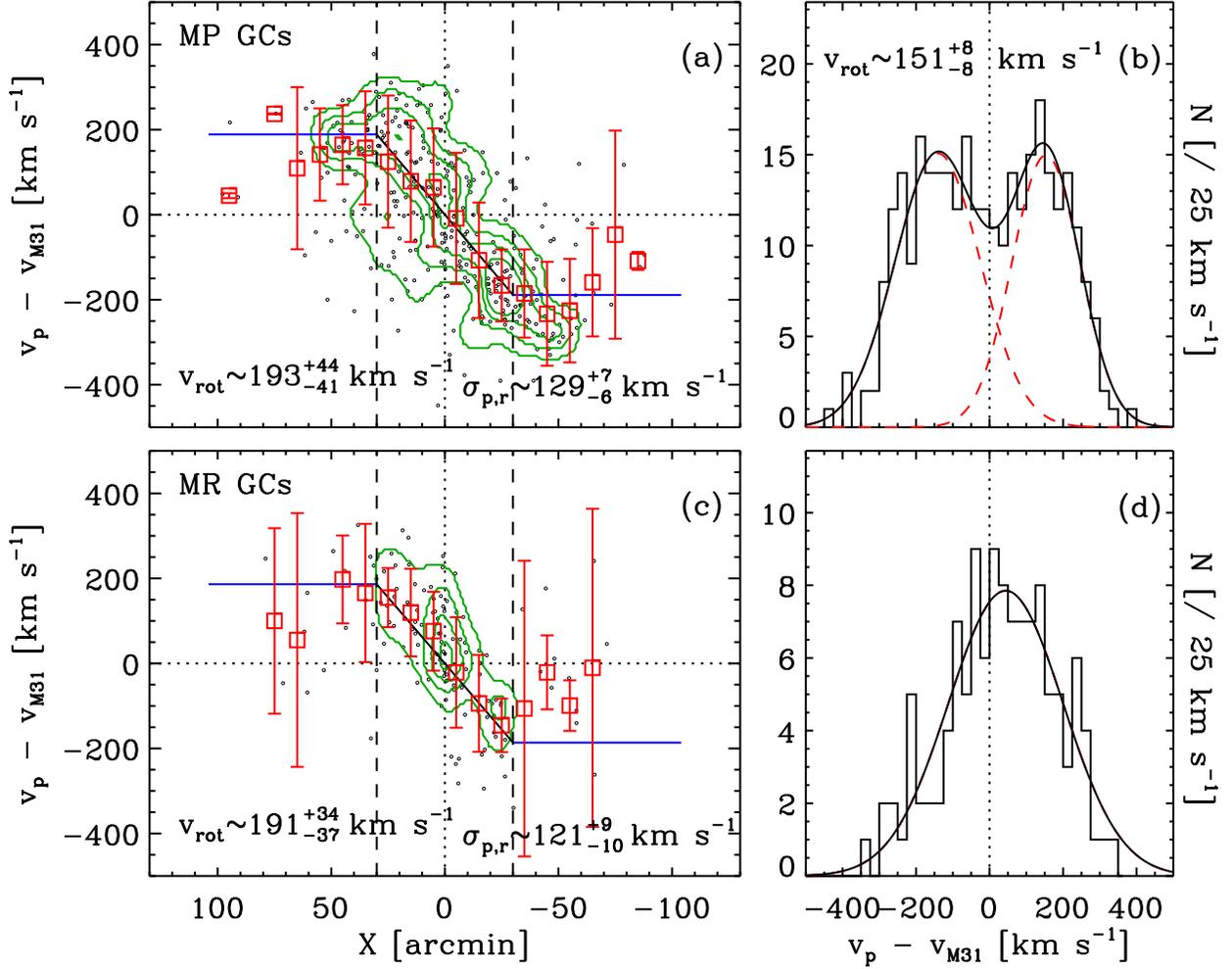}
\caption{Radial velocities (dots) versus the projected distances along
the major axis ($X$) and the velocity histogram for the MP (a,b)
and MR (c,d) GCs. Large open squares indicate the mean radial
velocity of the GCs in a distance bin of 10 $\arcmin$ along the
major axis. The vertical errorbar denotes the velocity dispersion
of the GCs in the distance bin. The contours indicate the number
densities in the space of radial velocity and $X$,
and their levels are the same in (a) and (c). In the
velocity histogram, the sum of individual Gaussian fits is shown by
a solid line, and the individual fits by dashed lines. Vertical
dashed lines indicate the boundaries for the fit of a solid-body
rotation. } \label{fig-rotfitfeh}
\end{figure}
\clearpage

\begin{figure}
\includegraphics [width=165mm] {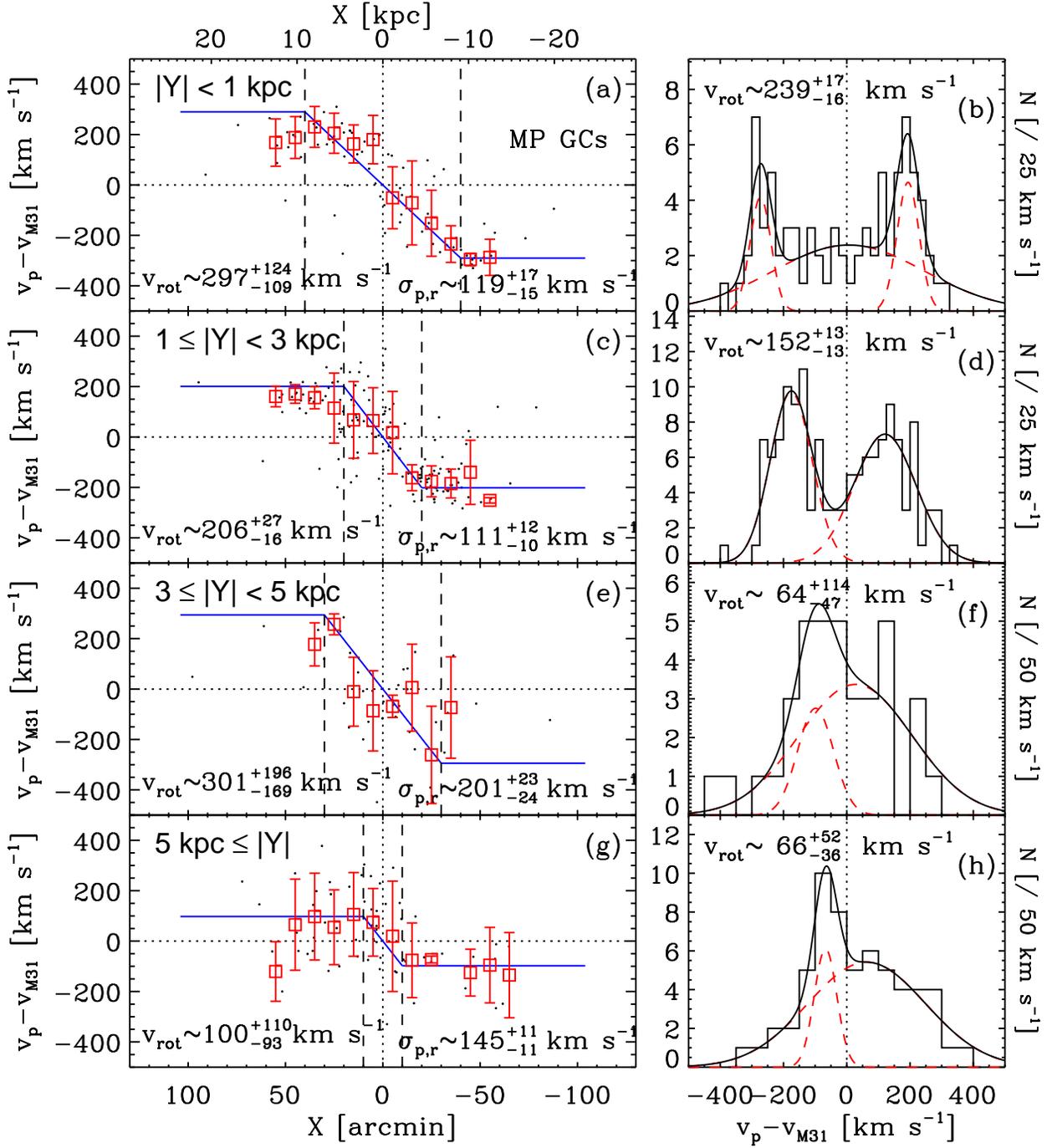} 
\caption{Same as Fig. \ref{fig-rotfit}, but for the metal-poor GCs.} \label{fig-rotfitmp}
\end{figure}
\clearpage

\begin{figure}
\includegraphics [width=165mm] {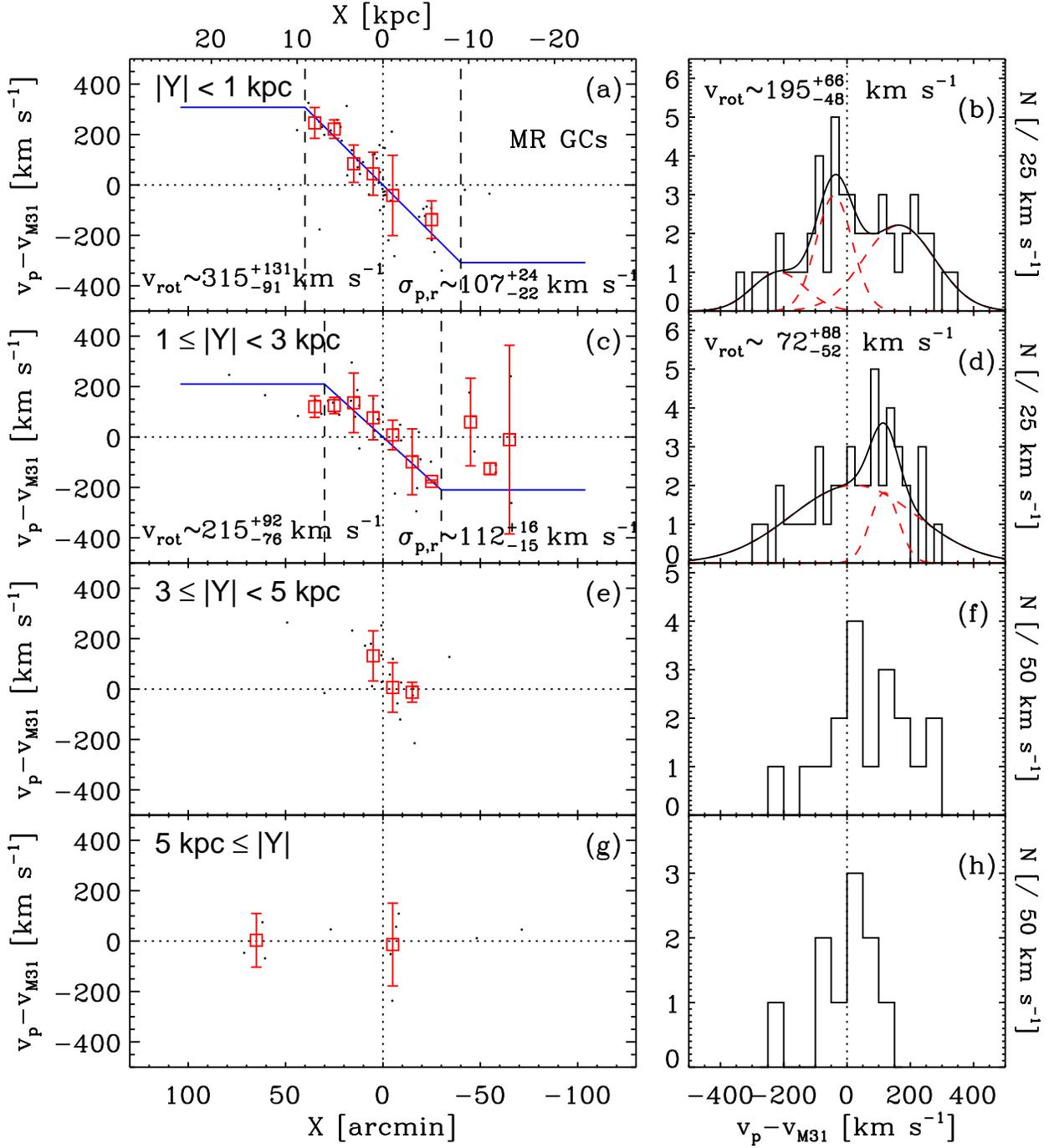} 
\caption{Same as Fig. \ref{fig-rotfit}, but for the metal-rich GCs.} \label{fig-rotfitmr}
\end{figure}
\clearpage

\begin{figure}
\includegraphics [width=165mm] {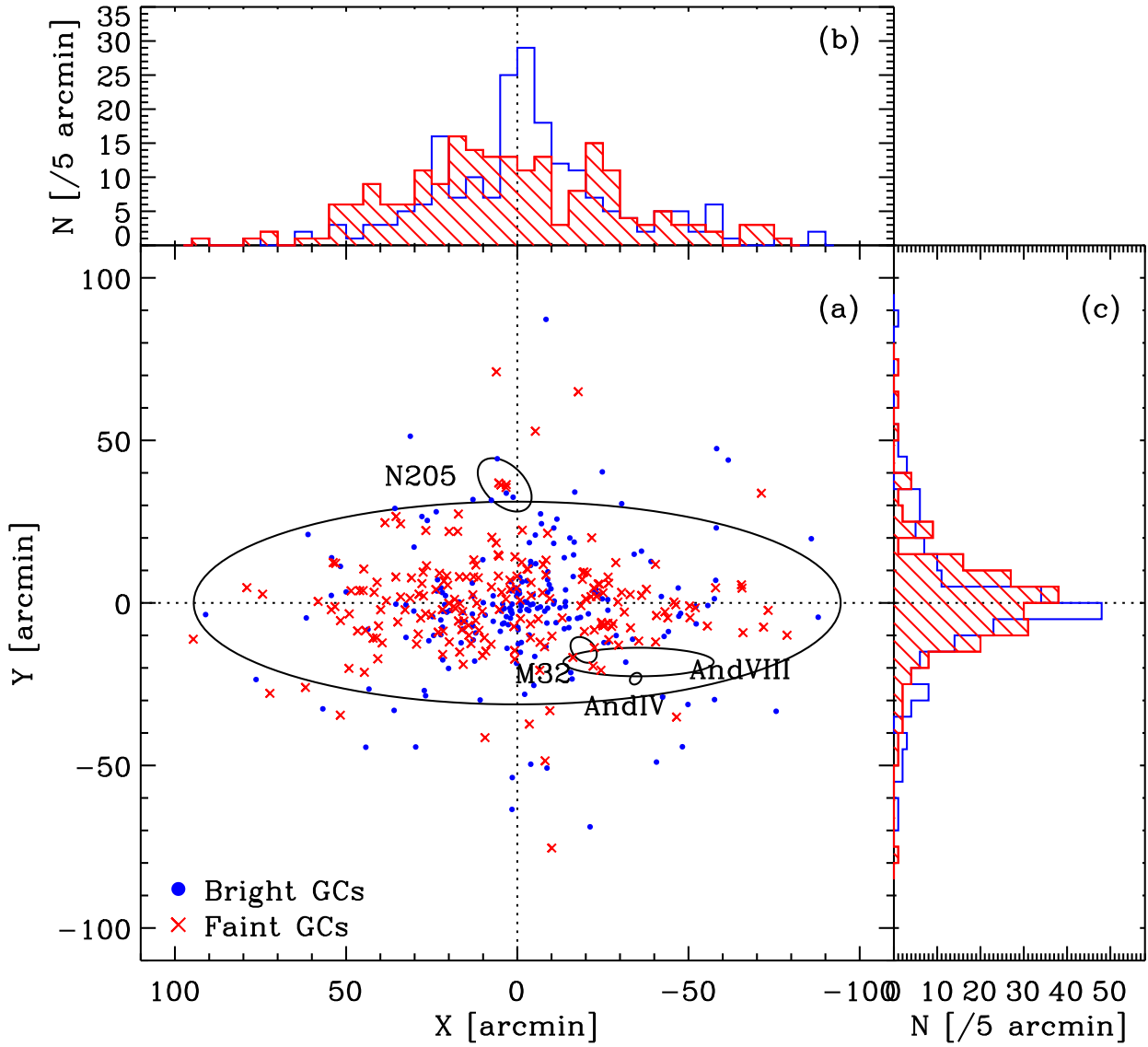}
\caption{Same as Fig. \ref{fig-spathistfeh}, but for the bright
(filled circles and open histogram) and faint (crosses and hatched
histogram) GCs. } \label{fig-spathistmag}
\end{figure}

\begin{figure}
\includegraphics [width=165mm] {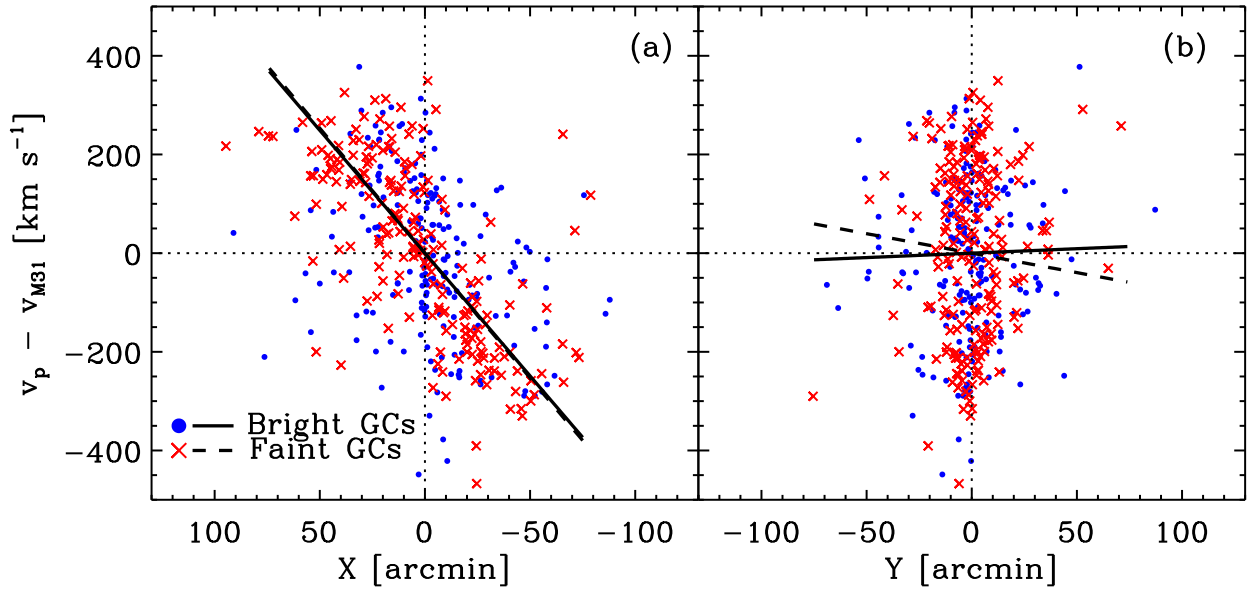}
\caption{Same as Fig. \ref{fig-spatvelfeh}, but for the bright
(filled circles and solid lines) and faint (crosses and dashed
lines) GCs} \label{fig-spatvelmag}
\end{figure}

\begin{figure}
\includegraphics [width=165mm] {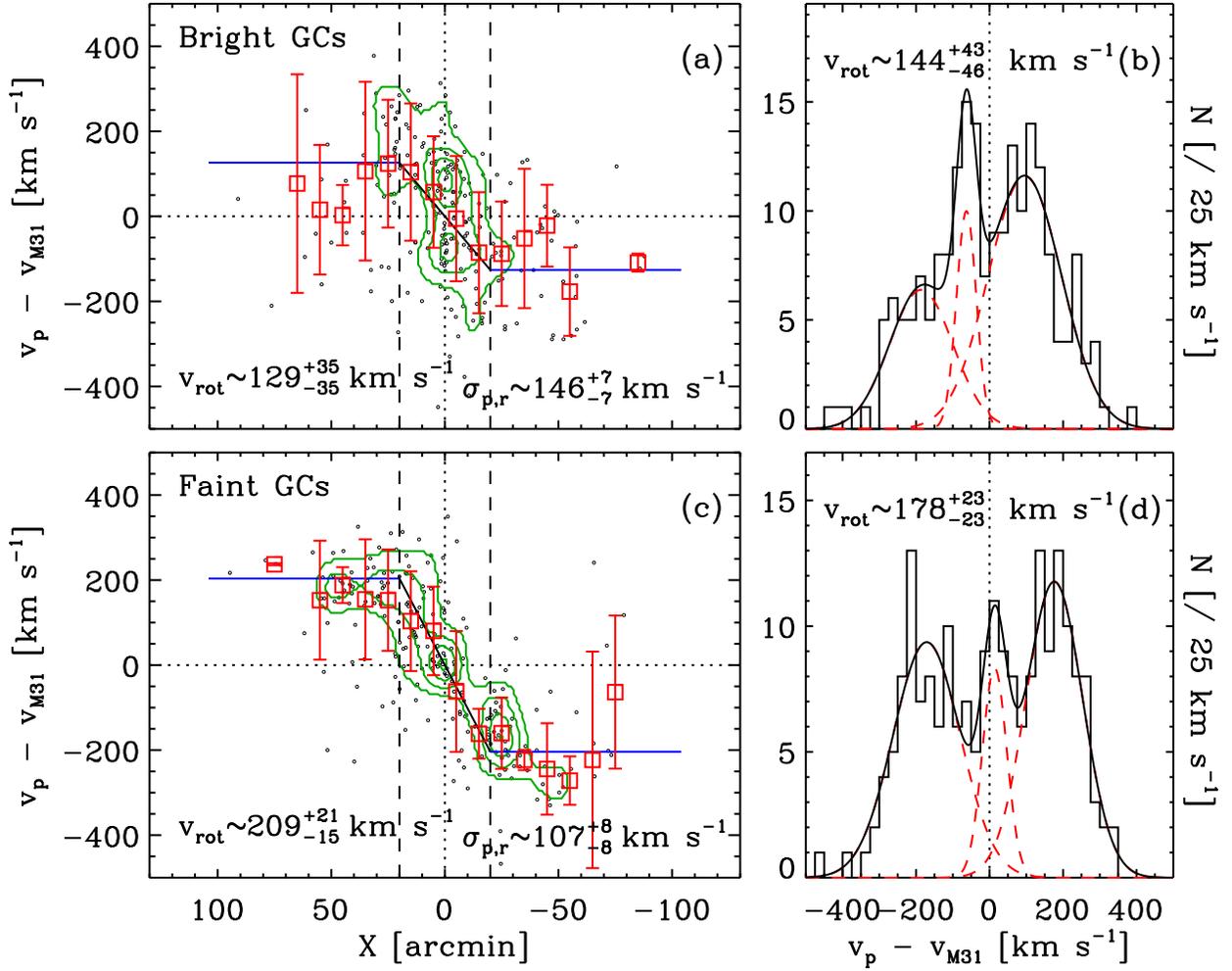}
\caption{Same as Fig. \ref{fig-rotfitfeh}, but for the bright
(a,b) and faint (c,d) GCs} \label{fig-rotfitmag}
\end{figure}

\begin{figure}
\includegraphics [width=165mm] {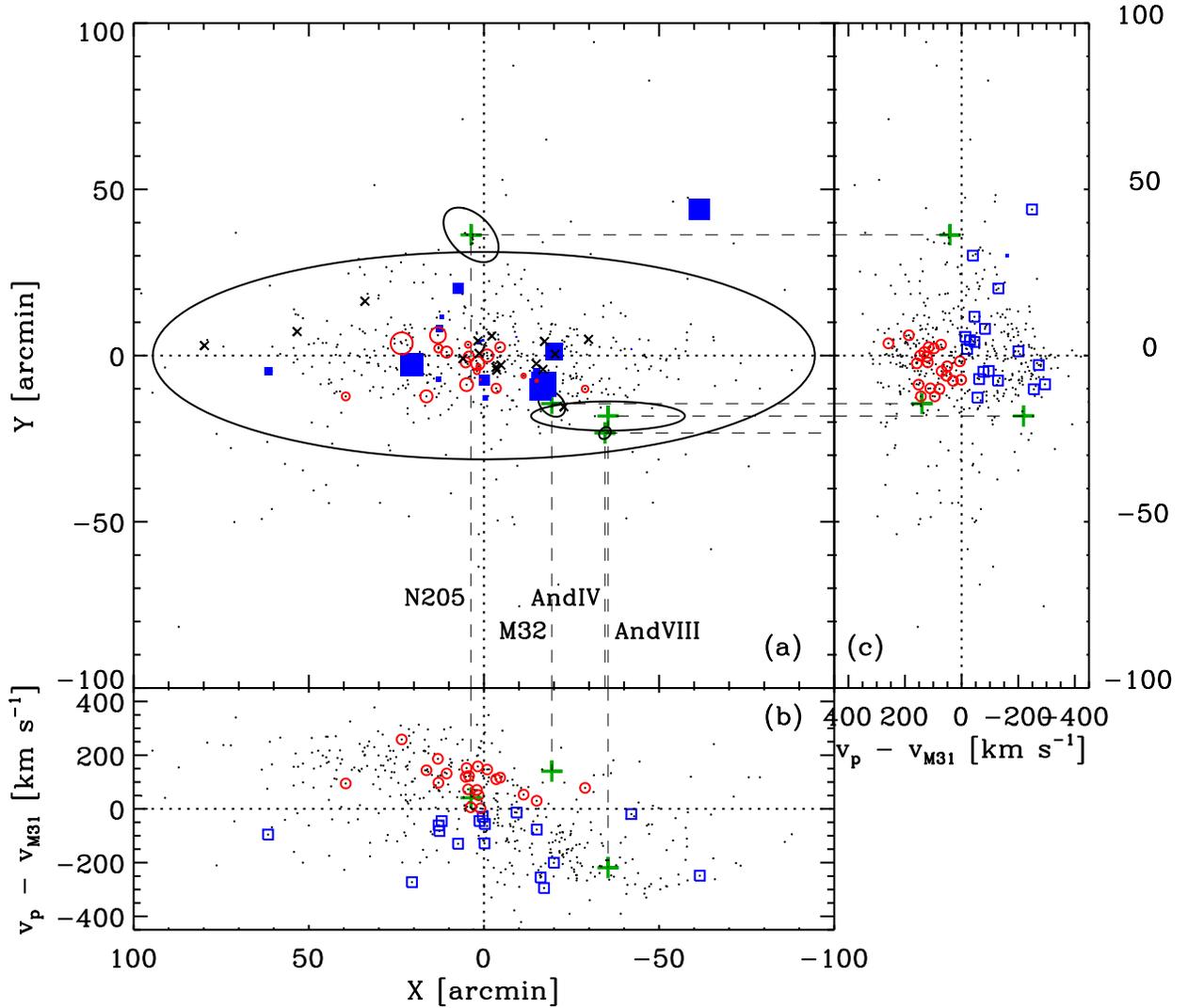}
\caption{Spatial distribution of X-ray emitting GCs (a), and
radial velocities versus the projected radii along the major axis
(b) and along the minor axis (c). We plot 39 genuine GCs with
X-ray detection and measured velocity using variable symbol
sizes according to their velocity deviations from the systemic
velocity of M31 (receding GCs with open circles and approaching
GCs with filled rectangles). One GC and 16 GC candidates with
X-ray detection but no measured velocity are plotted by crosses,
and all 504 GCs by dots. The largest ellipse represents the
optical extent of M31 based on the standard diameter measured
at a level of $25$ mag arcsec$^{-2}$ and ellipticity. Satellite
galaxies of M31 are indicated by thick pluses with small ellipses.
Receding and approaching X-ray emitting GCs with measured
velocity are plotted by open circles and rectangles,
respectively, in (b) and (c). } \label{fig-spatxgc}
\end{figure}

\begin{figure}
\includegraphics [width=165mm] {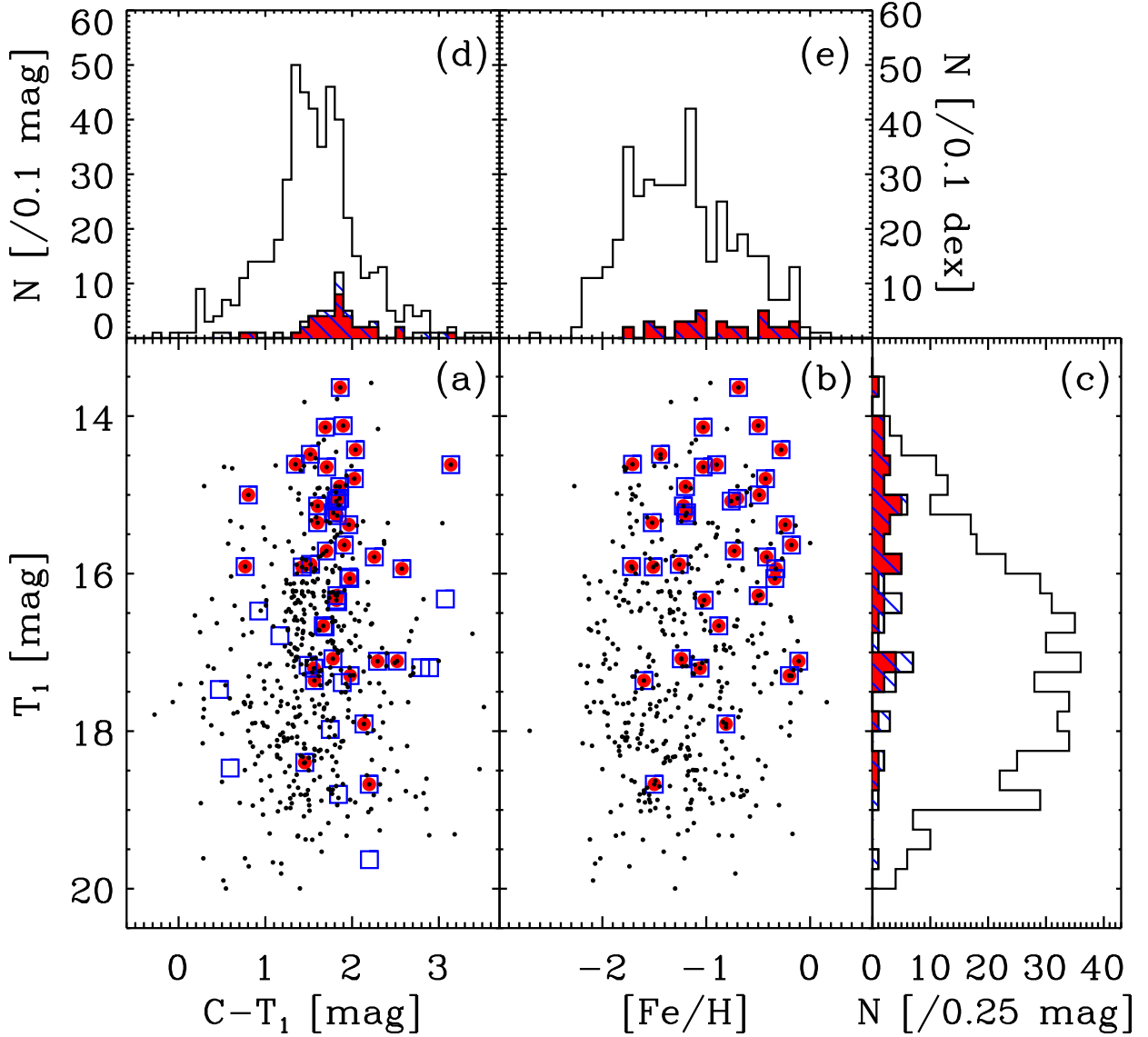}
\caption{$T_1$ magnitudes versus ($C-T_1$) colors (a) and versus
metallicities ([Fe/H]) (b), with histograms for $T_1$ magnitudes
(c), ($C-T_1$) colors (d), and metallicities (e) for X-ray
emitting GCs. All GCs are indicated by small dots, the GCs with
X-ray detection by open rectangles and hatched histogram, and the
GCs with X-ray detection and measured velocity by filled circles and
filled histogram. } \label{fig-photxgc}
\end{figure}

\begin{figure}
\includegraphics [width=165mm] {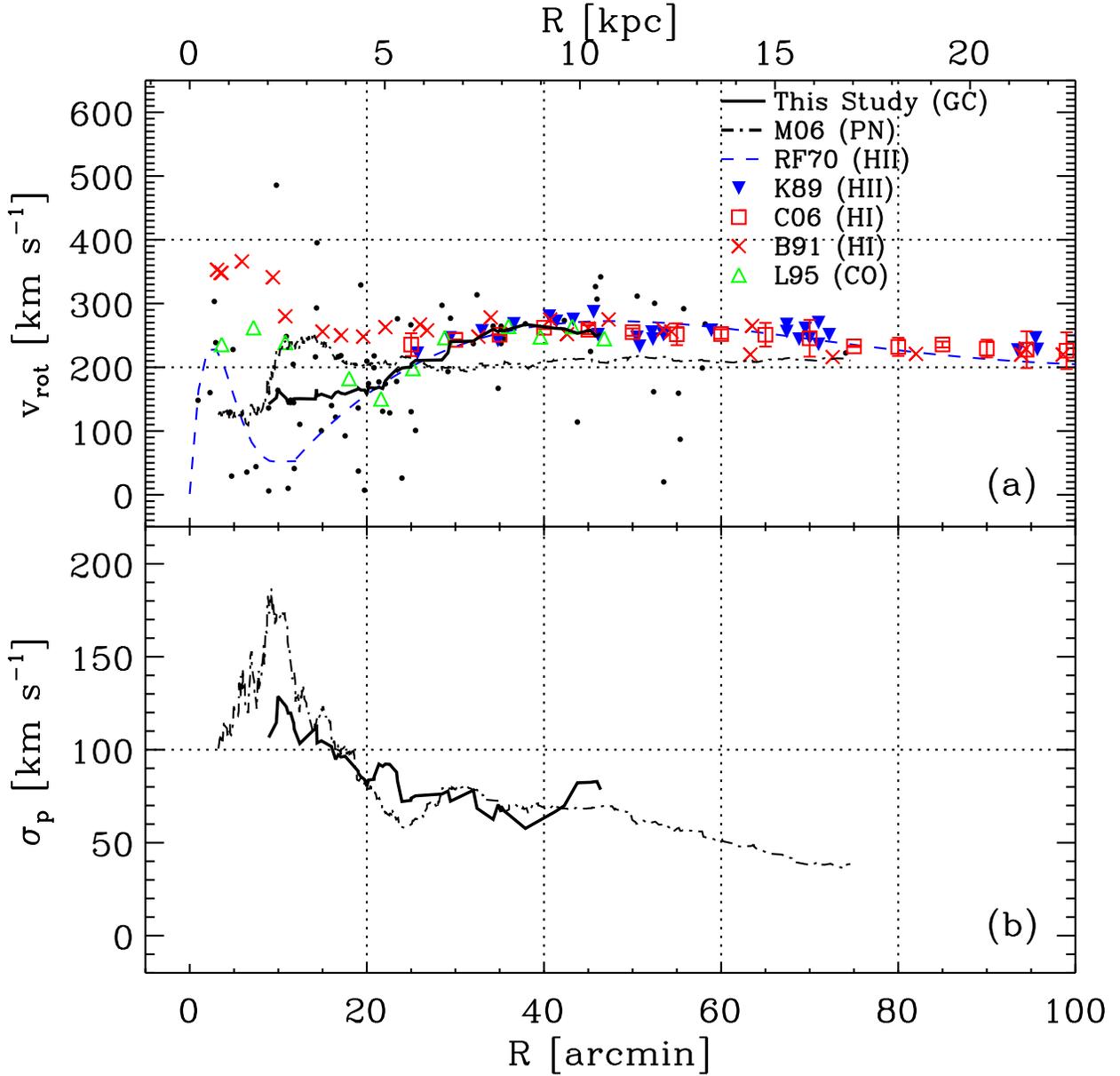}
\caption{Rotation curves (a) and velocity dispersion profiles (b)
derived using the data for the GCs (dots)  in this study (solid lines).
In comparison, we plot the rotation curve
and the dispersion profile derived in this study
using PNe data of \citet{mer06} by dot-dashed lines.
The rotation curves based on other tracers are also plotted:
dashed line and reversed triangles for HII regions \citep{rub70,kent89},
open squares and crosses for HI \citep{bra91,car06}, and
triangles for CO \citep{loi95}.
} \label{fig-rotcurve}
\end{figure}
\clearpage

\begin{figure}
\includegraphics [width=165mm] {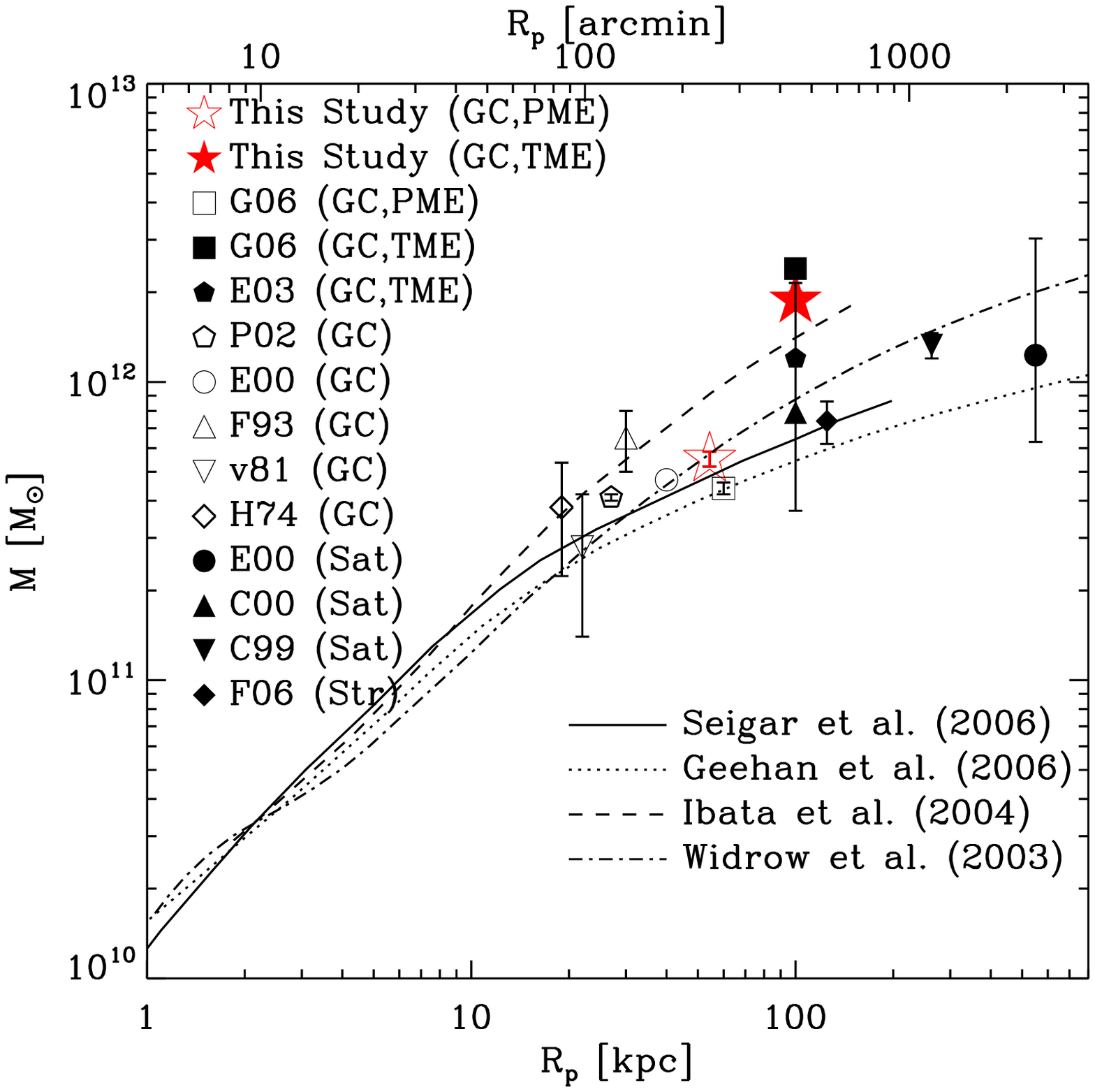}
\caption{Mass estimates for M31.
The masses derived in this study are indicated
by open (PME) and filled (TME) star symbols.
The masses derived using GCs are
annotated as `GC', using satellite galaxies as `Sat',
and using streams as `Str'.
The full references for the mass estimates are as follows:
G06 \citep{gal06}, E03 \citep{eva03}, P02 \citep{per02}, E00 \citep{eva00},
F93 \citep{fed93}, v81 \citep{van81}, H74 \citep{har74}, C00 \citep{cote00},
C99 \citep{cou99}, and F06 \citep{far06a}.
Model mass profiles of M31 are plotted by several lines:
solid line \citep[their model M1 B86]{sei06},
dotted line \citep[their formal best-fitting model]{gee06},
dashed line \citep[their high mass model based on the mass models of \citealt{kly02}]{iba04}, and
dot-dashed line \citep[their modified model A with a NFW halo]{wid03}.
}
\label{fig-mass}
\end{figure}
\clearpage

\begin{figure}
\includegraphics [width=165mm] {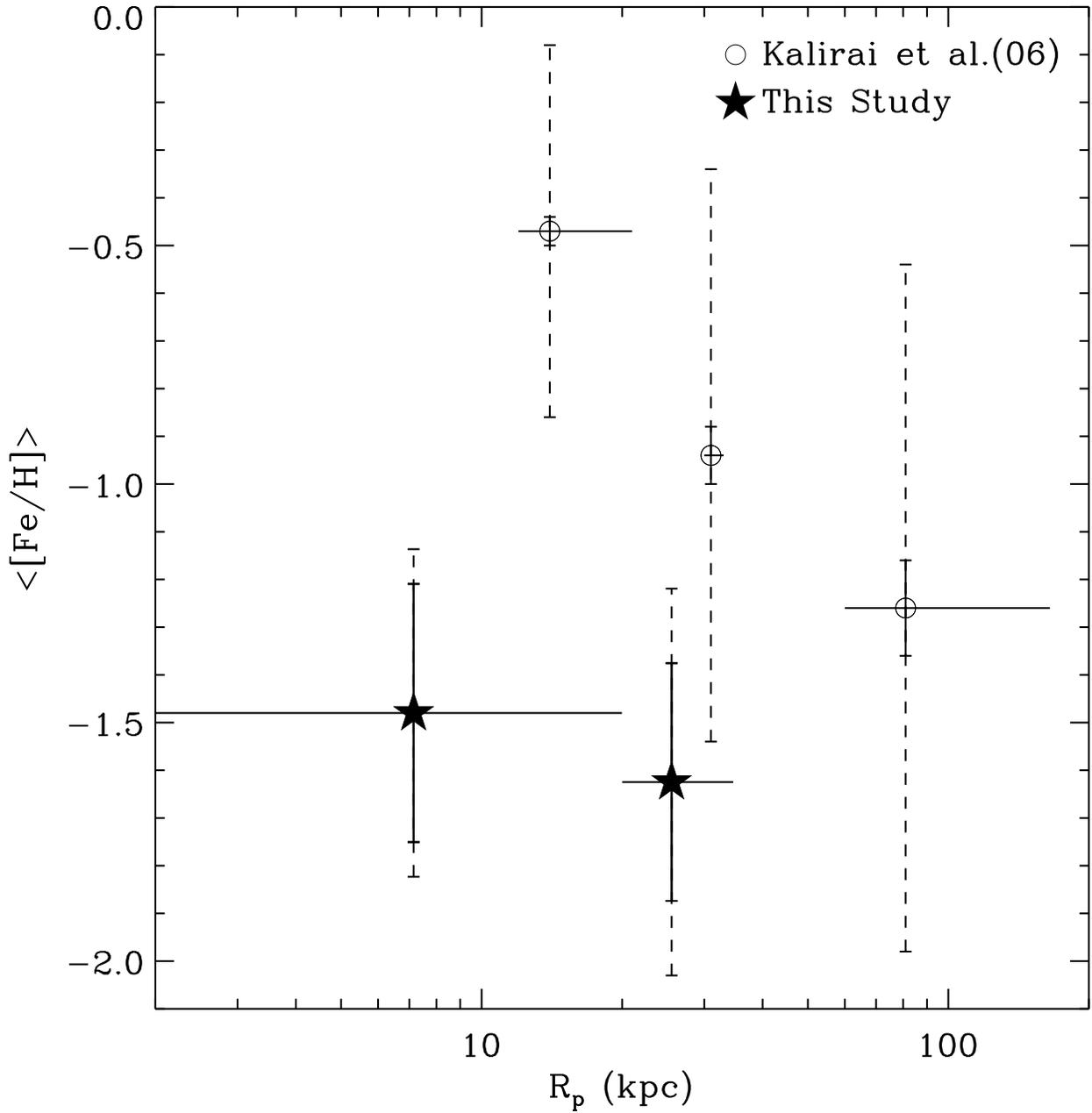}
\caption{Mean metallicity versus the projected galactocentric radius
for the metal-poor GCs (starlets) in comparison
with the red giant branch stars given by \citet{kal06} (circles).
The vertical error bars represent the mean errors (solid line) and dispersion
(dashed line) of the mean metallicity and the horizontal error bars
represent the radial coverage of the data.
}
\label{fig-metal}
\end{figure}
\clearpage

\begin{figure}
\includegraphics [width=165mm] {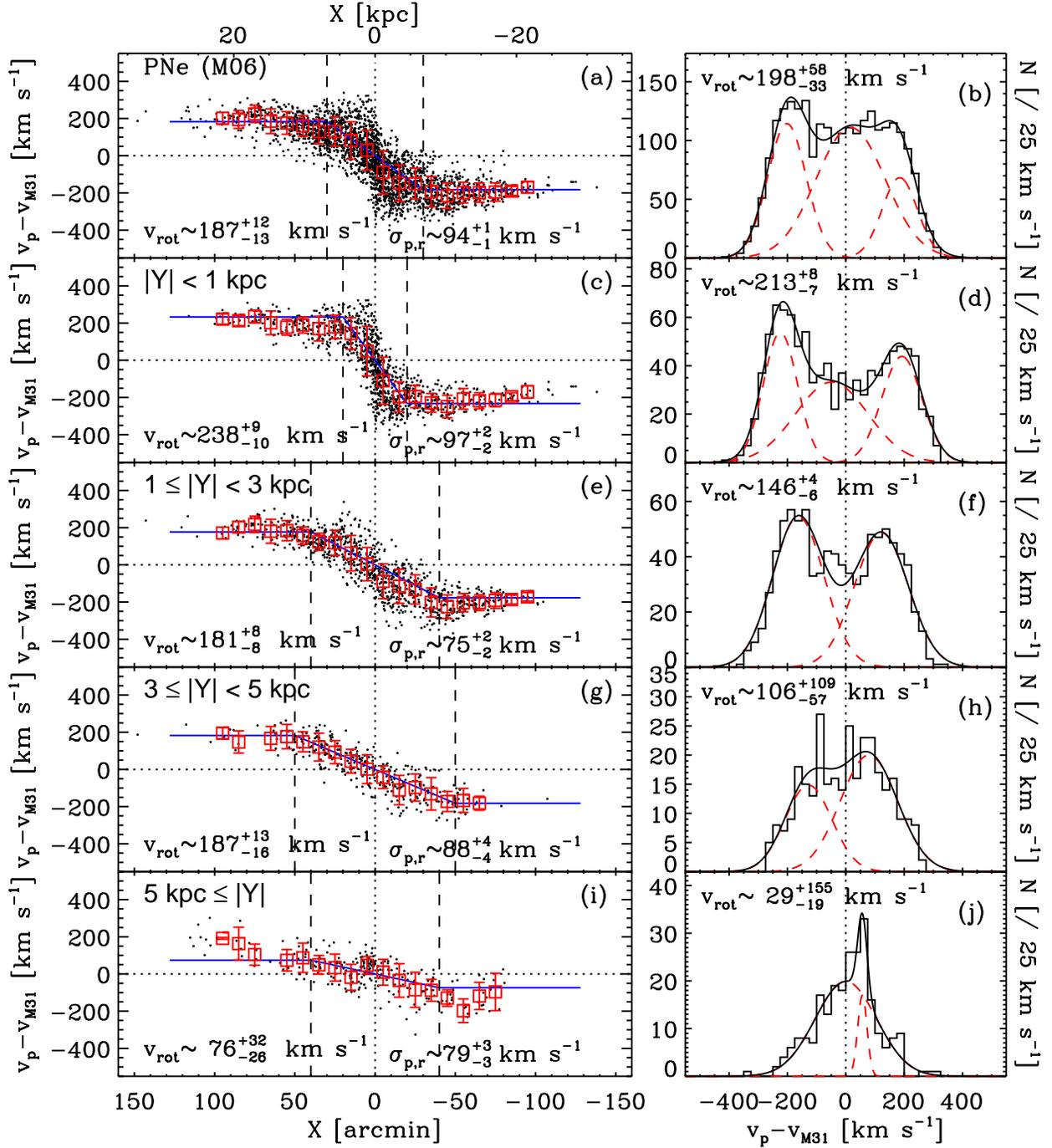}
\caption{Same as Figure 8, but for the PNe based on the data given by \citet{mer06}.
} \label{fig-pnrotfit}
\end{figure}





\begin{thebibliography}{}

\bibitem[Armandroff (1989)]{arm89}Armandroff, T. E. 1989. \aj, 97, 375

\bibitem[Ashman \& Bird (1993)]{ash93} Ashman, K.~M., \& Bird,
C.~M.\ 1993, \aj, 106, 2281

\bibitem[Ashman et al. (1994)]{ash94} Ashman, K.~M., Bird,
C.~M., \& Zepf, S.~E.\ 1994, \aj, 108, 2348

\bibitem[Bahcall \& Tremaine (1981)]{bah81} Bahcall, J.~N., \&
Tremaine, S.\ 1981, \apj, 244, 805

\bibitem[Barmby et al. (2000)]{bar00} Barmby, P., Huchra,
J.~P., Brodie, J.~P., Forbes, D.~A., Schroder, L.~L., \& Grillmair, C.~J.\
2000, \aj, 119, 727

\bibitem[Battistini et al. (1987)]{bat87} 
   Battistini, P., B\`onoli, F., Braccesi, A., Federici, L., Fusi Pecci, F.,
   Marano, B., \& B\"orngen, F. 1987, \aaps, 67, 447

\bibitem[Beasley et al. (2004)]{bea04} Beasley, M.~A., Brodie,
J.~P., Strader, J., Forbes, D.~A., Proctor, R.~N., Barmby, P., \& Huchra,
J.~P.\ 2004, \aj, 128, 1623

\bibitem[Beasley et al. (2005)]{bea05} Beasley, M.~A., Brodie,
J.~P., Strader, J., Forbes, D.~A., Proctor, R.~N., Barmby, P., \& Huchra,
J.~P.\ 2005, \aj, 129, 1412

\bibitem[Beers et al. (1990)]{beers90} Beers, T. C., Flynn, K., \& Gebhardt, K. 1990, \aj, 100, 32

\bibitem[Bellazzini et al. (1995)]{bel95} Bellazzini, M.,
Pasquali, A., Federici, L., Ferraro, F.~R., \& Fusi Pecci, F.~F.\ 1995, \apj, 439, 687

\bibitem[Braun (1991)]{bra91} Braun, R.\ 1991, \apj, 372, 54


\bibitem[Brown et al. (2003)]{bro03} Brown, T.~M., Ferguson,
H.~C., Smith, E., Kimble, R.~A., Sweigart, A.~V., Renzini, A., Rich, R.~M.,
\& VandenBerg, D.~A.\ 2003, \apjl, 592, L17

\bibitem[Brown et al. (2007)]{bro07} Brown, T.~M.,  et al. 2007, \aj, 658, L95

\bibitem[Burstein et al. (2004)]{bur04} Burstein, D., et al.\
2004, \apj, 614, 158

\bibitem[Carignan et al. (2006)]{car06} Carignan, C., Chemin,
L., Huchtmeier, W.~K., \& Lockman, F.~J.\ 2006, \apjl, 641, L109

\bibitem[Carignan et al. (2007)]{car07} Carignan, C., Chemin,
L., \& Foster, T.\ 2007, (astro-ph/0702609)

\bibitem[Chandar et al. (2002)]{cha02} Chandar, R., Bianchi,
L., Ford, H.~C., \& Sarajedini, A.\ 2002, \apj, 564, 712 

\bibitem[Chapman et al. (2006)]{cha06}Chapman, S. C., Ibata, R., Lewis, G. F., Ferguson, M. N.,
 Irwin, M., MaConnachie, A., \& Tanvir, N. 2006, \apj, 653, 255

\bibitem[Chiba \& Beers (2000)]{chi00}Chiba, M. \& Beers, T. 2000, \aj, 119, 2843

\bibitem[C{\^o}t{\'e} (1999)]{cot99}C{\^o}t{\'e}, P. 1999, \aj, 118, 406

\bibitem[C{\^o}t{\'e} et al. (2000)]{cote00} C{\^o}t{\'e}, P.,
Mateo, M., Sargent, W.~L.~W., \& Olszewski, E.~W.\ 2000, \apjl, 537, L91

\bibitem[Courteau \& van den Bergh(1999)]{cou99} Courteau,
S., \& van den Bergh, S.\ 1999, \aj, 118, 337

\bibitem[Crane et al. (1992)]{cra92} Crane, P.~C., Dickel,
J.~R., \& Cowan, J.~J.\ 1992, \apjl, 390, L9

\bibitem[de Vaucouleurs (1958)]{dev58} de Vaucouleurs, G.\
1958, \apj, 128, 465

\bibitem[de Vaucouleurs et al. (1991)]{dev91} de Vaucouleurs, G., de Vaucouleurs, A.,
Corwin, H. G. Jr., Buta, R. J., Paturel, H. G., \& Fouqu\'e, P. 1991, Third Reference Catalog of Bright Galaxies (New York: Springer)

\bibitem[Di Stefano et al. (2002)]{di02} Di Stefano, R.,
Kong, A.~K.~H., Garcia, M.~R., Barmby, P., Greiner, J., Murray, S.~S., \&
Primini, F.~A.\ 2002, \apj, 570, 618
 
\bibitem[Durrell, Harris, \& Pritchet (2001)]{dur01}Durrell, P. R., Harris, W. E., \& Pritchet, C. J. 2001, \aj, 121, 2557 

\bibitem[Eggen, Lynden-Bell, \& Sandage (1962)]{egg62} Eggen, O. J., Lynden-Bell, D. \& Sandage, A. R. 1962, \apj, 136, 929 

\bibitem[Evans \& Wilkinson (2000)]{eva00} Evans, N.~W., \&
Wilkinson, M.~I.\ 2000, \mnras, 316, 929

\bibitem[Evans et al. (2003)]{eva03} Evans, N.~W., Wilkinson,
M.~I., Perrett, K.~M., \& Bridges, T.~J.\ 2003, \apj, 583, 752

\bibitem[Fan et al. (2005)]{fan05} Fan, Z., Ma, J., Zhou, X.,
Chen, J., Jiang, Z., \& Wu, Z.\ 2005, \pasp, 117, 1236

\bibitem[Fardal et al. (2006a)]{far06a} Fardal, M.~A., Babul,
A., Geehan, J.~J., \& Guhathakurta, P.\ 2006a, \mnras, 366, 1012

\bibitem[Fardal et al. (2006b)]{far06b} Fardal, M.~A.,
Guhathakurta, P., Babul, A., \& McConnachie, A.~W.\ 2006b, \mnras,
submitted (astro-ph/0609050)

\bibitem[Federici et al. (1990)]{fed90} Federici, L., Marano,
B., \& Fusi Pecci, F.\ 1990, \aap, 236, 99

\bibitem[Federici et al. (1993)]{fed93} Federici, L., Bonoli,
F., Ciotti, L., Fusi-Pecci, F., Marano, B., Lipovetsky, V.~A., Niezvestny,
S.~I., \& Spassova, N.\ 1993, \aap, 274, 87

\bibitem[Ferguson et al. (2002)]{fer02} Ferguson, A.~M.~N.,
Irwin, M.~J., Ibata, R.~A., Lewis, G.~F., \& Tanvir, N.~R.\ 2002, \aj, 124,
1452

\bibitem[Font et al. (2006a)]{fon06a} Font, A.~S., Johnston,
K.~V., Guhathakurta, P., Majewski, S.~R., \& Rich, R.~M.\ 2006a, \aj, 131, 1436

\bibitem[Font et al. (2006b)]{fon06b} Font, A.~S., Johnston,
K.~V., Bullock, J. S., \& Robertson, B. E.  2006b, \apj, 638, 585

\bibitem[Freedman \& Madore (1990)]{fre90}Freedman, W. L., \& Madore, B. F. 1990,
\apj, 365, 186


\bibitem[Galleti et al.(2004)]{gal04} Galleti, S., Federici,
L., Bellazzini, M.,  Fusi Pecci, F., \& Macrina, S. \ 2004, \aap, 416, 917 

\bibitem[Galleti et al. (2006)]{gal06} Galleti, S., Federici,
L., Bellazzini, M., Buzzoni, A., \& Fusi Pecci, F.\ 2006, \aap, 456, 985

\bibitem[Geehan et al. (2006)]{gee06} Geehan, J.~J., Fardal,
M.~A., Babul, A., \& Guhathakurta, P.\ 2006, \mnras, 366, 996

\bibitem[Gilbert et al. (2006)]{gil06}Gilbert, K. M. et al. 2006, \apj, 652, 1188

\bibitem[Gilbert et al. (2007)]{gil07}Gilbert, K. M. et al. 2007, \apj, in press (astro-ph/0703029)

	
\bibitem[Guhathakurta et al. (2005)]{guh05}Guhathakurta, P., Ostheimer, J. C., Gilbert, K. M., Rich, R. M., Majewski, S. R., Kalirai, J. S., Reitzel, D. B., and Patterson, R. J. 2005,
unpublished (astro-ph/0502366)

\bibitem[Halliday et al. (2006)]{hal06}Halliday, C. et al. 2006, \mnras, 369, 97 

\bibitem[Hammer et al. (2007)]{ham07}Hammer, F., Puech, M., Chemin, L., Flores, H., \& Lehnert, M. D. 2007,
 \apj, 662, 322

\bibitem[Harris (1996)]{har96} Harris, W.~E. 1996, \aj, 112, 1487

\bibitem[Hartwick \& Sargent(1974)]{har74} Hartwick,
F.~D.~A., \& Sargent, W.~L.~W.\ 1974, \apj, 190, 283

\bibitem[Heisler et al.(1985)]{hei85} Heisler, J., Tremaine,
S., \& Bahcall, J.~N.\ 1985, \apj, 298, 8

\bibitem[Hodge (1992)]{hod92}Hodge, P. 1992, The Andromeda Galaxy,
Astrophysics and Space Science Library, Dordrecht:Kluwer

\bibitem[Hubble (1932)]{hub32}Hubble, E. P. 1932, \apj, 76, 44

\bibitem[Huchra et al. (1982)]{huc82} Huchra, J., Stauffer,
J., \& van Speybroeck, L.\ 1982, \apjl, 259, L57

\bibitem[Huchra et al. (1991)]{huc91} Huchra, J.~P., Brodie,
J.~P., \& Kent, S.~M.\ 1991, \apj, 370, 495

\bibitem[Hurley-Keller et al. (2004)]{hur04} Hurley-Keller, D., 
Morrison, H. L., Harding, P., \& Jacoby, G. H. 2004, \apj, 616, 804

\bibitem[Huxor et al. (2005)]{hux05} Huxor, A.~P., Tanvir,
N.~R., Irwin, M.~J., Ibata, R., Collett, J.~L., Ferguson, A.~M.~N.,
Bridges, T., \& Lewis, G.~F.\ 2005, \mnras, 360, 1007

\bibitem[Ibata et al. (2001)]{iba01} Ibata, R., Irwin, M.,
Lewis, G., Ferguson, A.~M.~N., \& Tanvir, N.\ 2001, \nat, 412, 49

\bibitem[Ibata et al. (2004)]{iba04} Ibata, R., Chapman, S.,
Ferguson, A.~M.~N., Irwin, M., Lewis, G., \& McConnachie, A.\ 2004, \mnras,
351, 117

\bibitem[Ibata et al. (2005)]{iba05} Ibata, R., Chapman, S.,
Ferguson, A.~M.~N., Lewis, G., Irwin, M., \& Tanvir, N. \ 2005, \apj,
634, 287

\bibitem[Ibata et al. (2007)]{iba07} Ibata, R., Martin, N. F., Irwin, M.,
Chapman, S., Ferguson, A.~M.~N.,  Lewis, G., \& McConnachie, A.\ 2007, in press
(astro-ph/0704.1318) 


\bibitem[Irwin et al. (2005)]{irw05}Irwin, M. J., Ferguson, A. M. N., Ibata, R. A., Lewis, G. F., \& Tanvir, N. R. 2005, \apj, 628, L105 
 
\bibitem[Jablonka et al. (1998)]{jab98} Jablonka, P., Bica,
E., Bonatto, C., Bridges, T.~J., Langlois, M., \& Carter, D.\ 1998, \aap,
335, 867

\bibitem[Kaaret (2002)]{kaa02} Kaaret, P.\ 2002, \apj, 578, 114

\bibitem[Kalirai et al. (2006)]{kal06} Kalirai, J. S., Gilbert, K. M., Guhathakurta, P., Majewski, S. R.,
 Ostheimer, J. C., Rich, R. M., Cooper, M. C., Reitzel, D. B., \& Patterson, R. J. 2006, \apj, 648, 389

\bibitem[Karachentsev et al. (2004)]{kar04} Karachentsev,
I.~D., Karachentseva, V.~E., Huchtmeier, W.~K., \& Makarov, D.~I.\ 2004,
\aj, 127, 2031

\bibitem[Kent (1989)]{kent89} Kent, S.~M.\ 1989, \pasp, 101, 489

\bibitem[Kim et al. (2006)]{kim06} Kim, E., Kim, D.-W.,
Fabbiano, G., Lee, M.~G., Park, H.~S., Geisler, D., \& Dirsch, B.\ 2006,
\apj, 647, 276

\bibitem[Kim et al. (2007)]{kim07} Kim, S. C., Lee, M. G., Geisler, D., Sarajedini, A.,
Park, H. S., Hwang, H. S., Harris, W. E., Seguel, J. C., \& von Hippel, T. \
 2007, \aj, 134, 706 (Paper I)

\bibitem[Klypin et al. (2002)]{kly02} Klypin, A., Zhao, H., \&
Somerville, R.~S.\ 2002, \apj, 573, 597

\bibitem[Kong et al. (2002)]{kong02} Kong, A.~K.~H., Garcia,
M.~R., Primini, F.~A., Murray, S.~S., Di Stefano, R., \& McClintock, J.~E.\
2002, \apj, 577, 738

\bibitem[Lee et al. (1993)]{lee93} Lee, M.~G., Freedman, W., 
Mateo, M., Thompson, I., Roth, M., \& Ruiz, M.-T.\ 1993, \aj, 106, 1420 
 

\bibitem[Loinard et al. (1995)]{loi95} Loinard, L., Allen,
R.~J., \& Lequeux, J.\ 1995, \aap, 301, 68

\bibitem[Mackey et al. (2007)]{mac07} Mackey, A.~D., et al.\
2007, \apjl, 655, L85

\bibitem[Majewski et al. (2007)]{maj07} Majewski, S.~R., et
al.\ 2007, \apjl, submitted (astro-ph/0702635)

\bibitem[McConnachie et al. (2004)]{mcc04} McConnachie, A.~W.,
Irwin, M.~J., Lewis, G.~F., Ibata, R.~A., Chapman, S.~C., Ferguson,
A.~M.~N., \& Tanvir, N.~R.\ 2004, \mnras, 351, L94

\bibitem[McConnachie et al. (2005)]{mcc05} McConnachie, A.~W.,
Irwin, M.~J., Ferguson, A.~M.~N., Ibata, R.~A., Lewis, G.~F., \& Tanvir,
N.\ 2005, \mnras, 356, 979

\bibitem[Merrett et al. (2003)]{mer03} Merrett, H.~R., et al.\
2003, \mnras, 346, L62

\bibitem[Merrett et al. (2006)]{mer06} Merrett, H.~R., et al.\
2006, \mnras, 369, 120


\bibitem[Morrison et al. (2003)]{mor03} Morrison, H.~L.,
Harding, P., Hurley-Keller, D., \& Jacoby, G.\ 2003, \apjl, 596, L183

\bibitem[Morrison et al. (2004)]{mor04} Morrison, H.~L.,
Harding, P., Perrett, K., \& Hurley-Keller, D.\ 2004, \apj, 603, 87
\bibitem[Mould \& Kristian (1986)]{mou86}Mould, J. , \& Kristian, J. 1986, \apj, 305, 591

\bibitem[Perrett et al. (2002)]{per02} Perrett, K.~M.,
Bridges, T.~J., Hanes, D.~A., Irwin, M.~J., Brodie, J.~P., Carter, D.,
Huchra, J.~P., \& Watson, F.~G.\ 2002, \aj, 123, 2490

\bibitem[Perrett et al. (2003)]{per03} Perrett, K.~M., Stiff,
D.~A., Hanes, D.~A., \& Bridges, T.~J.\ 2003, \apj, 589, 790 

\bibitem[Pietsch et al. (2005)]{pie05} Pietsch, W., Freyberg,
M., \& Haberl, F.\ 2005, \aap, 434, 483

\bibitem[Pritchet \& van den Bergh (1994)]{pri94}Pritchet, C. J., \& van den Bergh, S. 1994, \aj, 107, 1730

\bibitem[Puzia et al. (2005)]{puz05} Puzia, T.~H., Perrett,
K.~M., \& Bridges, T.~J.\ 2005, \aap, 434, 909

\bibitem[Renda et al. (2005a)]{ren05a}Renda, A., Kawata, D., Fenner, Y., \& Gibson, B. 2005a, \mnras, 356, 1071

\bibitem[Renda et al. (2005b)]{ren05b}Renda, A., Gibson, B., Mouhcine, M., Ibata, R. A., Kawata, D., 
Flynn, C., \& Brook, C. B. 2005b, \mnras, 363, 16L

\bibitem[Richstone \& Tremaine (1984)]{ric84} Richstone,
D.~O., \& Tremaine, S.\ 1984, \apj, 286, 27

\bibitem[Rubin \& Ford (1970)]{rub70} Rubin, V.~C., \& Ford,
W.~K.~J.\ 1970, \apj, 159, 379

\bibitem[Sarazin et al. (2003)]{sar03} Sarazin, C.~L., Kundu,
A., Irwin, J.~A., Sivakoff, G.~R., Blanton, E.~L., \& Randall, S.~W.\ 2003,
\apj, 595, 743

\bibitem[Sargent et al. (1977)]{sar77} 
   Sargent, W. L. W., Kowal, C. T., Hartwick, F. D. A., \& van den Bergh, S.
   1977, \aj, 82, 947

\bibitem[Searle \& Zinn (1978)]{sea78}Searle, L., \& Zinn, R. 1978, \apj, 225, 357 

\bibitem[Seigar et al. (2006)]{sei06} Seigar, M.~S., Barth,
A.~J., \& Bullock, J.~S.\ 2006, \apj, submitted (astro-ph/0612228)

\bibitem[Sivakoff et al. (2006)]{siv06} Sivakoff, G.~R., et
al.\ 2006, \apj, 669, 1246 

\bibitem[Sofue \& Rubin (2001)]{sof01} Sofue, Y., \& Rubin,
V.\ 2001, \araa, 39, 137

\bibitem[Supper et al. (2001)]{sup01} Supper, R., Hasinger,
G., Lewin, W.~H.~G., Magnier, E.~A., van Paradijs, J., Pietsch, W., Read,
A.~M., \& Tr{\"u}mper, J.\ 2001, \aap, 373, 63

\bibitem[Trudolyubov \& Priedhorsky (2004)]{tru04}
Trudolyubov, S., \& Priedhorsky, W.\ 2004, \apj, 616, 821

\bibitem[Trudolyubov et al. (2005)]{tru05} Trudolyubov, S.,
Kotov, O., Priedhorsky, W., Cordova, F., \& Mason, K.\ 2005, \apj, 634, 314

\bibitem[Trudolyubov et al. (2006)]{tru06} Trudolyubov, S.,
Priedhorsky, W., \& Cordova, F.\ 2006, \apj, 645, 277 

\bibitem[van den Bergh (1969)]{van69} van den Bergh, S.\ 1969,
\apjs, 19, 145

\bibitem[van den Bergh (1981)]{van81} van den Bergh, S.\ 1981,
\pasp, 93, 428

\bibitem[van den Bergh (2000)]{van00} van den Bergh, S.\ 2000,
The Galaxies of the Local Group, Cambridge Universe Press 

\bibitem[Walterbos \& Kennicutt (1987)]{wal87}Walterbos, R. , \& Kennicutt, R. 1987, \aaps, 69, 311

\bibitem[White \& Rees (1978)]{whi78}White, S. D. M. , \& Rees, M. J. 1978, \mnras, 183, 341 

\bibitem[Widrow et al. (2003)]{wid03} Widrow, L.~M., Perrett,
K.~M., \& Suyu, S.~H.\ 2003, \apj, 588, 311

\bibitem[Williams et al. (2004)]{wil04} Williams, B.~F.,
Garcia, M.~R., Kong, A.~K.~H., Primini, F.~A., King, A.~R., Di Stefano, R.,
\& Murray, S.~S.\ 2004, \apj, 609, 735

\end{thebibliography}
\end{document}